\documentstyle[eqsecnum,prd,aps,epsf]{revtex}
\def\beq{\begin{equation}}
\def\eeq{\end{equation}}
\def\beqa{\begin{eqnarray}}
\def\eeqa{\end{eqnarray}}
\def\bI{\hbox{$\,I\!\!\!\!-$}}
\def\a{\alpha}
\def\b{\beta}
\def\k{\kappa}
\def\p{\partial}
\def\e{\epsilon}

\def\r{\rho}
\def\vp{\varphi}
\def\O{\Omega}
\def\d{\delta}
\def\D{\Delta}
\def\th{\theta}
\def\Th{\Theta}
\def\t{\tilde}

\def\b{\bar}
\def\n{\nabla}
\def\ra{\rightarrow}
\def\one{\hbox{${}^{(1)}\!$}}
\def\two{\hbox{${}^{(2)}\!$}}
\def\three{\hbox{${}^{(3)}\!$}}
\def\four{\hbox{${}^{(4)}\!$}}
\def\five{\hbox{${}^{(5)}\!$}}
\def\six{\hbox{${}^{(6)}\!$}}
\def\bxi{\mbox{\boldmath $\xi$}}
\def\br{\mbox{\boldmath $r$}}
\def\bu{\mbox{\boldmath $u$}}
\def\bv{\mbox{\boldmath $v$}}
\def\bx{\mbox{\boldmath $x$}}
\def\bX{\mbox{\boldmath $X$}}
\def\bth{\mbox{\boldmath $\th$}}
\def\bvp{\mbox{\boldmath $\varphi$}}
\def\bJ{\mbox{\boldmath $J$}}
\def\bR{\mbox{\boldmath $R$}}

\begin{document}
\draft
%\preprint{YITP--99--??}

\title{Equilibrium sequences of irrotational binary polytropic stars : \\
The case of double polytropic stars}

\author{Keisuke Taniguchi$^{*}$
%\footnote{e-mail: Keisuke.Taniguchi@obspm.fr}
 and Takashi Nakamura$^{**}$
%\footnote{e-mail: takashi@yukawa.kyoto-u.ac.jp}
}
\address{${}^{*}$D\'epartement d'Astrophysique Relativiste
  et de Cosmologie, UPR 8629 du C.N.R.S., Obserbatoire de Paris,
  F-92195 Meudon Cedex, France \\
  ${}^{**}$Yukawa Institute for Theoretical Physics,
  Kyoto University, Kyoto 606-8502, Japan}

\date{\today}
%\date{March, 2000}

\maketitle
\begin{abstract}

Solutions to equilibrium sequences of irrotational binary polytropic
stars in Newtonian gravity are expanded in a power of $\e=a_0/R$,
where $R$ and $a_0$ are the orbital separation of the binary system
and the radius of each star for $R=\infty$. For each order of $\e$, we
should solve ordinary differential equations for arbitrary polytropic
indices $n$. We show solutions for polytropic indices $n= 0.5, 1, 1.5$
and $2$ up to $\e^6$ orders. Our semi-analytic solutions can be used
to check the validity of numerical solutions.

\end{abstract}

\pacs{PACS number(s): 04.30.Db, 04.25.Dm, 97.80.Fk}

%%%%%%%%%%%%%%%%%%%%%%
\section{Introduction}
%%%%%%%%%%%%%%%%%%%%%%

Coalescing binary neutron stars (BNSs) are considered to be one of the
most promising sources of gravitational waves for laser
interferometers such as TAMA300\cite{TAMA300}, GEO600\cite{GEO600},
VIRGO\cite{VIRGO} and LIGO\cite{LIGO,Thorne94}. If we have accurate
theoretical templates of inspiraling phase of BNSs, we can determine
the mass and the spin of neutron stars from the gravitational wave
signals in the inspiraling phase\cite{Thorne95}. We may also extract
the informations on the equation of state of a neutron star from the
signals in the premerging phase\cite{Lindblom92}. For this purpose it
is important to complete theoretical templates of gravitational waves
in the premerging phase as well as in the inspiraling
phase. Moreover, the study of BNSs in the premerging phase is
motivated by the need to provide the realistic initial condition for
the merging phase simulations\cite{Shibata99,SU99,ON99}.

In order to obtain accurate theoretical templates of gravitational
waves from BNSs around the premerging phase\cite{Rasio99}, a general
relativistic hydrostatic problem with compressible equation of state
must be numerically solved. In solving this problem, we suppose that
BNSs reach quasi-equilibrium states because the timescale of the
orbital decay driven by the radiation reaction is much longer than the
orbital period of BNSs until their innermost stable circular orbit
(ISCO). (The formalism for solving quasi-equilibrium figures of
irrotational binary neutron stars in general relativity is given in
the references \cite{BGM97,Asada98,Shibata98,Teukolsky98}.)
Furthermore, we regard the internal state of a neutron star as an
irrotational or nearly irrotational one because the viscosity of a
neutron star is not large enough for synchronization even near the
ISCO\cite{Kochanek92,BC92}.

Recently Bonazzola, Gourgoulhon and Marck have numerically studied the
irrotational binary neutron stars in general relativity with the
conformally flat condition\cite{BGM99a,BGM99b,BGGM99}. More recently,
Ury\=u and Eriguchi have also numerically solved the same problem as
that of Bonazzola, Gourgoulhon and Marck\cite{UE99}. In order to check
the validity of these numerical calculations, it is necessary to
compare them with analytic or semi-analytic ones. However, we have not
obtained such analytic solutions in general relativity yet. There are
solutions of irrotational BNSs only in the first post-Newtonian (1PN)
approximation of general relativity. For example, Lombardi, Rasio and
Shapiro have semi-analytically studied irrotational BNSs by using the
energy variational method\cite{Lombardi97}, and one of the authors of
the present paper (KT) has analytically calculated equilibrium
sequences of irrotational BNSs by using the tensor virial
method\cite{Taniguchi99}. The former one restricts the internal motion
within the plane orthogonal to the rotational axis assuming the shape
of the star to an ellipsoidal one in order to treat compressible
equation of state. While in the latter one the velocity component
along the orbital axis is included at 1PN order and the shape of the
star at 1PN order is not restricted to an ellipsoidal one although the
fluid is incompressible.

The situation is also the same in Newtonian gravity, that is, even in
Newtonian gravity useful analytic or semi-analytic solutions for
equilibrium sequences of irrotational binary systems are not obtained.
Let us consider numerically constructed stationary structures of
irrotational binary stars computed by Ury\=u and Eriguchi in Newtonian
gravity\cite{UE98a,UE98b}. One may think that semi-analytic solutions
by Lai, Rasio and Shapiro\cite{LRS94} may be used to check the
validity of numerical solutions. However, in numerical solutions of
Ury\=u and Eriguchi the velocity component along the orbital axis
exists while in those of Lai, Rasio and Shapiro such a component is
assumed to be zero from the beginning. Therefore new analytic or
semi-analytic solutions are needed to check numerical solutions even
in Newtonian gravity. Such a check of numerical solutions is
extremely important because in the numerical calculation, there is a
possibility to obtain another solution although the binding energy of
a binary neutron star is almost the same value, and to lead a
different conclusion\cite{MMW99,MW99}.

In the previous paper\cite{TN99}, we showed such a new, almost
analytic solution to an equilibrium of irrotational binary polytropic
stars for the polytropic index $n=1$ in Newtonian gravity by expanding
all physical quantities in a power of $\e \equiv a_0/R$, where $R$ and
$a_0$ are the orbital separation of the binary system and the radius
of each star for $R=\infty$. In that paper, we have extended the
method developed by Chandrasekhar more than 65 years ago for
corotating fluids\cite{Ch33,Kovetz68} to the one for irrotational
fluids. In this paper, we show semi-analytic solutions for arbitrary
polytropic indices by numerically solving ordinary differential
equations.

This paper is organized as follows. In \S 2, we formulate the method
to solve the irrotational binary polytropic stars. In \S 3, the
physical values, i.e., the central density of a star, the orbital
angular velocity, the total energy and total angular momentum of the
binary system are calculated. The numerical results are presented in
\S 4. Section 5 is devoted to summary and discussions.

Although a binary system consists of two stars, we pay particular
attention to one of two stars. We call it star 1 whose mass is $M_1$
and the companion one star 2 whose mass is $M_2$. In this paper, we
adopt two corotating coordinate systems. First one is a Cartesian
coordinate system $\bX$ whose origin is located at the center of mass
of the binary system. For calculational convenience, we choose the
orbital axis as $X_3$, and we take the direction of $X_1$ from the
center of mass of star 2 to that of star 1. The second coordinate
system is the spherical one $\br =(r,\th,\vp)$ whose origin is located
at the center of mass of star 1. We use units of $G=1$.

%%%%%%%%%%%%%%%%%%%%%
\section{Formulation}
%%%%%%%%%%%%%%%%%%%%%

Since we treat irrotational fluids in Newtonian gravity, the basic
equations for obtaining equilibrium configurations of binary systems
are the equation of state, the Euler equation, the equation of
continuity and the Poisson equation:
\beqa
  &&P=K \r^{1 +{1 \over n}}, \\
  &&\n \biggl[ \int {dP \over \r} -U +{1 \over 2} v^2 -\bv \cdot
  ({\bf \O} \times \br) \biggr] =0, \label{Eq:Euler} \\
  &&\n \cdot \bv =-\bigl( \bv -{\bf \O} \times \br \bigr)
  \cdot {\n \r \over \r}, \label{Eq:continuity} \\
  &&\D U =-4 \pi \r, \label{Eq:Poisson}
\eeqa
where $P$, $\r$, $n$, $U$ and $\O$ are the pressure, the density, the
polytropic index, the gravitational potential and the orbital angular
velocity, respectively. $K$ is a constant related to entropy and $\bv$
represents the velocity field in the inertial frame. The gravitational
potential $U$ is separated into two parts, i.e., the contribution from
star 1 to itself $U^{1 \ra 1}$ and that from star 2 to star 1 $U^{2
\ra 1}$. We can also express the gravitational potential contributed
from star 1 to star 2 as $U^{1 \ra 2}$. These gravitational potentials
are written as
\beqa
  U^{2 \ra 1} &=&{M_2 \over R} \sum_{l=0}^{\infty} (-1)^l \Bigl( {r
    \over R} \Bigr)^l P_l (\sin \th \cos \vp) \nonumber \\
  &&+{3 \bI_{11}' \over 2R^3} \Bigl[ 1 -3\Bigl( {r \over R}
  \Bigr) P_1 (\sin \th \cos \vp) +O(R^{-2}) \Bigr]
  +{\rm higher~order~terms}, \label{U21} \\
%%%%%
  U^{1 \ra 2} &=&{M_1 \over R} \sum_{l=0}^{\infty} \Bigl( {r' \over R} 
  \Bigr)^l P_l (\sin \th' \cos \vp') \nonumber \\
  &&+{3\bI_{11} \over 2R^3} \Bigl[ 1 + 3\Bigl( {r' \over R} \Bigr)
  P_1 (\sin \th' \cos \vp') +O(R^{-2}) \Bigr] +{\rm higher~order~terms},
\eeqa
where the superscript $'$ means the term concerned with star 2 and
$P_l$ denotes the Legendre function of order $l$. $\bI_{11}$ is the
reduced quadrupole moment defined by
\beqa
  \bI_{11} &=&{1 \over 3} \bigl( 2I_{11} -I_{22} -I_{33} \bigr), \nonumber
  \\
  &=&{2 \over 3} \int_{star~1} d^3 x \r r^2 P_2 (\sin \th \cos \vp).
\eeqa
In the irrotational fluid case, we can express $\bv$
as a gradient of a scalar function $\Phi$, i.e.,
\beq
  \bv =\n \Phi.
\eeq

Following the {\it Lane-Emden equation} we first express the density
as
\beq
  \r =\r_c \Th^n(\xi, \th, \vp)
\eeq
with $\xi=r/\a$.
Here the definitions of $\a$ and $\a'$ are
\beqa
  \a &\equiv& \biggl[ {K(1+n) \r_c^{{1 \over n} -1} \over 4\pi}
  \biggr]^{1/2}, \\
  \a' &\equiv& \biggl[ {K'(1+n') {\r_c'}^{{1 \over n'} -1} \over 4\pi}
  \biggr]^{1/2},
\eeqa
with $\r_c$ being the central density.
Let us define $\e \equiv a_0/R$, where $R$ and $a_0$ are the orbital
separation of the binary system and the radius of star 1 for
$R=\infty$\cite{TN99}. The radius of a spherical star $a_0$ is given
by $a_0=\a \xi_1$ as usual. Also we define $\e' \equiv a_0'/R$ for
star 2. Note that in the case of Newtonian gravity, we have two
degrees of freedom for describing the differences of mass, i.e.,
$\r_c$ and $\a$. Therefore, we can express the differences of mass by
changing only $\r_c$ with fixing $\a$. In this case, we can make the
radius of each star and each expansion parameter coincide. Although we
have adopted such a situation in Ref. \cite{TN99}, in the present
paper we express equations concerned with star 2 by using $\e'$, $\a'$
and so on.

We expand $\Th$ in a power series of a parameter $\e$ as\footnote{Note
that we express the Lane-Emden function for star 2 as $\b{\Th}$ and
expand it in a power series of a parameter $\e'$ as $\b{\Th}=\sum_i
{\e'}^i \b{\Th}_i$.}
\beq
  \Th=\sum_{i=0}^{\infty} \e^i \Th_i.
\eeq
Since the shape of star 1 is
spherical when $R$ is large, the lowest order term of $\Th$ is the
solution of the {\it Lane-Emden equation}. Then we expand $\Th_i$ by
spherical harmonics as
\beq
  \Th_i =\sum_{l,m} {}^{(i)}\! \psi_{lm} (\xi) Y_l^m (\th,\vp).
\eeq

Now we consider the orbital motion of star 1. In the spherical
coordinate system, it becomes
\beq
  {\bf \O} \times \br =\O R \bigl( \t{\bf \O} \times \bxi
  \bigr)_{orb} +\O \a \bigl( \t{\bf \O} \times \bxi \bigr)_{fig},
  \label{Eq:orbit}
\eeq
where
\beqa
  \bigl( \t{\bf \O} \times \bxi \bigr)_{orb}&=&{1 \over 1+p}
  (\sin \th \sin \vp, ~\cos \th \sin \vp, ~\cos \vp), \\
  \bigl( \t{\bf \O} \times \bxi \bigr)_{fig} &=&(0, ~0, ~\xi \sin \th),
\eeqa
and $p \equiv M_1/M_2$. The first term on the right-hand side of Eq.
(\ref{Eq:orbit}) comes from the orbital motion of the center of mass
of star 1 and the second term comes from the fluid motion around the
center of mass of star 1. The orbital motion of star 2 becomes
\beq
  {\bf \O} \times \br' =\O R \bigl( \t{\bf \O} \times \bxi
  \bigr)'_{orb} +\O \a' \bigl( \t{\bf \O} \times \bxi \bigr)'_{fig},
\eeq
where
\beqa
  \bigl( \t{\bf \O} \times \bxi \bigr)'_{orb}&=&-{p \over 1+p}
  (\sin \th' \sin \vp', ~\cos \th' \sin \vp', ~\cos \vp'), \\
  \bigl( \t{\bf \O} \times \bxi \bigr)'_{fig} &=&(0, ~0, ~\xi' \sin \th'),
\eeqa
in the spherical coordinate system whose origin is located at the center
of mass of star 2. Here we have expressed the spherical
coordinate system for star 2 as $\br'=(r',\th',\vp')$ with $\xi'=r'/\a'$.

Next, we rewrite the equation of continuity (\ref{Eq:continuity}) as
\beq
  \D \Phi =-n (\n \Phi -{\bf \O} \times \br) \cdot {\n \Th \over
  \Th}. \label{Eq:Phi}
\eeq
The condition for $\Phi$ at the stellar surface is
\beq
  (\n \Phi -{\bf \O} \times \br) \cdot (\n \Th) \bigl|_{surf} =0,
\eeq
since $\Th=0$ at the surface. We expand $\Phi$ also as
\beq
  \Phi =\sum_{i=0}^{\infty} \e^i \Phi_i.
\eeq
When $R$ is large, the shape of star 1 is spherical and star 1 has
only the orbital motion of the center of mass in the inertial frame
with no intrinsic spin so that the gradient of the lowest order term
of $\Phi$ should agree with the orbital motion of the center of mass
of star 1, i.e.,
\beq
  \n \Phi_0 =\O R (\t{\bf \O} \times \bxi)_{orb}.
\eeq
Considering $\Phi_0$, we normalize $\Phi$ as
\beq
  \t{\Phi} ={\e \Phi \over \O \a a_0}. \label{Eq:normalize}
\eeq
We again expand $\t{\Phi}_i$ by spherical harmonics as
\beq
  \t{\Phi}_i =\sum_{l,m} {}^{(i)}\! \phi_{lm} (\xi) \t{Y}_l^m (\th,
  \vp),
\eeq
where $\t{Y}_l^m$ is defined by using the spherical harmonics $Y_l^m$ as
\beq
  \t{Y}_l^m (\th,\vp) \equiv C_{lm} {\p \over \p \vp}
  Y_l^m (\th,\vp)
\eeq
with constants $C_{lm}$.
Note that the gradient of the lowest order term of $\Phi$ has order
$\e^{-1}$ except for the implicit dependence in $\O$. When we normalize
$\Phi$ using Eq. (\ref{Eq:normalize}), the lowest order term of $\t{\Phi}$
has order $\e^0$. We must keep this difference in mind.

The orbital angular velocity is derived from the force balance equation,
i.e., the first tensor virial relation, defined by\cite{LRS93} 
\beq
  \int_{star~1} d^3 x {\p P \over \p x_1} =0,
\eeq
where $x_1 =r\sin \th \cos \vp$.
This equation is expressed by using Eq. (\ref{Eq:Euler}) as
\beq
  0 =\int_{star~1} d^3 x \r \Bigl[ {\p U^{1 \ra 1} \over \p x_1}
	+{\p U^{2 \ra 1} \over \p x_1} -{1 \over 2} {\p \over \p x_1}
	(\n \Phi)^2 +{\p \over \p x_1}
  \Bigl\{ (\n \Phi) \cdot ({\bf \O} \times \br) \Bigr\} \Bigr].
  \label{Eq:firstTV}
\eeq
If we substitute Eq. (\ref{Eq:orbit}),
$\Th_0$ and $\Phi_0$ into Eq. (\ref{Eq:firstTV}), we obtain the
orbital angular velocity in the lowest order as
\beq
  \O_0^2 ={M_{tot} \over R^3},
\eeq
where $M_{tot}=M_1 +M_2$. Note here that we also expand $\O^2$  as
\beq
  \O^2= \sum_{i=0}^{\infty} \e^i \O_i^2.
\eeq

Finally, we express the gravitational potential by rewriting
   Eq. (\ref{Eq:Euler}) as
\beq
  U =K(1+n) \r^{1 \over n} +{1 \over 2} v^2
  -\bv \cdot ({\bf \O} \times \br) +U_0, \label{Eq:U}
\eeq
where $U_0$ is constant.
Substituting Eq. (\ref{Eq:U}) into the Poisson equation
(\ref{Eq:Poisson}), we obtain the equation to  determine 
the equilibrium figure as
\beq
  \a^2 \D \Th =-\Th^n -{1 \over 8\pi \r_c} \D \Bigl[ (\n \Phi)^2 -2
  (\n \Phi) \cdot ({\bf \O} \times \br) \Bigr]. \label{Eq:Theta}
\eeq

Now a solution can be obtained iteratively. Firstly, $\Th_i$ is
determined by demanding that the gravitational potential and its
normal derivative are continuous at the stellar surface\cite{Ch33},
that is,
\beqa
  U_{int} (\xi=\Xi) &=&U_{ext} (\xi=\Xi), \\
  {\p U_{int} \over \p \xi} (\xi=\Xi) &=&{\p U_{ext} \over \p \xi}
  (\xi=\Xi),
\eeqa
where $\Xi(\th,\vp)$
expresses the surface,
\beq
  \Th (\xi=\Xi(\th,\vp)) =0.
\eeq
Substituting $\Th_i$ and Eq. (\ref{Eq:orbit}) into Eqs. (\ref{Eq:Phi})
and (\ref{Eq:firstTV}), we obtain $\Phi_i$ and $\O_i^2$. Secondly we
substitute  $\Phi_i$ and $\O_i^2$ into Eq. (\ref{Eq:Theta}) and derive
$\Th_{i+1}$. We have continued this procedure up to order $\epsilon^6$
in this paper.

%%%%%%%%%%%%%%%%%%%%%%%%%%%%%%%%%%%%%%%%%%%
\section{Derivation of the Physical Values}
%%%%%%%%%%%%%%%%%%%%%%%%%%%%%%%%%%%%%%%%%%%

In this section, we calculate physical values such as the baryon mass
of star 1, the orbital angular velocity, the total energy and total
angular momentum of the binary system. We will just show the main
results. All the details will be shown in the Appendix.

%%%%%%%%%%%%%%%%%%%%%%%%%%%%%%%%%%%%%
\subsection{Mass and Central Density}
%%%%%%%%%%%%%%%%%%%%%%%%%%%%%%%%%%%%%

The baryon mass of star 1 is calculated by integrating the density
over the volume of star 1 as
\beqa
  M_1 &=& \int_{star~1} d^3 x \r, \\
  &=&\a^3 \r_c \int d\xi d\th d\vp \xi^2 \sin \th \Bigl[ \Th_0^n +\e^3 
  n \Th_0^{n-1} \Th_3 +\e^4 n \Th_0^{n-1} \Th_4 +\e^5 n \Th_0^{n-1}
  \Th_5 \nonumber \\
  &&\hspace{100pt}+\e^6 n \Bigl\{ \Th_0^{n-1} \Th_6 +{1 \over 2} (n-1)
  \Th_0^{n-2} \Th_3^2 \Bigr\} \Bigr].
\eeqa
Note here that the integration range for $\xi$ is $[0,~\Xi(\th, \vp)]$.

For $n>0$ case, the mass of star 1 becomes
\beqa
  M_1 &=&-4\pi \r_c \a^3 \xi_1^2 \Th_{0,\xi} (\xi_1)
  \Bigl[ 1 +\e^6 \Bigl( {1 \over \Th_{0,\xi} (\xi_1)} \Bigr)
  \Bigl( {(\three \psi_2 (\xi_1))^2 \over 10\Th_{0,\xi} (\xi_1)}n
  \Th_0^{n-1} (\xi_1)+\six \psi_{0,\xi} (\xi_1) \Bigr) \Bigr],
  \label{Eq:mass1}\\
  &=&-4\pi \r_c \a^3 \xi_1^2 \Th_{0,\xi} (\xi_1)
  \Bigl[ 1 +\e^6 \Bigl( {-1 \over \xi_1 \Th_{0,\xi} (\xi_1)} \Bigr)
  \Bigl( \t{c}_6 +\six \psi_0 (\xi_1) \Bigr) \Bigr]. \label{Eq:mass2}
\eeqa
where we define
\beq
  {}_{,\xi} \equiv {d \over d\xi}.
\eeq
We have used Eq. (\ref{Eq:tilda_c6}) in deriving Eq. (\ref{Eq:mass2})
from Eq. (\ref{Eq:mass1}). Here $\t{c}_6$ is a constant, and one can
see the expression of $\t{c}_6$ in Appendix \ref{order:6th}. We
assume that the baryon mass $M_1=-4\pi \r_{c0} \a_0^3 \xi_1^2
\Th_{0,\xi} (\xi_1)$ is conserved throughout the equilibrium sequences
of the binary system, where $\r_{c0}$ and $\a_0$ denote the values for
a spherical star. Therefore, the central density and $\a$ are
expressed as
\beqa
  \r_c &=&\r_{c0} \Bigl( 1-\e^6 {\d \r_c \over \r_c} \Bigr), \\
  \a &=&\Bigl[ {K(1+n) \r_{c0}^{{1 \over n}-1} \over 4\pi} \Bigr]^{1/2}
  \Bigl( 1 -\e^6 {1-n \over 2n} {\d \r_c \over \r_c} \Bigr),
\eeqa
where we define\footnote{The expression of $\d \r_c /\r_c$ seems to
diverge if $n < 1$ when one see Eq. (\ref{Eq:mass1}).
However physically $\d \r_c /\r_c$ should be finite.
In reality we determined $\d \r_c /\r_c$ by using the virial
relation as shown in the Appendix \ref{sec:change_rho_c}.}
\beq
  {\d \r_c \over \r_c} \equiv {2n \over 3-n}
  \Bigl( {-1 \over \xi_1 \Th_{0,\xi}(\xi_1)}
  \Bigr) \Bigl( \t{c}_6 +\six \psi_0 (\xi_1) \Bigr).
\eeq

%%%%%%%%%%%%%%%%%%%%%%%%%%%%%%%%%%%%%
\subsection{Orbital angular velocity}
%%%%%%%%%%%%%%%%%%%%%%%%%%%%%%%%%%%%%

We obtain the orbital angular velocity from Eq. (\ref{Eq:firstTV}) as
\beqa
  \O^2 &=&{M_{tot} \over R^3} \Bigl[ 1 +{9 \over 2R^2} \Bigl(
  \e^3 {\bar{\bI}_{11} \over M_1} +{\e'}^3 {\bar{\bI}_{11}' \over M_2}
  \Bigr) +O(\e^7) \Bigr], \nonumber \\
  &=&{M_{tot} \over a_0^3} \e^3 \Bigl[ 1 +{9\e^2 \over 2a_0^2} \Bigl(
  \e^3 {\bar{\bI}_{11} \over M_1} +{\e'}^3 {\bar{\bI}_{11}' \over M_2}
  \Bigr) +O(\e^7) \Bigr],
\eeqa
where $\bar{\bI}_{11}$ denotes $\bI_{11}/\e^3$.

%%%%%%%%%%%%%%%%%%%%%%%%%
\subsection{Total energy}
%%%%%%%%%%%%%%%%%%%%%%%%%

The total energy is written as
\beq
  E=\Pi_{tot} +(W_{self})_{tot} +(W_{int})_{tot} +T_{tot},
\eeq
where $\Pi_{tot}$, $(W_{self})_{tot}$, $(W_{int})_{tot}$ and $T_{tot}$ 
denote the total internal energy, the total self-gravity energy, the
total interaction energy and the total kinetic energy, respectively.
We calculate each energy in the following.

%%%%%%%%%%%%%%%%%%%%%%%%%%%%%%%
\subsubsection{Internal energy}
%%%%%%%%%%%%%%%%%%%%%%%%%%%%%%%

The definition of the internal energy of star 1 is
\beq
  \Pi_1 =n \int_{star~1} d^3 x P.
\eeq

For $n>0$ case, we obtain
\beq
  \Pi_1 ={n \over 1+n} {M_1^2 \over a_0 \xi_1^3 (\Th_{0,\xi} (\xi_1))^2}
  \Bigl[ \Bigl( 1 -\e^6 {5-n \over 2n} {\d \r_c \over \r_c} \Bigr)
  \int_0^{\xi_1} d\xi \xi^2 \Th_0^{1+n}
  +\e^6 (1+n) \int_0^{\xi_1} d\xi \xi^2 \Bigl\{ \Th_0^n 
  \six \psi_0 +{1 \over 10} n \Th_0^{n-1} (\three \psi_2)^2 \Bigr\}
  \Bigr].
\eeq
Therefore, the total internal energy is expressed as
\beqa
  \Pi_{tot} &=&{n \over 1+n} {M_1^2 \over a_0 \xi_1^3
  (\Th_{0,\xi} (\xi_1))^2}
  \Bigl[ \Bigl( 1 -\e^6 {5-n \over 2n} {\d \r_c \over \r_c} \Bigr)
  \int_0^{\xi_1} d\xi
  \xi^2 \Th_0^{1+n} +\e^6 (1+n) \int_0^{\xi_1} d\xi \xi^2 \Bigl\{ \Th_0^n 
  \six \psi_0 +{1 \over 10} n \Th_0^{n-1} (\three \psi_2)^2 \Bigr\}
  \Bigr] \nonumber \\
  &&+{n' \over 1+n'} {M_2^2 \over a_0' {\xi_1'}^3
  (\bar{\Th}_{0,\xi'} (\xi_1'))^2}
  \Bigl[ \Bigl( 1-{\e'}^6 {5-n' \over 2n'} {\d \r_c' \over \r_c'} \Bigr)
  \int_0^{\xi_1'} d\xi' {\xi'}^2 \bar{\Th}_0^{1+n'} \nonumber \\
  &&\hspace{50pt}+{\e'}^6 (1+n')
  \int_0^{\xi_1'} d\xi' {\xi'}^2 \Bigl\{ \bar{\Th}_0^{n'} 
  \six \bar{\psi}_0 +{1 \over 10} n' \bar{\Th}_0^{n'-1}
  (\three \bar{\psi}_2)^2 \Bigr\}
  \Bigr], \nonumber \\
\eeqa
where the superscript $'$ means the terms concerned with star 2,
and $~\bar{}~$ also means the functions concerned with star 2.

%%%%%%%%%%%%%%%%%%%%%%%%%%%%%%%%%%%
\subsubsection{Self-gravity energy}
%%%%%%%%%%%%%%%%%%%%%%%%%%%%%%%%%%%

The definition of the self-gravity energy of star 1 is
\beq
  W_{self,1} =-{1 \over 2} \int_{star~1} d^3 x \r U^{1 \ra 1}.
\eeq

For $n>0$ case, we obtain
\beq
  W_{self,1} =-{1+n \over 2n} \Pi_1 +\e^6 \pi \r_c \Bigl[ {3\mu
  \bar{\bI}_{11} \over \xi_1^2 (1+p)} -2\a^2 M_1 \t{c}_6 \Bigr] +2\pi
  \r_c \a^2 M_1 \xi_1 \Th_{0,\xi} (\xi_1),
\eeq
where $\mu$ is defined as
\beq
  \mu \equiv {M_{tot} \over 4\pi \r_c \a^3 \xi_1}
  =-\Bigl( {1+p \over p} \Bigr) \xi_1 \Th_{0,\xi} (\xi_1).
\eeq
The self-gravity energy of star 2 is
\beq
  W_{self,2} =-{1+n' \over 2n'} \Pi_2 +{\e'}^6 \pi \r_c' \Bigl[ {3\mu'
  p \bar{\bI}_{11}' \over {\xi_1'}^2 (1+p)} -2{\a'}^2 M_2 \t{c}_6' \Bigr]
  +2\pi \r_c' {\a'}^2 M_2 \xi_1' \bar{\Th}_{0,\xi'} (\xi_1').
\eeq
Here, $\t{c}_6'$ is a constant similar to $\t{c}_6$, and
\beq
  \mu' \equiv {M_{tot} \over 4\pi \r_c' {\a'}^3 \xi_1'}
  =-(1+p) \xi_1' \bar{\Th}_{0,\xi'} (\xi_1').
\eeq
Therefore, the total self-gravity energy becomes
\beqa
  (W_{self})_{tot} &=&-{1+n \over 2n} \Pi_1 +\e^6 {M_1^2 \over 2a_0} \Bigl(
  {3 \over 2p} {\bar{\bI}_{11} \over M_1 a_0^2} +{\t{c}_6 \over \xi_1
  \Th_{0,\xi} (\xi_1)} \Bigr)
  -{M_1^2 \over 2a_0} \Bigl( 1 -\e^6 {1 \over n} {\d \r_c \over \r_c}
  \Bigr) \nonumber \\
  &&-{1+n' \over 2n'} \Pi_2 +{\e'}^6 {M_2^2 \over 2a_0'} \Bigl( {3p \over 2}
  {\bar{\bI}_{11}' \over M_2 {a_0'}^2}
  +{\t{c}_6' \over \xi_1' \bar{\Th}_{0,\xi'} (\xi_1')} \Bigr)
  -{M_2^2 \over 2a_0'} \Bigl( 1 -{\e'}^6 {1 \over n'} {\d \r_c' \over \r_c'}
  \Bigr).
\eeqa

%%%%%%%%%%%%%%%%%%%%%%%%%%%%%%%%%%
\subsubsection{Interaction energy}
%%%%%%%%%%%%%%%%%%%%%%%%%%%%%%%%%%

The definition of the interaction energy of star 1 is
\beqa
  W_{int,1} &=&-{1 \over 2} \int_{star~1} d^3 x \r U^{2 \ra 1}, \\
  &=&-{M_1 M_2 \over 2a_0} \e -\e^3 {3M_1 M_2 \over 4a_0^3} \Bigl(
  \e^3 {\bar{\bI}_{11} \over M_1} +{\e'}^3 {\bar{\bI}_{11}' \over M_2}
  \Bigr).
\eeqa
Therefore, the total interaction energy becomes
\beq
  (W_{int})_{tot} =-{M_1 M_2 \over a_0} \e -\e^3 {3M_1 M_2 \over
  2a_0^3} \Bigl( \e^3 {\b{\bI}_{11} \over M_1} +{\e'}^3 {\b{\bI}_{11}'
  \over M_2} \Bigr).
\eeq

%%%%%%%%%%%%%%%%%%%%%%%%%%%%%%
\subsubsection{Kinetic energy}
%%%%%%%%%%%%%%%%%%%%%%%%%%%%%%

The definition of the kinetic energy of star 1 is
\beqa
  T_1 &=&{1 \over 2} \int_{star~1} d^3 x \r v_1^2, \\
  &=&{M_1 M_2 \over 2a_0 (1+p)} \e \Bigl[ 1+{9\e^2 \over 2a_0^2}
  \Bigl( \e^3 {\b{\bI}_{11} \over M_1} +{\e'}^3 {\b{\bI}_{11}' \over M_2}
  \Bigr) \Bigr].
\eeqa
The kinetic energy of star 2 is also given as
\beq
  T_2 ={M_1 M_2 p \over 2a_0 (1+p)} \e \Bigl[ 1+{9\e^2 \over
  2a_0^2} \Bigl( \e^3 {\b{\bI}_{11} \over M_1} +{\e'}^3 {\b{\bI}_{11}'
  \over M_2} \Bigr) \Bigr].
\eeq
Then the total kinetic energy becomes
\beq
  T_{tot} ={M_1 M_2 \over 2a_0} \e +\e^3 {9M_1 M_2 \over 4a_0^3}
  \Bigl( \e^3 {\b{\bI}_{11} \over M_1} +{\e'}^3 {\b{\bI}_{11}' \over M_2}
  \Bigr).
\eeq

%%%%%%%%%%%%%%%%%%%%%%%%%%%%%%%%%%%
\subsection{Total angular momentum}
%%%%%%%%%%%%%%%%%%%%%%%%%%%%%%%%%%%

The total angular momentum is calculated from the equation;
\beq
  \bJ =\int_{star~1,~star~2} d^3 x \r (\bR \times \bv),
\eeq
where $\bR$ for star 1 is written as
\beq
  \bR={R \over 1+p} (\sin \th \cos \vp,~\cos \th \cos \vp,~-\sin \vp)
  +r (1,~0,~0),
\eeq
and for star 2 is
\beq
  \bR' =-{pR \over 1+p} (\sin \th' \cos \vp',~\cos \th' \cos \vp',
  ~-\sin \vp') +r' (1,~0,~0)
\eeq
in the spherical coordinate system. Note that since the velocity field
$\bv=\n \Phi$ is written in the spherical coordinate system, it is
convenient to calculate $\bR \times \bv$ in the spherical
one. However, we should be careful to integrate it over the volume of
each star, because in the spherical coordinate system, the unit
vectors depend on the coordinates. For example, the unit vector
$(1,0,0)$ in the {\it Cartesian} coordinate system becomes $(\sin
\th \cos \vp,\cos \th \cos \vp,-\sin \vp)$ in the {\it spherical}
one.

As a result, we obtain the total angular momentum;
\beq
  \bJ =(0,~0,~J)
\eeq
in the {\it Cartesian} coordinate system. Here $J$ is calculated as
\beq
  J ={M_1 M_2 \over M_1 +M_2} R^2 \O \Bigl[ 1 +{\rm higher~order~terms~
  than~} O(\e^6) \Bigr].
\eeq

%%%%%%%%%%%%%%%%%%%%%%%%%%%%
\subsection{Virial Relation}
%%%%%%%%%%%%%%%%%%%%%%%%%%%%

By using the above energies, we can calculate the virial relation
which is written as
\beq
  {3 \over n} \Pi_1 +{3 \over n'} \Pi_2 +(W_{self})_{tot}
  +(W_{int})_{tot} +2T_{tot} =0.
\eeq
This equation is used to check the solutions as shown in the previous
letter\cite{TN99}. However in this paper we use it to determine
$\d \r_c /\r_c$ (see Appendix \ref{sec:change_rho_c}).

%%%%%%%%%%%%%%%%%%%%%%%%%%%
\section{Numerical results} \label{sec:results}
%%%%%%%%%%%%%%%%%%%%%%%%%%%

In the following sections, we set $M_1 =M_2 =M$, $n=n'$, $\a=\a'$,
$\r_c=\r_c'$ and so on, i.e., the identical
star binary for simplicity, and investigate the cases of different
polytropic indices.

First, we show the Lane-Emden function of star 1 up to $O(\e^6)$.
It is written as
\beq
  \Th=\Th_0 +\e^3 \Th_3 +\e^4 \Th_4 +\e^5 \Th_5 +\e^6 \Th_6,
\eeq
where
\beqa
  \Th_3 &=&\three \psi_2 (\xi) P_2 (\sin \th \cos \vp), \\
  \Th_4 &=&\four \psi_3 (\xi) P_3 (\sin \th \cos \vp), \\
  \Th_5 &=&\five \psi_4 (\xi) P_4 (\sin \th \cos \vp), \\
  \Th_6 &=&\six \psi_0 (\xi) +\six \psi_2 (\xi) P_2 (\sin \th \cos
  \vp) +\six \psi_{22} (\xi) P_2^2 (\cos \th) \cos 2\vp \nonumber \\
  &&+\six \psi_4 (\xi) P_4 (\sin \th \cos \vp) +\six \psi_5 (\xi) P_5
  (\sin \th \cos \vp).
\eeqa
The functions $\Th_0$, \hbox{${}^{(i)}\!$}$\psi_j$ and their derivatives
for each polytropic index are given in Tables \ref{table1}--\ref{table8}.

Next, we give the velocity potential up to $O(\e^6)$.
We can write it as
\beq
  \t{\Phi} =\t{\Phi}_0 +\e^4 \t{\Phi}_4 +\e^5 \t{\Phi}_5 +\e^6
  \t{\Phi}_6,
\eeq
where
\beqa
  \t{\Phi}_0 &=&{1 \over 1+p} \xi \sin \th \sin \vp, \label{Eq:phi0} \\
  \t{\Phi}_4 &=&\four \phi_2 (\xi) P_2^2 (\cos \th) \sin 2\vp, \\
  \t{\Phi}_5 &=&\five \phi_3 (\xi) \Bigl[ P_3^1 (\cos \th) \sin \vp
  -{1 \over 2} P_3^3 (\cos \th) \sin 3\vp \Bigr], \\
  \t{\Phi}_6 &=&\six \phi_4 (\xi) \Bigl[ P_4^2 (\cos \th) \sin 2\vp
  -{1 \over 4} P_4^4 (\cos \th) \sin 4\vp \Bigr].
\eeqa
The functions $\four \phi_2$, $\five \phi_3$ and $\six \phi_4$ and
their derivatives are shown in Tables \ref{table9}--\ref{table12}.
The derivatives are related to the velocity field in the inertial frame.
Note here that in Eq. (\ref{Eq:phi0}) we write the dependence of the
mass ratio $p=M_1/M_2$ although we consider only $p=1$ case in this
section.

Here we directly show the existence of the velocity component along the
orbital axis.
In this paper, we take $X_3$ as the orbital axis. Since the origin of
the corotating coordinate system $\bX$ is located at the center of mass
of the binary system, it is convenient to consider another Cartesian
coordinate system which origin is located at the center of mass of
star 1. We call it $\bx=(x_1,x_2,x_3)$. The relations between $\bx$
and the spherical coordinate system $\br$ are
\beqa
  x_1 &=&r \sin \th \cos \vp, \\
  x_2 &=&r \sin \th \sin \vp, \\
  x_3 &=&r \cos \th,
\eeqa
as usual. Then, the velocity component along the orbital axis in the
inertial frame is written as
\beq
  v_3 ={\p \Phi \over \p x_3}.
\eeq
In the corotating frame, we can express the velocity field $\bu$ as
\beqa
  \bu &=&\n \Phi -{\bf \O} \times \br, \\
  &=&-\O \a (\t{\bf \O} \times \bxi)_{fig} +\e^4 \n \Phi_4 +\e^5 \n \Phi_5
  +\e^6 \n \Phi_6,   \label{Eq:velo_comp_corot} \\
  &=&-\O \a (\t{\bf \O} \times \bxi)_{fig} +\O a_0 \Bigl[ \e^3 \t{\n}
  \t{\Phi}_4 +\e^4 \t{\n} \t{\Phi}_5 +\e^5 \t{\n} \t{\Phi}_6 \Bigr],
\eeqa
where $\t{\n}=\a \n$ is a nabla defined by using $(\xi,\th,\vp)$ and
given in Eq. (\ref{Eq:nomnabla}).
Then, the velocity component along the orbital axis in the corotating
frame becomes
\beq
  u_3 =\O a_0 \Bigl[ \e^3 {\p \t{\Phi}_4 \over \p \t{x}_3} +\e^4 {\p
  \t{\Phi}_5 \over \p \t{x}_3} +\e^5 {\p \t{\Phi}_6 \over \p \t{x}_3}
  \Bigr], \label{Eq:velo_corot_3}
\eeq
where we have used the normalized coordinate system,
\beqa
  \t{x}_1 &=&\xi \sin \th \cos \vp, \\
  \t{x}_2 &=&\xi \sin \th \sin \vp, \\
  \t{x}_3 &=&\xi \cos \th,
\eeqa
for the calculational convenience.
Note that since $\n \Phi$ is proportional to $\O/\e$, the equation
(\ref{Eq:velo_comp_corot}) does not express the real dependence on $\e$.

Next, we calculate the term for $\t{\Phi}_4$ in Eq. (\ref{Eq:velo_corot_3}).
Then we obtain
\beqa
  {\p \t{\Phi}_4 \over \p \t{x}_3} &=&\cos \th {\p \t{\Phi}_4
  \over \p \xi} -{\sin \th \over \xi} {\p \t{\Phi}_4 \over \p \th},
  \nonumber \\
  &=&3 \Bigl( {d\four \phi_2 \over d\xi} -{2\four \phi_2 \over \xi}
  \Bigr) \sin^2 \th \cos \th \sin 2\vp.
\eeqa
If $\four \phi_2(\xi)$ is proportional to $\xi^2$, the velocity component
along the orbital axis disappears\footnote{For the $n=0$ case, i.e., the
incompressible fluid case, $\four \phi_2$ is proportional to $\xi^2$.}.
However, $\four \phi_2 (\xi)$ is not proportional to $\xi^2$ at all for
$n>0$. Moreover, we can show the velocity components along the
orbital axis for $\Phi_5$ and $\Phi_6$ as
\beqa
  {\p \t{\Phi}_5 \over \p \t{x}_3} &=&6 \Bigl[ \Bigl( {d\five
  \phi_3 \over d\xi} -{3\five \phi_3 \over \xi} \Bigr) (1 -5\sin^2 \th
  \cos^2 \vp) +{2\five \phi_3 \over \xi} \Bigr] \sin \th \cos \th \sin \vp,
  \\
%%%%%
  {\p \t{\Phi}_6 \over \p \t{x}_3} &=&30 \Bigl[ \Bigl( {d\six
  \phi_4 \over d\xi} -{4\six \phi_4 \over \xi} \Bigr) (3 -7\sin^2 \th
  \cos^2 \vp) +{6\six \phi_4 \over \xi} \Bigr] \sin^2 \th \cos \th
  \sin \vp \cos \vp.
\eeqa
In these equations, we can see that even if $\five \phi_3 \propto
\xi^3$ and $\six \phi_4 \propto \xi^4$, there remain the velocity
components along the orbital axis. The velocity component along the
orbital axis is order $\e^3$ higher. However, if one would like to pay
attention to the internal state of neutron stars in the binary system,
it should be taken into account.

We give also the other components of the velocity field as
\beqa
  u_1 &=&\O x_2 +\O a_0 \Bigl[ \e^3 {\p \t{\Phi}_4 \over \p \t{x}_1}
  +\e^4 {\p \t{\Phi}_5 \over \p \t{x}_1} +\e^5 {\p \t{\Phi}_6 \over
  \p \t{x}_1} \Bigr], \label{Eq:velo_corot_1} \\
  u_2 &=&-\O x_1 +\O a_0 \Bigl[ \e^3 {\p \t{\Phi}_4 \over \p \t{x}_2}
  +\e^4 {\p \t{\Phi}_5 \over \p \t{x}_2} +\e^5 {\p \t{\Phi}_6 \over
  \p \t{x}_2} \Bigr], \label{Eq:velo_corot_2}
\eeqa
where
\beqa
  {\p \t{\Phi}_4 \over \p \t{x}_1} &=&\sin \th \cos \vp {\p \t{\Phi}_4
  \over \p \xi} +{\cos \th \cos \vp \over \xi} {\p \t{\Phi}_4 \over \p \th}
  -{\sin \vp \over \xi \sin \th} {\p \t{\Phi}_4 \over \p \vp}, \\
  &=&6 \Bigl[ \Bigl( {d \four \phi_2 \over d\xi} -{2\four \phi_2 \over \xi}
  \Bigr) \sin^2 \th \cos^2 \vp +{\four \phi_2 \over \xi} \Bigr]
  \sin \th \sin \vp, \\
%%%%%
  {\p \t{\Phi}_4 \over \p \t{x}_2} &=&\sin \th \sin \vp {\p \t{\Phi}_4
  \over \p \xi} +{\cos \th \sin \vp \over \xi} {\p \t{\Phi}_4 \over \p \th}
  +{\cos \vp \over \xi \sin \th} {\p \t{\Phi}_4 \over \p \vp}, \\
  &=&6 \Bigl[ \Bigl( {d \four \phi_2 \over d\xi} -{2\four \phi_2 \over \xi}
  \Bigr) \sin^2 \th \sin^2 \vp +{\four \phi_2 \over \xi} \Bigr]
  \sin \th \cos \vp, \\
%%%%%
  {\p \t{\Phi}_5 \over \p \t{x}_1} &=&6\Bigl[ \Bigl( {d\five \phi_3 \over
  d\xi} -{3\five \phi_3 \over \xi} \Bigr) (1 -5\sin^2 \th \cos^2 \vp)
  -{8 \five \phi_3 \over \xi} \Bigr] \sin^2 \th \sin \vp \cos \vp, \\
%%%%%
  {\p \t{\Phi}_5 \over \p \t{x}_2} &=&6\Bigl[ \Bigl( {d\five \phi_3 \over
  d\xi} -{3\five \phi_3 \over \xi} \Bigr) (1 -5\sin^2 \th \cos^2 \vp)
  \sin^2 \th \sin^2 \vp +{\five \phi_3 \over \xi} \Bigl\{ 1 +\sin^2 \th
  (2\sin^2 \vp -5\cos^2 \vp) \Bigr\} \Bigr], \\
%%%%%
  {\p \t{\Phi}_6 \over \p \t{x}_1} &=&30 \Bigl[ \Bigl( {d\six \phi_4 \over
  d\xi} -{4\six \phi_4 \over \xi} \Bigr) (3 -7\sin^2 \th \cos^2 \vp)
  \sin^3 \th \sin \vp \cos^2 \vp +{3\six \phi_4 \over \xi}
  (1 -5\sin^2 \th \cos^2 \vp) \sin \th \sin \vp \Bigr], \\
%%%%%
  {\p \t{\Phi}_6 \over \p \t{x}_2} &=&30 \Bigl[ \Bigl( {d\six \phi_4 \over
  d\xi} -{4\six \phi_4 \over \xi} \Bigr) (3 -7\sin^2 \th \cos^2 \vp)
  \sin^3 \th \sin^2 \vp \cos \vp \nonumber \\
  &&\hspace{20pt}+{\six \phi_4 \over \xi} \Bigl\{
  3 +\sin^2 \th (7\cos^2 \vp -6\sin^2 \vp) \Bigr\} \sin \th \cos \vp \Bigr].
\eeqa

In Table \ref{table13}, we show the first zero point of $\Th_0$, the
orbital separation at the contact point, the energies, the quadrupole
moment, and the change in the central density. In this table, we have
used
\beq
  \t{E} \equiv {E \over M^2/a_0} =\t{E}_{self} +\e \t{E}_{point}
  +\e^6 \t{E}_{quad}.
\eeq
Each energy is given in Table \ref{table14}, where we have defined as
\beqa
  \t{\Pi}_{tot} &=&\t{\Pi}_{self} +\e^6 \t{\Pi}_{quad}, \\
%%%%%
  (\t{W}_{self})_{tot} &=&(\t{W}_{self})_{self}
  +\e^6 (\t{W}_{self})_{quad}, \\
%%%%%
  (\t{W}_{int})_{tot} &=&\e (\t{W}_{int})_{point}
  +\e^6 (\t{W}_{int})_{quad}, \\
%%%%%
  \t{T}_{tot} &=&\e \t{T}_{point} +\e^6 \t{T}_{quad}.
\eeqa
Accordingly, we can express
\beqa
  \t{E}_{self} &=&\t{\Pi}_{self} +(\t{W}_{self})_{self}, \\
  \t{E}_{point} &=&(\t{W}_{int})_{point} +\t{T}_{point}, \\
  \t{E}_{quad} &=&\t{\Pi}_{quad} +(\t{W}_{self})_{quad}
  +(\t{W}_{int})_{quad} +\t{T}_{quad}.
\eeqa

The total angular momentum for the identical star binary becomes
\beq
  J ={M \over 2} R^2 \O,
\eeq
and the orbital angular velocity is calculated as
\beq
  \O^2 ={2M \over a_0^3} \e^3 \Bigl[ 1 +{9\e^5 \over a_0^2}
  {\b{\bI}_{11} \over M} +O(\e^7) \Bigr].
\eeq
We show the total energy, the total angular momentum and the orbital
angular velocity along the equilibrium sequences of the binary system
in Table \ref{table15}.

\begin{figure}[ht]
 \begin{center}
  \epsfysize 7cm
  \leavevmode
  \epsfbox{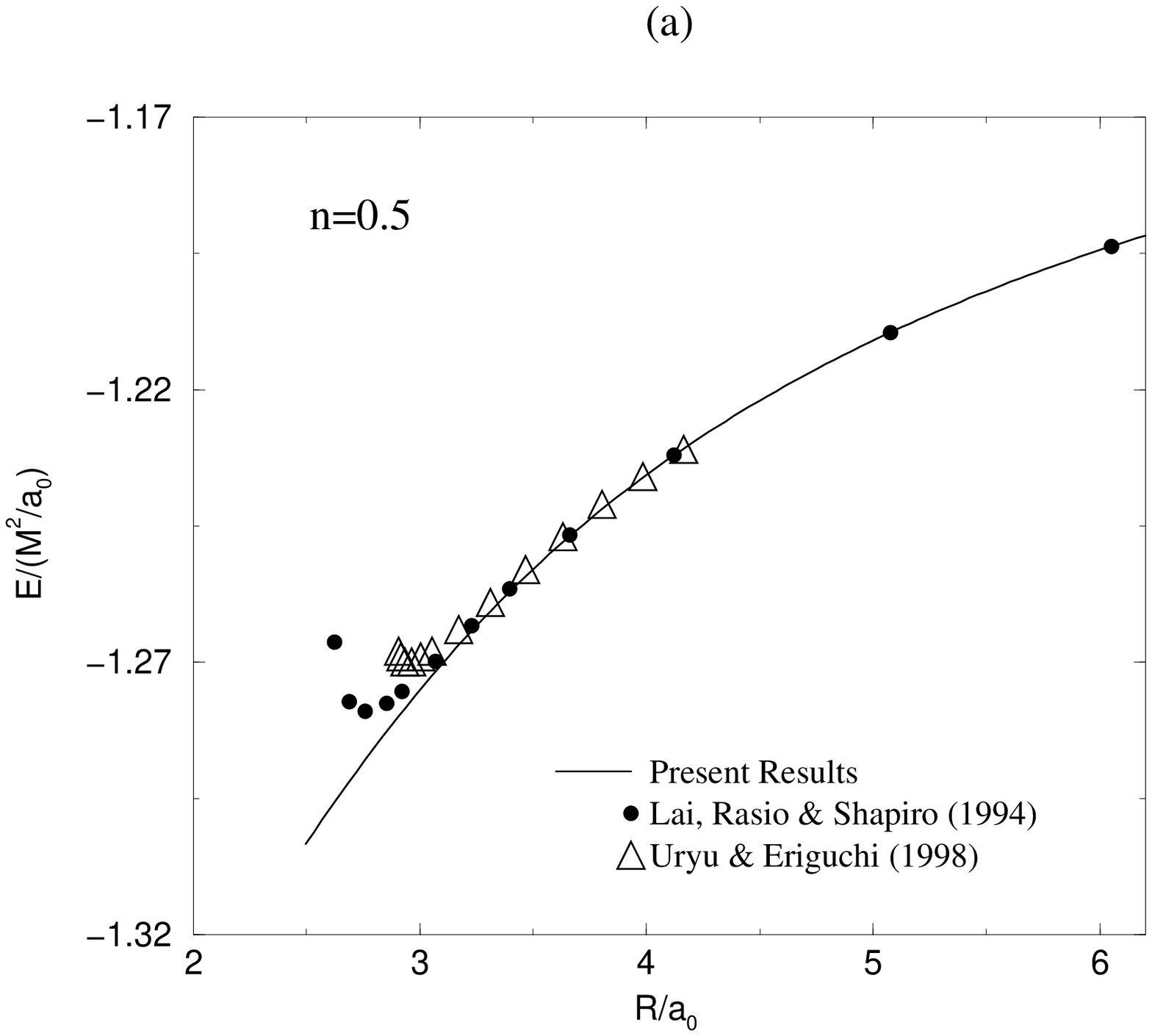}
%%%%%
 \hspace{5mm}
  \epsfysize 7cm
  \leavevmode
  \epsfbox{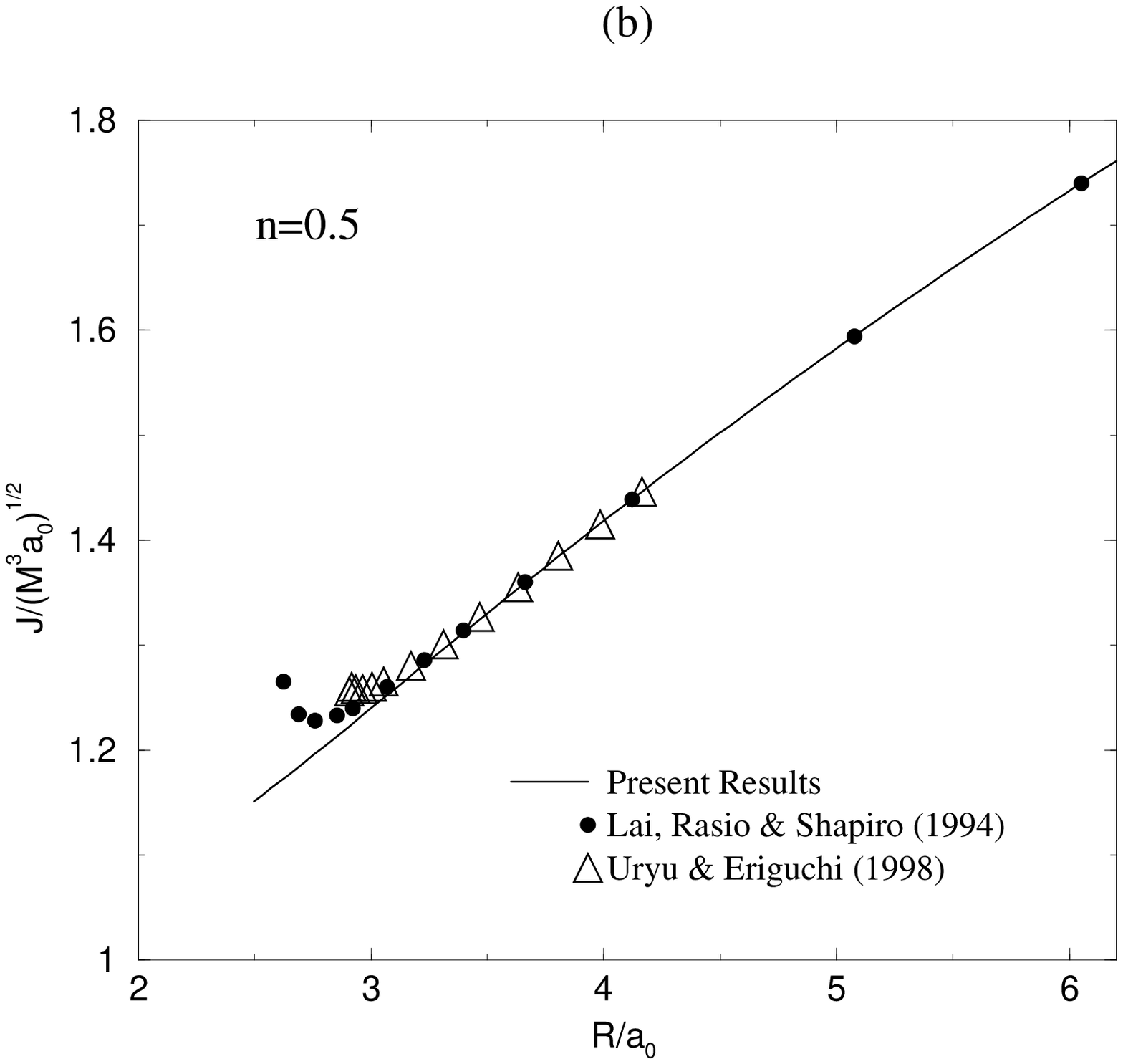}
%%%%%
 \vspace{5mm}
  \epsfysize 7cm
  \leavevmode
  \epsfbox{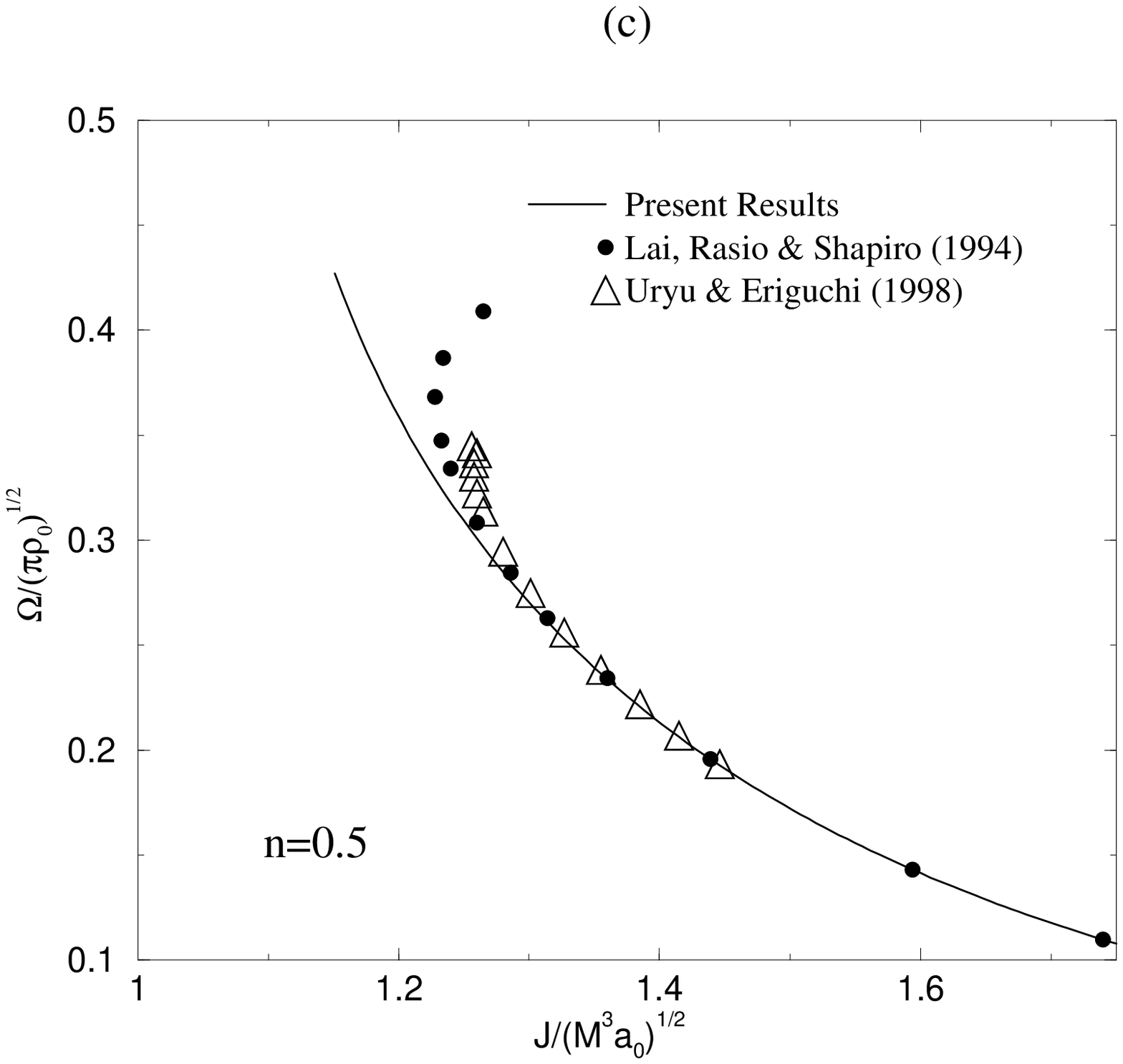}
 \end{center}
\caption{(a) The total energy and (b) the total angular momentum as
functions of the orbital separation, and (c) the orbital angular
velocity as functions of the total angular momentum for the polytropic
index $n=0.5$. Solid line denotes our results. Filled circles and
open triangles are the results of Lai, Rasio and Shapiro (1994) and
Ury\=u and Eriguchi (1998).
}
 \label{fign05}
\end{figure}

\begin{figure}[ht]
 \begin{center}
  \epsfysize 7cm
  \leavevmode
  \epsfbox{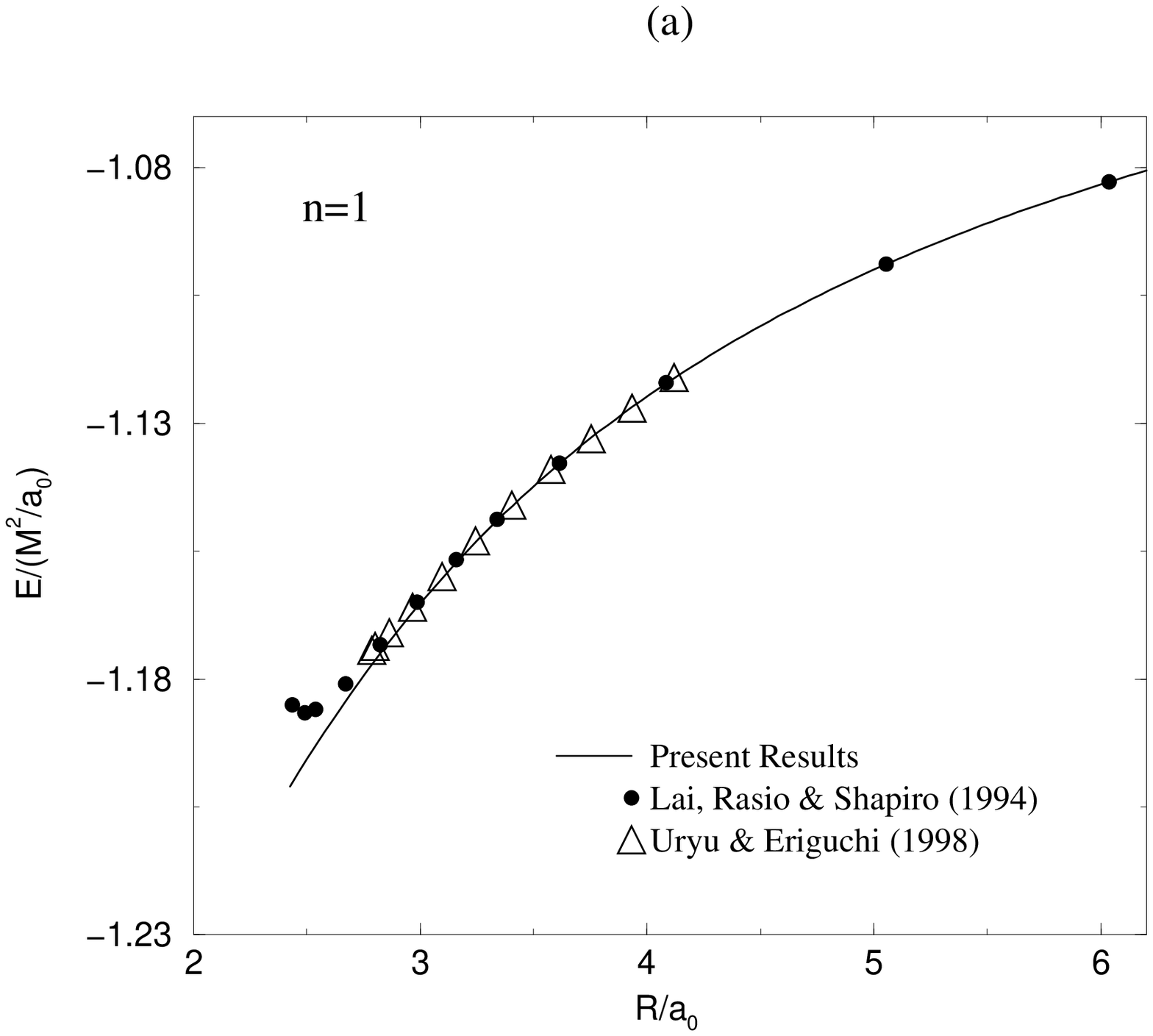}
%%%%%
 \hspace{5mm}
  \epsfysize 7cm
  \leavevmode
  \epsfbox{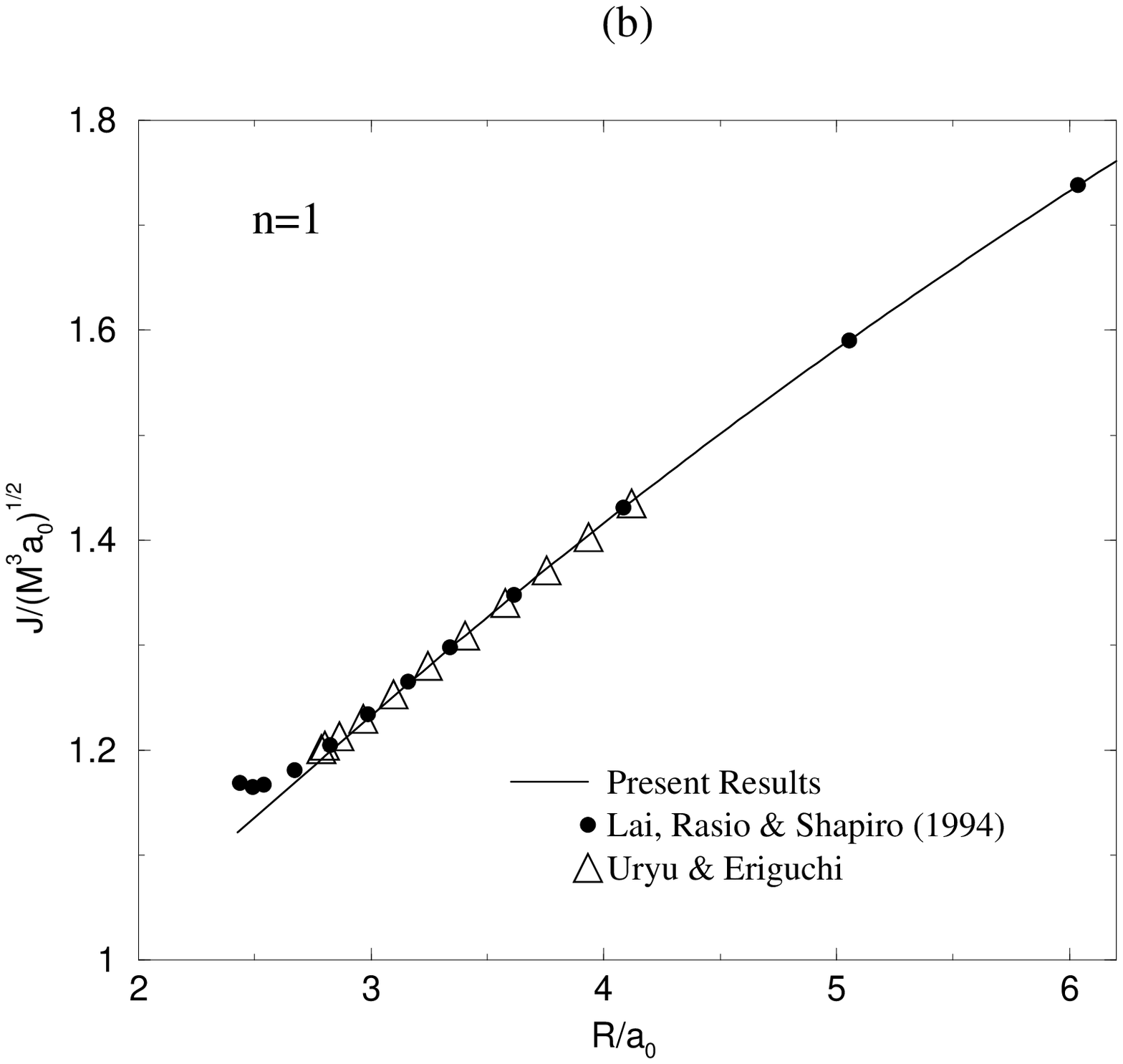}
%%%%%
 \vspace{5mm}
  \epsfysize 7cm
  \leavevmode
  \epsfbox{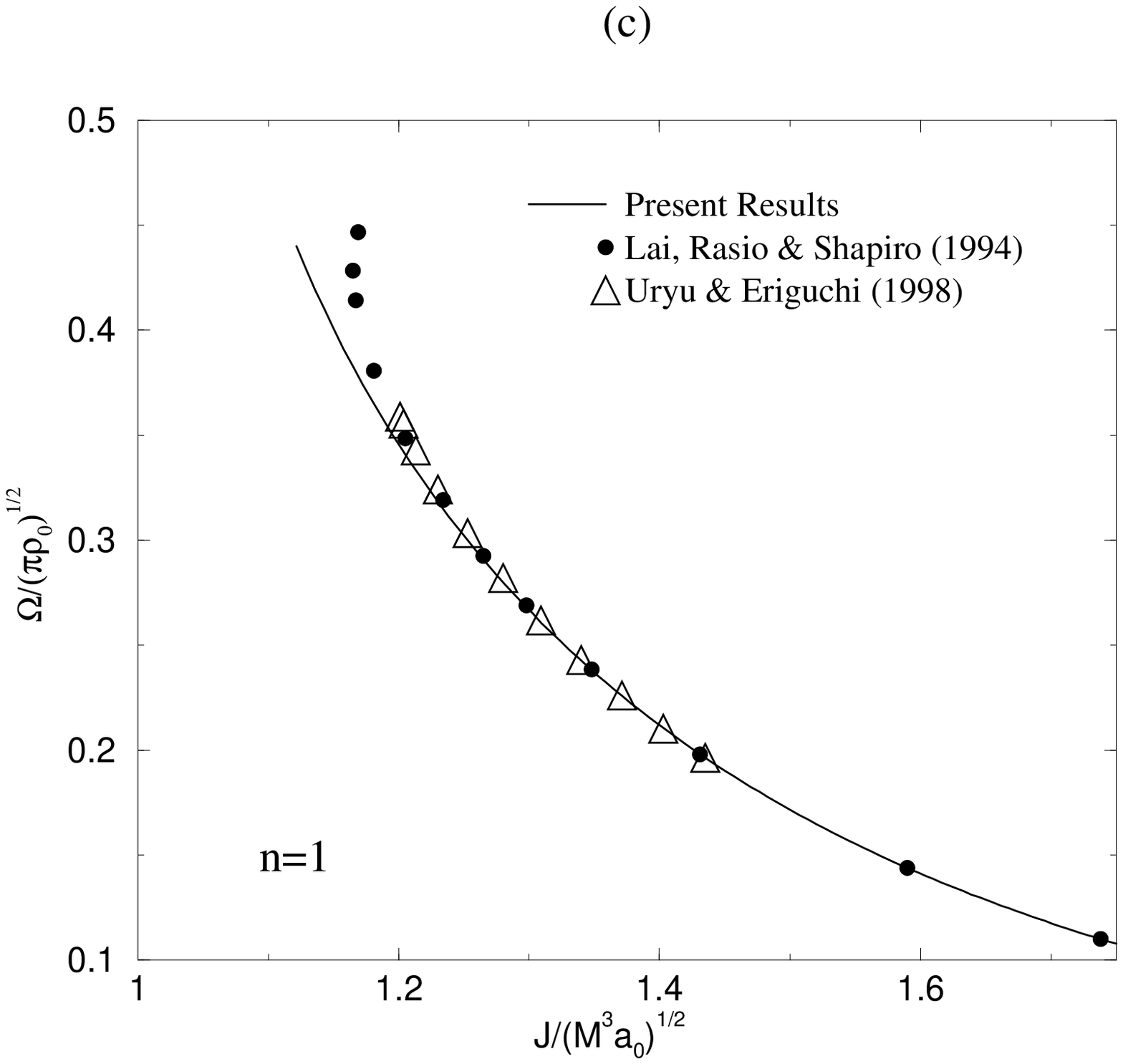}
 \end{center}
\caption{(a) The total energy and (b) the total angular momentum as
functions of the orbital separation, and (c) the orbital angular
velocity as functions of the total angular momentum for the polytropic
index $n=1$. Solid line denotes our results. Filled circles and
open triangles are the results of Lai, Rasio and Shapiro (1994) and
Ury\=u and Eriguchi (1998).
}
 \label{fign1}
\end{figure}

\begin{figure}[ht]
 \begin{center}
  \epsfysize 7cm
  \leavevmode
  \epsfbox{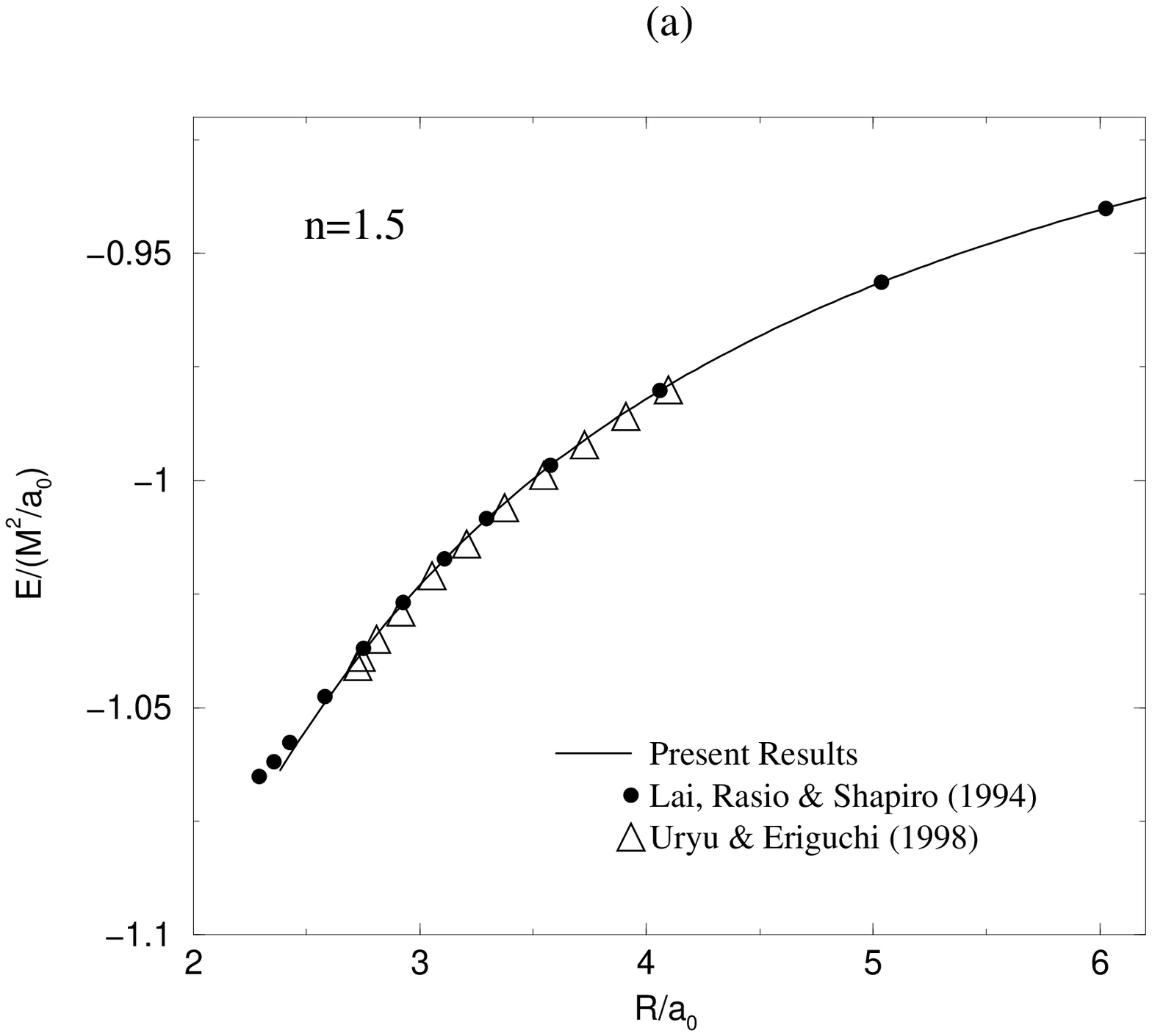}
%%%%%
 \hspace{5mm}
  \epsfysize 7cm
  \leavevmode
  \epsfbox{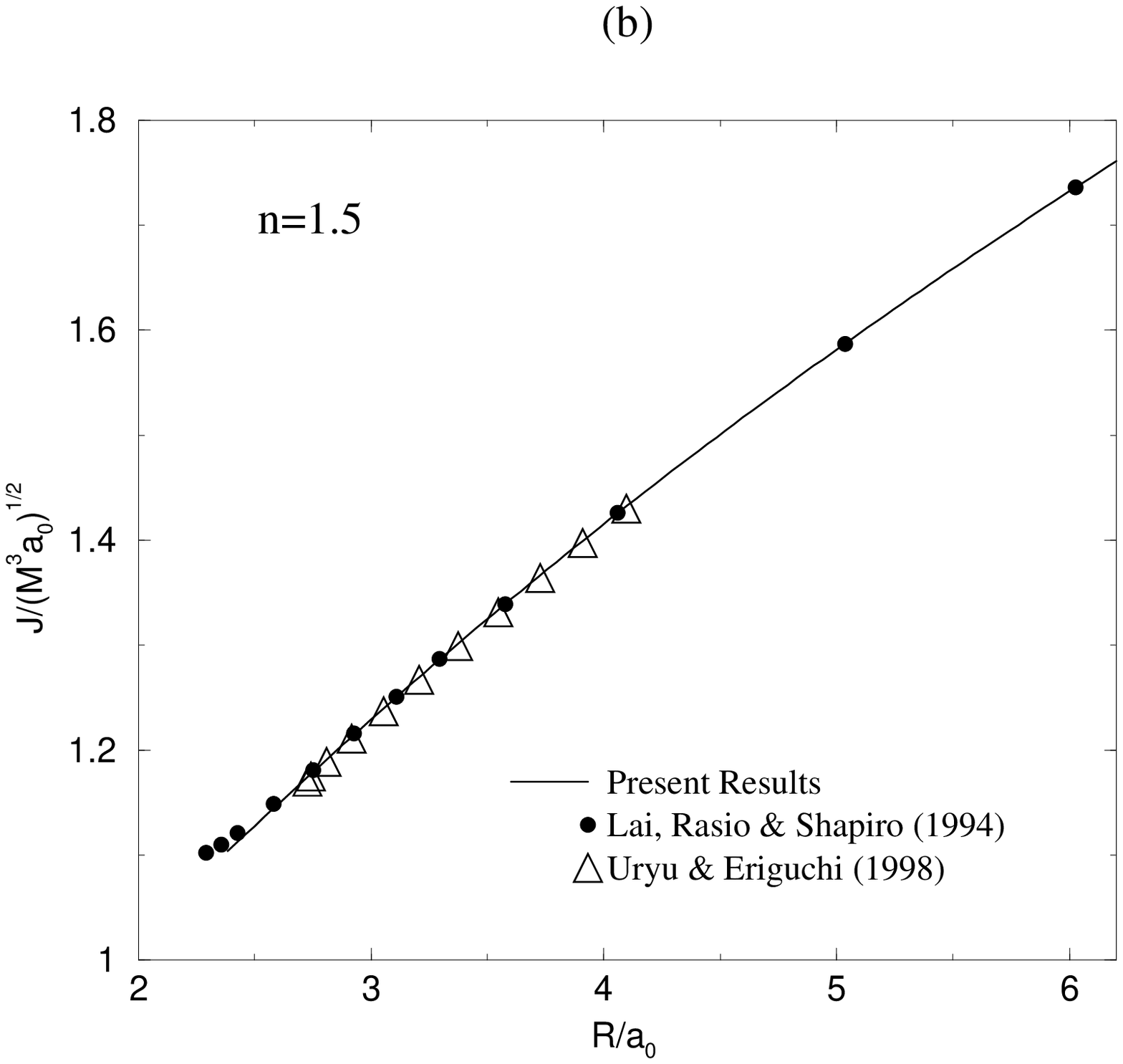}
%%%%%
 \vspace{5mm}
  \epsfysize 7cm
  \leavevmode
  \epsfbox{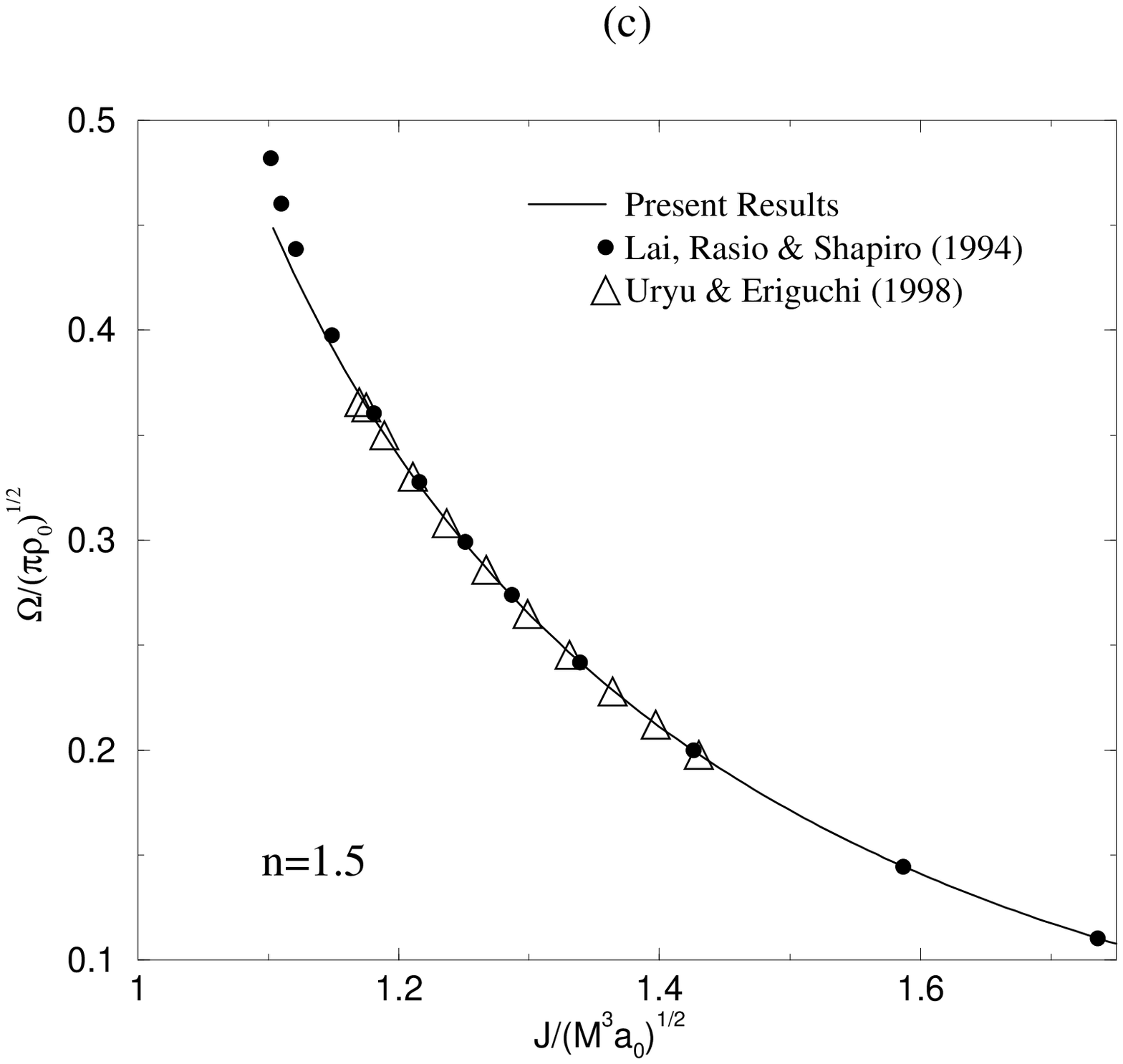}
 \end{center}
\caption{(a) The total energy and (b) the total angular momentum as
functions of the orbital separation, and (c) the orbital angular
velocity as functions of the total angular momentum for the polytropic
index $n=1.5$. Solid line denotes our results. Filled circles and
open triangles are the results of Lai, Rasio and Shapiro (1994) and
Ury\=u and Eriguchi (1998).
}
 \label{fign15}
\end{figure}

\begin{figure}[ht]
 \begin{center}
  \epsfysize 7cm
  \leavevmode
  \epsfbox{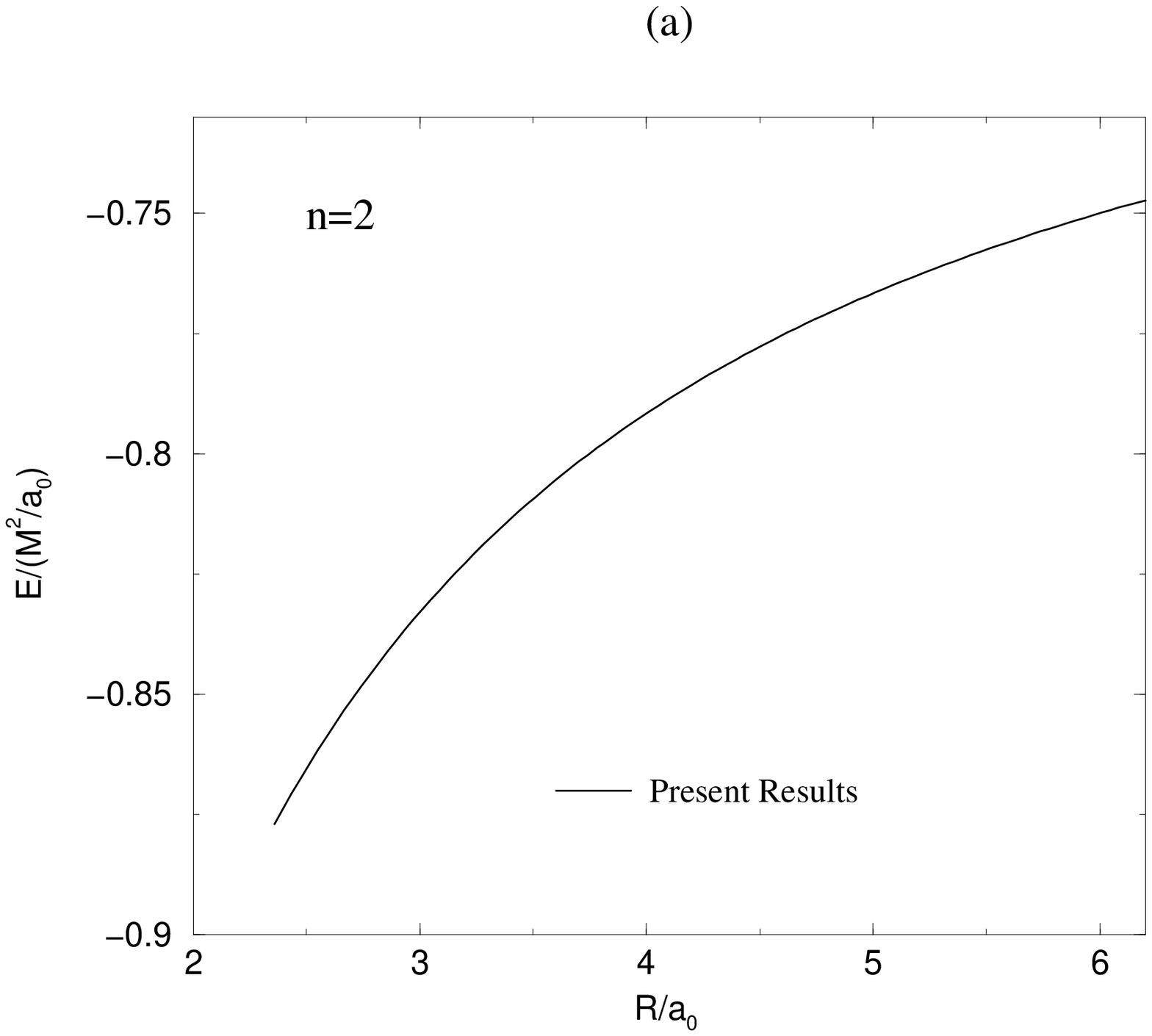}
%%%%%
 \hspace{5mm}
  \epsfysize 7cm
  \leavevmode
  \epsfbox{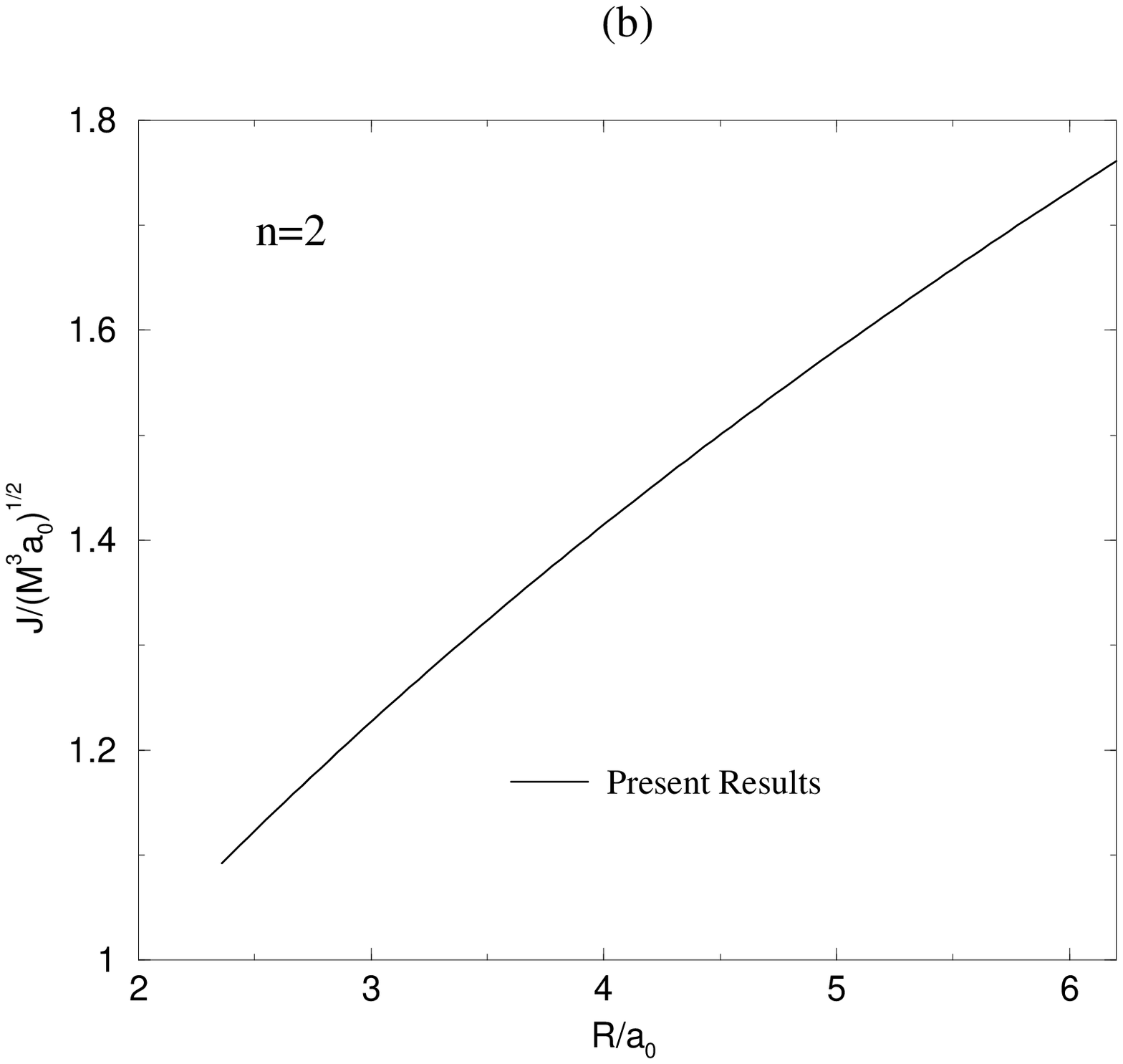}
%%%%%
 \vspace{5mm}
  \epsfysize 7cm
  \leavevmode
  \epsfbox{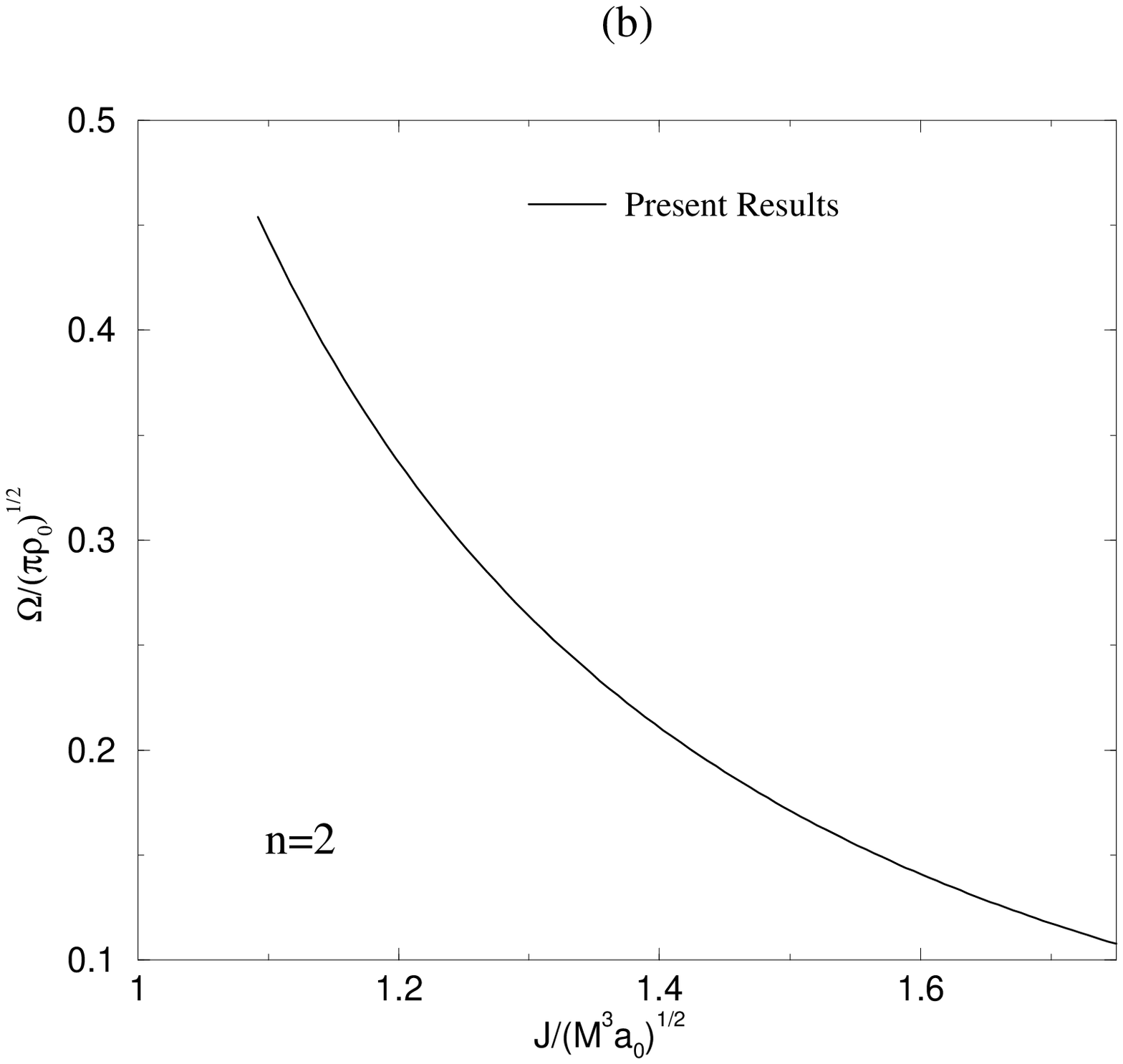}
 \end{center}
\caption{(a) The total energy and (b) the total angular momentum as
functions of the orbital separation, and (c) the orbital angular
velocity as functions of the total angular momentum for the polytropic
index $n=2$. Solid line denotes our results.
}
 \label{fign2}
\end{figure}

In Figs. \ref{fign05}--\ref{fign2}, we show (a) the total energy and
(b) the total angular momentum as functions of the orbital separation,
and (c) the orbital angular velocity as functions of the total angular
momentum for each polytropic index, where we have used
\beq
  \r_0 \equiv {3M \over 4\pi a_0^3}
\eeq
In these figures, we compare our results (solid lines) with those of
numerical calculations computed by Ury\=u and Eriguchi\cite{UE98b}
(open triangles) and those of semi-analytic ones proposed by Lai,
Rasio and Shapiro\cite{LRS94} (filled circles).  For the smaller
orbital separation, we must expand the physical values up to higher
order than $O(\e^6)$ in order to include the effect of the spin of
each star. In the case of the irrotational binary system, there is no
intrinsic spin for the large distant stars. However, when the stars
reach close range, the small spins are produced by the deformations of
the stars. For example, the effect of these spins appears in the total
energy at $O(\e^9)$\footnote{This is because the spin kinetic energy
is given by the volume integral of $(\e^4 \n \Phi_4)^2 \sim \O^2 \e^6
(\t{\n} \t{\Phi}_4)^2$, where $\O^2$ is $O(\e^3)$.  The definition of
$\t{\n}$ is in Eq. (\ref{Eq:nomnabla}).}.  The effect of the spin in
the case of the smaller polytropic index is larger than that in the
case of the larger polytropic index. The reason is that the matter is
not concentrated in the case of the smaller polytropic index, and the
quadrupole moment becomes larger (see Table \ref{table13}).
Therefore, at the smaller orbital separation, the deviation between
our results and those of Ury\=u and Eriguchi for $n=0.5$
(Fig. \ref{fign05}) is larger than that for $n=1.5$
(Fig. \ref{fign15}).

\begin{figure}[ht]
 \begin{center}
  \epsfxsize 10cm
  \leavevmode
  \epsfbox{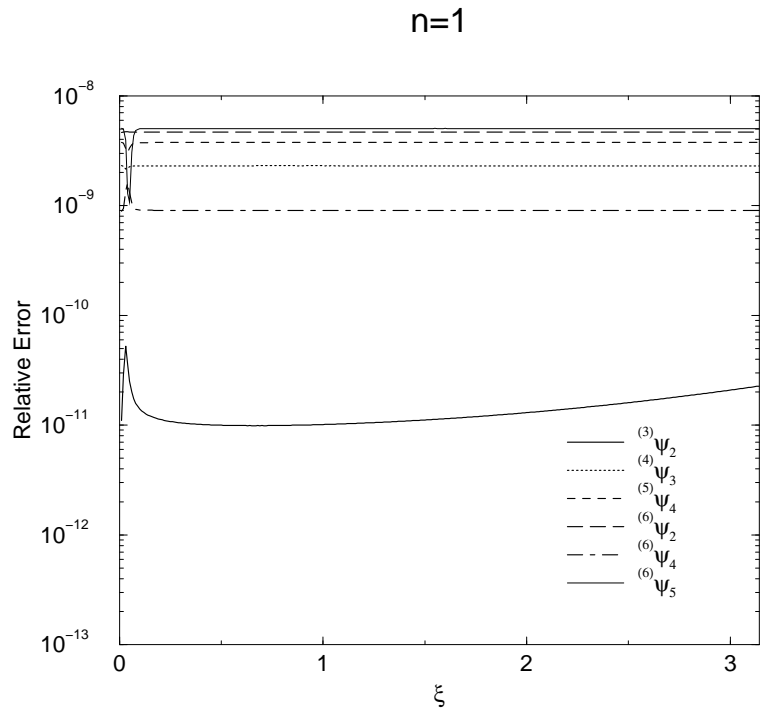}
 \end{center}
\caption{The relative errors for the functions presented the deformations
of the figure for the polytropic index $n=1$. Thick solid, dotted, dashed,
long-dashed, dot-dashed and thin solid lines denote the relative errors
for $\three \psi_2$, $\four \psi_3$, $\five \psi_4$, $\six \psi_2$,
$\six \psi_4$ and $\six \psi_5$, respectively.
}
 \label{errorfig}
\end{figure}

Finally, we represent the accuracy of our numerical calculations by
comparing numerical and analytic solutions for $n=1$ cases\cite{TN99}.
In Fig. \ref{errorfig}, we show the absolute values of the relative error
for $\three \psi_2$, $\four \psi_3$, $\five \psi_4$, $\six \psi_2$,
$\six \psi_4$ and $\six \psi_5$, where we define the relative error by
\beq
  \biggl| {{\rm Numerical~solution} \over {\rm Analytic~solution}} -1
  \biggr|.
\eeq
We can see from Fig. \ref{errorfig} that the relative errors are
within $10^{-8}$. Therefore, we regard that our results have enough
accuracy.

%%%%%%%%%%%%%%%%%%%%%%%%%%%%%%%%%
\section{Summary and Discussions}
%%%%%%%%%%%%%%%%%%%%%%%%%%%%%%%%%

%%%%%%%%%%%%%%%%%%%%
\subsection{Summary}
%%%%%%%%%%%%%%%%%%%%

In this paper, we have calculated the equilibrium solutions of
irrotational binary polytropic stars in Newtonian gravity by expanding
all physical quantities in a power of $\e$. We have presented the
results of the cases of several polytropic indices ($n=0.5, 1, 1.5$,
and $2$). In particular, we
have shown the velocity fields by solving the equation of
continuity. It is found that there exists the small velocity component
along the orbital axis (see \S \ref{sec:results}). It agrees
with the numerical calculations performed by Ury\=u and
Eriguchi\cite{UE98b} and Bonazzola, Gourgoulhon and
Marck\cite{BGM99a,BGM99b,BGM99c}.

Furthermore, we have given the figures and tables of the total energy,
total angular momentum and orbital angular velocity along the
equilibrium sequences for each polytropic indices. We can see from
these figures that our results agree with those of Lai, Rasio and
Shapiro\cite{LRS94} and Ury\=u and Eriguchi\cite{UE98b} for $R/a_0
>3$.

Since our solutions are correct if $\e \ll 1$, they can be used to
check the validity of numerical solutions. For any numerical codes,
one can ask to solve an equilibrium for large $R$, and compare
numerically derived velocity distribution and so on with our
semi-analytic solutions.

However, since we expanded physical quantities up to $O(\e^6)$, it may
not be enough to discuss about the behavior of solutions for small
$R$. In order to apply our solutions in the case of small $R$ and to
check the validity of numerical codes in this case, further higher
order calculations are needed.

%%%%%%%%%%%%%%%%%%%%%%%%
\subsection{Discussions}
%%%%%%%%%%%%%%%%%%%%%%%%

It is important to compare the velocity field which we obtain by solving
the continuity equation with that given by Lai, Rasio and
Shapiro. We can see the velocity field they give in their
papers\cite{LRS94,LRS93} or in the Chandrasekhar's textbook\cite{Ch69}.
In the irrotational case, it becomes
\beqa
  (u_{LRS})_1 &=&{2a_1^2 \over a_1^2 +a_2^2} \O x_2, \\
  (u_{LRS})_2 &=&-{2a_2^2 \over a_1^2 +a_2^2} \O x_1, \\
  (u_{LRS})_3 &=&0,
\eeqa
in the corotating frame. Here $a_i$ denotes the length of the principal
axis parallel to the $x_i$-axis. We can rewrite these component of
the velocity field as
\beqa
  (u_{LRS})_1 &=&\O x_2 +\Bigl( {a_1^2 -a_2^2 \over a_1^2 +a_2^2}
  \Bigr) \O x_2, \\
  (u_{LRS})_2 &=&-\O x_1 +\Bigl( {a_1^2 -a_2^2 \over a_1^2 +a_2^2}
  \Bigr) \O x_1.
\eeqa
The second terms in the above equations are order $\e^3$ because the
deviation between $a_1$ and $a_2$ is produced by the tidal force and
the effect of the tidal force is order $\e^3$.

On the other hand, if we restrict the form of the function in our velocity
field as $\four \phi_2 =(b_2/\xi_1) \xi^2$, $\five \phi_3 =(\a b_3/\xi_1)
\xi^3$ and $\six \phi_4 =(\a^2 b_4/\xi_1) \xi^4$ as in the case of the
incompressible fluid, we can express our velocity field by using
Eqs. (\ref{Eq:velo_corot_1}), (\ref{Eq:velo_corot_2}) and
(\ref{Eq:velo_corot_3}) as
\beqa
  (u_{present})_1 &=&\O x_2 +\O \bigl[ 6\e^3 b_2 x_2 -48\e^4 b_3 x_1 x_2
  +90 \e^5 b_4 x_2 (r^2 -5x_1^2) \bigr], \\
  (u_{present})_2 &=&-\O x_1 +\O \bigl[ 6\e^3 b_2 x_1 +6\e^4 b_3 (r^2
  -5x_1^2 +2x_2^2) +30\e^5 b_4 r(3r^2 +7x_1^2 -6x_2^2) \bigr], \\
  (u_{present})_3 &=&\O \bigl[ 12\e^4 b_3 x_2 x_3 +180\e^5 b_4 x_1 x_2 x_3
  \bigr],
\eeqa
where $r^2=x_1^2 +x_2^2 +x_3^2$.
Therefore, we find that the form of the velocity field we obtain in the
case of the incompressible fluid coincides with that given by Lai, Rasio
and Shapiro up to $O(\e^3)$ including the value of $b_2$.
This means that the velocity field of Lai, Rasio and Shapiro is correct
only in the case of the ``incompressible'' equation of state and
``ellipsoidal'' figures,
because the velocity field at order $\e^3$ is produced by the
ellipsoidal deformation of star 1 (see Appendix \ref{order:4th}).

Finally, we discuss about the configuration of each star. When we pay
attention to star 1, the equation for the stellar surface is written as
\beqa
  \Xi (\th, \vp) &=&\xi_1 +\e^3 S_3(\th, \vp) +\e^4 S_4(\th, \vp)
  +\e^5 S_5(\th, \vp) +\e^6 S_6(\th, \vp), \\
  &=&\xi_1 +\e^3 {\three \psi_2(\xi_1) \over |\Th_{0,\xi} (\xi_1)|}
  P_2 (\sin \th \cos \vp) +\e^4 {\four \psi_3 (\xi_1) \over
  |\Th_{0,\xi} (\xi_1)|} P_3 (\sin \th \cos \vp) +\e^5 {\five \psi_4(\xi_1)
  \over |\Th_{0,\xi} (\xi_1)|} P_4 (\sin \th \cos \vp) \nonumber \\
  &&+\e^6 {1 \over |\Th_{0,\xi} (\xi_1)|} \Bigl[ {\three \psi_2 (\xi_1)
  \over |\Th_{0,\xi} (\xi_1)|} \Bigl( {\three \psi_2 (\xi_1) \over \xi_1}
  +{d\three \psi_2 \over d\xi} (\xi_1) \Bigr)
  \Bigl\{ {18 \over 35} P_4 (\sin \th \cos \vp) +{2 \over 7} P_2 (\sin \th
  \cos \vp) +{1 \over 5} \Bigr\} \nonumber \\
  &&\hspace{50pt}+\six \psi_0 (\xi_1) +\six \psi_2 (\xi_1) P_2 (\sin \th
  \cos \vp) +\six \psi_{22} (\xi_1) P_2^2 (\cos \th) \cos 2\vp \nonumber \\
  &&\hspace{50pt}+\six \psi_4 (\xi_1) P_4 (\sin \th \cos \vp)
  +\six \psi_5 (\xi_1) P_5 (\sin \th \cos \vp) \Bigr],
  \label{Eq:stellar_surface}
\eeqa
where $S_i$ are defined in Appendix \ref{Ap:1} and we have used the
relation (\ref{Eq:ap_relation}).
Accordingly, we can express the length of the principal axis. Although
the real length of the axis is written as $\a \Xi$, we show the results
divided by $\a$.

\vspace{0.3cm}
\noindent
(1) The opposite direction to star 2:
\beqa
  \Xi \Bigl( {\pi \over 2}, 0 \Bigr) &=&\xi_1 +\e^3 {\three \psi_2 (\xi_1)
  \over |\Th_{0,\xi}
  (\xi_1)|} +\e^4 {\four \psi_3 (\xi_1) \over |\Th_{0,\xi} (\xi_1)|}
  +\e^5 {\five \psi_4 (\xi_1) \over |\Th_{0,\xi} (\xi_1)|} \nonumber \\
  &&+\e^6 {1 \over |\Th_{0,\xi} (\xi_1)|} \Bigl[ {\three \psi_2 (\xi_1)
  \over |\Th_{0,\xi} (\xi_1)|} \Bigl( {\three \psi_2 (\xi_1) \over \xi_1}
  +{d\three \psi_2 \over d\xi} (\xi_1) \Bigr) +\six \psi_0 (\xi_1)
  +\six \psi_2 (\xi_1) +3 \six \psi_{22} (\xi_1) \nonumber \\
  &&\hspace{40pt}+\six \psi_4 (\xi_1) +\six \psi_5 (\xi_1) \Bigr],
\eeqa

\noindent
(2) The direction to star 2:
\beqa
  \Xi \Bigl( {\pi \over 2}, \pi \Bigr) &=&\xi_1 +\e^3 {\three \psi_2
  (\xi_1) \over |\Th_{0,\xi}
  (\xi_1)|} -\e^4 {\four \psi_3 (\xi_1) \over |\Th_{0,\xi} (\xi_1)|}
  +\e^5 {\five \psi_4 (\xi_1) \over |\Th_{0,\xi} (\xi_1)|} \nonumber \\
  &&+\e^6 {1 \over |\Th_{0,\xi} (\xi_1)|} \Bigl[ {\three \psi_2 (\xi_1)
  \over |\Th_{0,\xi} (\xi_1)|} \Bigl( {\three \psi_2 (\xi_1) \over \xi_1}
  +{d\three \psi_2 \over d\xi} (\xi_1) \Bigr) +\six \psi_0 (\xi_1)
  +\six \psi_2 (\xi_1) +3 \six \psi_{22} (\xi_1) \nonumber \\
  &&\hspace{40pt}+\six \psi_4 (\xi_1) -\six \psi_5 (\xi_1) \Bigr],
\eeqa

\noindent
(3) The (opposite) direction to the orbital motion:
\beqa
  \Xi \Bigl( {\pi \over 2}, {\pi \over 2} \Bigr) &=&\Xi \Bigl(
  {\pi \over 2}, {3\pi \over 2} \Bigr),
  \nonumber \\
  &=&\xi_1 -\e^3 {\three \psi_2 (\xi_1) \over 2|\Th_{0,\xi}
  (\xi_1)|} +\e^5 {3\five \psi_4 (\xi_1) \over 8|\Th_{0,\xi} (\xi_1)|}
  \nonumber \\
  &&+\e^6 {1 \over |\Th_{0,\xi} (\xi_1)|} \Bigl[ {\three \psi_2 (\xi_1)
  \over 4|\Th_{0,\xi} (\xi_1)|} \Bigl( {\three \psi_2 (\xi_1) \over \xi_1}
  +{d\three \psi_2 \over d\xi} (\xi_1) \Bigr) +\six \psi_0 (\xi_1)
  -{1 \over 2}\six \psi_2 (\xi_1) -3 \six \psi_{22} (\xi_1) \nonumber \\
  &&\hspace{40pt}+{3 \over 8}\six \psi_4 (\xi_1) \Bigr],
\eeqa

\noindent
(4) The direction parallel to the rotational axis:
\beqa
  \Xi(0, 0) &=&\Xi(-\pi, 0), \nonumber \\
  &=&\xi_1 -\e^3 {\three \psi_2 (\xi_1) \over 2|\Th_{0,\xi}
  (\xi_1)|} +\e^5 {3\five \psi_4 (\xi_1) \over 8|\Th_{0,\xi} (\xi_1)|}
  \nonumber \\
  &&+\e^6 {1 \over |\Th_{0,\xi} (\xi_1)|} \Bigl[ {\three \psi_2 (\xi_1)
  \over 4|\Th_{0,\xi} (\xi_1)|} \Bigl( {\three \psi_2 (\xi_1) \over \xi_1}
  +{d\three \psi_2 \over d\xi} (\xi_1) \Bigr) +\six \psi_0 (\xi_1)
  -{1 \over 2}\six \psi_2 (\xi_1) +{3 \over 8}\six \psi_4 (\xi_1) \Bigr].
\eeqa
We can see from these equations and Tables \ref{table1} -- \ref{table8}
that the axis to star 2 is the longest, and the deviation between
the axis to star 2 and that opposite to star 2 appears at order $\e^4$.
On the contrary, the deviation between the axis to the orbital motion
and that parallel to the rotational axis appears at order $\e^6$,
and the difference is the effect of the deformation induced by the spin
of the figure ($\six \psi_{22}$).

When we see the quadrupole moments in Eq. (\ref{Eq:stellar_surface}),
we find that the coeffient of the higher order term is not large as
\beq
  \e^3 {\three \psi_2 (\xi_1) \over |\Th_{0,\xi} (\xi_1)|}
  \Bigl[ 1 +\e^3 \Bigl\{ {2 \over 7|\Th_{0,\xi} (\xi_1)|} \Bigl(
  {\three \psi_2 (\xi_1) \over \xi_1} +{d\three \psi_2 \over d\xi}
  (\xi_1) \Bigr) +{\six \psi_2 (\xi_1) \over \three \psi_2 (\xi_1)} \Bigr\}
  \Bigr] P_2 (\sin \th \cos \vp).
\eeq
Therefore we can expect the convergence of these terms. However, there
is another quadrupole moment induced by the spin of the figure. The
term $\e^6 (\six \psi_{22} (\xi_1)/|\Th_{0,\xi} (\xi_1)|) P_2^2 (\cos
\th) \cos 2\vp$ seems to be as effective as the leading quadrupole
term for the smaller orbital separation. This means that the terms
concerned with the spin of the figure which appear at
order $\e^9$ in the total energy may change the behavior of the total
energy. Furthermore, the hexadecapole moments in
Eq. (\ref{Eq:stellar_surface}),
\beq
  \e^5 {\five \psi_4 (\xi_1) \over |\Th_{0,\xi} (\xi_1)|}
  \Bigl[ 1 +\e \Bigl\{ {18 \over 35\five \psi_4 (\xi_1)} {\three \psi_2
  (\xi_1) \over |\Th_{0,\xi} (\xi_1)|} \Bigl( {\three \psi_2
  (\xi_1) \over \xi_1} +{d\three \psi_2 \over d\xi} (\xi_1) \Bigr) 
  +{\six \psi_4 (\xi_1) \over \five \psi_4 (\xi_1)} \Bigr\} \Bigr]
  P_4 (\sin \th \cos \vp),
\eeq
does not seem to converge at order $\e^6$.
However, since these terms will appear at
order $\e^{10}$ in the total energy, they do not have so much effect.
Anyway, if we discuss about the behavior of the total energy, the total
angular momentum and so on for the smaller orbital separation such as
$R/a_0 <3$, we must calculate at least up to order $\e^9$.

%%%%%%%%%%%%%%%%
\acknowledgments
%%%%%%%%%%%%%%%%
We would like to thank K. Ioka and K. Nakao for useful discussions.
This work was partly supported by a Grant-in-Aid for Scientific
Research Fellowship (No.9402: KT) and Grant-in-Aid of Scientific
Research (No.11640274, 09NP0801: TN) of the Japanese Ministry of Education,
Science, Sports and Culture.

\appendix
%%%%%%%%%%%%%%%%%%%%%%%%%%%%%%%%%%%%%%%%%%%%%%%%%%%%%%%%%%%%%%%%
\section{Summary of the equations and their boundary conditions}
\label{Ap:1}
%%%%%%%%%%%%%%%%%%%%%%%%%%%%%%%%%%%%%%%%%%%%%%%%%%%%%%%%%%%%%%%%

In the following appendices, we derive equations order by order of
$\e$, and describe in detail. In this appendix, we summarize the equations
which should be solved numerically with their boundary conditions
in each order.

First of all, we give the total equations which include all terms up to
$O(\e^6)$. The equations for determination of the velocity potential
and stellar configuration are written as
\beqa
  \t{\D} \t{\Phi} &=& -n \Bigl[ \t{\n} \t{\Phi} -(\t{\bf \O} \times
  \bxi)_{orb} -{\e \over \xi_1} (\t{\bf \O} \times \bxi)_{fig} \Bigr]
  \cdot {\t{\n} \Th \over \Th}, \\
%%%%%
  \t{\D} \Th &=&-\Th^n -{\O^2 \xi_1^2 \over 8\pi \r_c \e^2} \t{\D}
  \Bigl[ (\t{\n} \t{\Phi})^2 -2(\t{\n} \t{\Phi}) \cdot \Bigl\{ (\t{\bf 
  \O} \times \bxi)_{orb} +{\e \over \xi_1} (\t{\bf \O} \times
  \bxi)_{fig} \Bigr\} \Bigr],
\eeqa
where
\beqa
  \t{\D} &=&\a^2 \D ={1 \over \xi^2} {\p \over \p \xi} \Bigl( \xi^2
  {\p \over \p \xi} \Bigr) +{1 \over \xi^2 \sin \th} {\p \over \p \th}
  \Bigl( \sin \th {\p \over \p \th} \Bigr) +{1 \over \xi^2 \sin^2 \th}
  {\p^2 \over \p \vp^2}, \\
%%%%%
  \t{\n} &=&\a \n ={\p \over \p \xi} \hat{\bxi} +{1 \over \xi}
  {\p \over \p \th} \hat{\bth} +{1 \over \xi \sin \th} {\p \over \p \vp}
  \hat{\bvp}. \label{Eq:nomnabla}
\eeqa

The boundary condition for the velocity field is
\beq
  \Bigl[ \t{\n} \t{\Phi} -(\t{\bf \O} \times \bxi)_{orb}
  -{\e \over \xi_1} (\t{\bf \O} \times \bxi)_{fig} \Bigr] \cdot
  (\t{\n} \Th) \Bigl|_{surf} =0.
\eeq

The internal and external gravitational potentials which should be
matched at the stellar surface are
\beqa
  \t{U}_{int} &\equiv& {U^{1 \ra 1} \over 4\pi \r_c \a^2} \nonumber \\
  &=&\Th +{\O^2 \xi_1^2 \over 8\pi \r_c \e^2} \Bigl[ (\t{\n}
  \t{\Phi})^2 -2(\t{\n} \t{\Phi}) \cdot \Bigl\{ (\t{\bf \O} \times
  \bxi)_{orb} +{\e \over \xi_1} (\t{\bf \O} \times \bxi)_{fig} \Bigr\} 
  \Bigr] +\t{U}_0 \nonumber \\
  &&-{\mu \over 1+p} \e \sum_{l=0}^{\infty} (-1)^l \e^l \Bigl( {\xi
  \over \xi_1} \Bigr)^l P_l (\sin \th \cos \vp) -{3\mu \over 2M_{tot}} 
  \Bigl( {\bI_{11}' \over a_0^2} \Bigr) \e^3 \Bigl[ 1 -3\e {\xi \over
  \xi_1} P_1 (\sin \th \cos \vp)+ \cdots \Bigr], \\
%%%%%
  \t{U}_{ext} &=& {\k_0 \over \xi} +\e \sum_{l,m} {\k_{1:lm}
  \over \xi^{l+1}} Y_l^m +\e^2 \sum_{l,m} {\k_{2:lm} \over
  \xi^{l+1}} Y_l^m +\e^3 \sum_{l,m} {\k_{3:lm} \over
  \xi^{l+1}} Y_l^m +\e^4 \sum_{l,m} {\k_{4:lm} \over
  \xi^{l+1}} Y_l^m +\e^5 \sum_{l,m} {\k_{5:lm} \over
  \xi^{l+1}} Y_l^m \nonumber \\
  &&+\e^6 \sum_{l,m} {\k_{6:lm} \over \xi^{l+1}} Y_l^m + \cdots,
\eeqa
where $\k_0$ and $\k_{i:lm}$ are multipole moments.
$\t{U}_0$ is constant and expanded as
\beq
  \t{U}_0 =c_0 +\e c_1 +\e^2 c_2 +\e^3 c_3 +\e^4 c_4 +\e^5 c_5 +\e^6 c_6.
\eeq
The velocity potential and configuration function are expanded up to
$O(\e^6)$ as
\beqa
  \t{\Phi} &=&\t{\Phi}_0 +\e \t{\Phi}_1 +\e^2 \t{\Phi}_2 +\e^3
  \t{\Phi}_3 +\e^4 \t{\Phi}_4 +\e^5 \t{\Phi}_5 +\e^6 \t{\Phi}_6, \\
  \Th &=& \Th_0 +\e \Th_1 +\e^2 \Th_2 +\e^3 \Th_3 +\e^4 \Th_4 +\e^5
  \Th_5 +\e^6 \Th_6.
\eeqa

The boundary conditions for $\Th$ are
\beqa
  \t{U}_{int}(\Xi) &=&\t{U}_{ext} (\Xi), \label{Eq:bc1} \\
  {\p \t{U}_{int} \over \p \xi} (\Xi) &=&{\p \t{U}_{ext} \over \p \xi} 
  (\Xi), \label{Eq:bc2}
\eeqa
where $\Xi$ denotes the first zero point of the function $\Th$, i.e.,
the stellar surface and is formally expressed as
\beq
  \Xi(\th,\vp) =\xi_1 +\e S_1(\th,\vp) +\e^2 S_2(\th,\vp)
  +\e^3 S_3(\th,\vp) +\e^4 S_4(\th,\vp) +\e^5 S_5(\th,\vp)
  +\e^6 S_6(\th,\vp).
\eeq
The regularity conditions at the center of each star is
\beq
  {\p \Th \over \p \xi} (\xi=0) =0,
\eeq
and also we normalize the central value of $\Th$ as
\beq
  \Th(\xi=0) =1.
\eeq

In the following, we show the equation for the velocity potential, its
boundary condition, the equation for determination of the figure and
its boundary conditions (\ref{Eq:bc1}) and (\ref{Eq:bc2}).

%%%%%%%%%%%%%%%%%%%%%%
\subsection{0th order}
%%%%%%%%%%%%%%%%%%%%%%
The equation for the velocity potential is
\beq
  \t{\D} \t{\Phi}_0 =-n \Bigl[ \t{\n} \t{\Phi}_0
  -(\t{\bf \O} \times \bxi)_{orb} \Bigr] \cdot {\t{\n} \Th_0 \over \Th_0},
\eeq
and its boundary condition is
\beq
  \Bigl[ \t{\n} \t{\Phi}_0 -(\t{\bf \O} \times \bxi)_{orb} \Bigr] \cdot
  (\t{\n} \Th_0) \Bigl|_{\xi_1} =0.
\eeq
The equation for determination of the figure is
\beq
  {1 \over \xi^2} {d \over d\xi} \Bigl( \xi^2 {d\Th_0 \over d\xi} \Bigr)
  =-\Th_0^n,
\eeq
and its boundary conditions at the stellar surface are
\beqa
  &&c_0 ={\k_0 \over \xi_1}, \\
  &&{d \Th_0 \over d\xi} (\xi_1) =-{\k_0 \over \xi_1^2}.
\eeqa
The regularity condition at the center of the star is
\beq
  {d \Th_0 \over d\xi} (\xi=0) =0,
\eeq
and the normalization of $\Th_0$ at the center of the star is
\beq
  \Th_0 (\xi=0) =1.
\eeq

%%%%%%%%%%%%%%%%%%%%%%
\subsection{1st order}
%%%%%%%%%%%%%%%%%%%%%%
The equation for the velocity potential is
\beq
  \t{\D} \t{\Phi}_1 =-n \Bigl[ \t{\n} \t{\Phi}_1
  -{1 \over \xi_1} (\t{\bf \O} \times \bxi)_{fig} \Bigr] \cdot
  {\t{\n} \Th_0 \over \Th_0}, \\
\eeq
and its boundary condition is
\beq
  (\t{\n} \t{\Phi}_1) \cdot (\t{\n} \Th_0) \bigl|_{\xi_1} =0.
\eeq
The equation for determination of the figure is
\beq
  \t{\D} \Th_1 =-n\Th_0^{n-1} \Th_1,
\eeq
and its boundary conditions at the stellar surface are
\beqa
  &&c_1 -{\mu (3+2p) \over 2(1+p)^2} =-{\k_0 \over \xi_1^2} S_1
  +\sum_{l,m} {\k_{1:lm} \over \xi_1^{l+1}} Y_l^m, \\
  &&S_1 {d^2 \Th_0 \over d\xi^2}(\xi_1) +{\p \Th_1 \over \p \xi}(\xi_1)
  ={2\k_0 \over \xi_1^3} S_1 -\sum_{l,m} (l+1)
  {\k_{1:lm} \over \xi_1^{l+2}} Y_l^m.
\eeqa
The regularity condition at the center of the star is
\beq
  {\p \Th_1 \over \p \xi} (\xi=0) =0,
\eeq
and the normalization of $\Th_1$ at the center of the star is
\beq
  \Th_1 (\xi=0) =0,
\eeq
because we take $\Th_0 (\xi=0)=1$.

%%%%%%%%%%%%%%%%%%%%%%
\subsection{2nd order}
%%%%%%%%%%%%%%%%%%%%%%
The equation for the velocity potential is
\beq
  \t{\D} \t{\Phi}_2 =-n (\t{\n} \t{\Phi}_2) \cdot
  {\t{\n} \Th_0 \over \Th_0},
\eeq
and its boundary condition is
\beq
  (\t{\n} \t{\Phi}_2) \cdot (\t{\n} \Th_0) \bigl|_{\xi_1} =0.
\eeq
The equation for determination of the figure is
\beq
  \t{\D} \Th_2 =-n \Th_0^{n-1} \Th_2,
\eeq
and its boundary conditions at the stellar surface are
\beqa
  &&c_2 =-{\k_0 \over \xi_1^2} S_2 +\sum_{l,m}
  {\k_{2:lm} \over \xi_1^{l+1}} Y_l^m, \\
  &&S_2 {d^2 \Th_0 \over d\xi^2}(\xi_1) +{\p \Th_2 \over \p \xi}(\xi_1)
  ={2\k_0 \over \xi_1^3} S_2 -\sum_{l,m} (l+1)
  {\k_{2:lm} \over \xi_1^{l+2}} Y_l^m.
\eeqa
The regularity condition at the center of the star is
\beq
  {\p \Th_2 \over \p \xi} (\xi=0) =0,
\eeq
and the normalization of $\Th_2$ at the center of the star is
\beq
  \Th_2 (\xi=0) =0,
\eeq
because we take $\Th_0 (\xi=0)=1$.

%%%%%%%%%%%%%%%%%%%%%%
\subsection{3rd order}
%%%%%%%%%%%%%%%%%%%%%%
The equation for the velocity potential is
\beq
  \t{\D} \t{\Phi}_3 =-n (\t{\n} \t{\Phi}_3) \cdot
  {\t{\n} \Th_0 \over \Th_0},
\eeq
and its boundary condition is
\beq
  (\t{\n} \t{\Phi}_3) \cdot (\t{\n} \Th_0) \bigl|_{\xi_1} =0.
\eeq
The equation for determination of the figure is
\beq
  \t{\D} \Th_3 =-n\Th_0^{n-1} \Th_3,
\eeq
and its boundary conditions at the stellar surface are
\beqa
  &&c_3 -{\mu \over 1+p} P_2(\sin \th \cos \vp) =-{\k_0 \over \xi_1^2} S_3
  +\sum_{l,m} {\k_{3:lm} \over \xi_1^{l+1}} Y_l^m, \\
  &&S_3 {d^2 \Th_0 \over d\xi^2}(\xi_1) +{\p \Th_3 \over \p \xi}(\xi_1)
  -{2\mu \over \xi_1 (1+p)} P_2(\sin \th \cos \vp) ={2\k_0 \over \xi_1^3}
  S_3 -\sum_{l,m} (l+1) {\k_{3:lm} \over \xi_1^{l+2}} Y_l^m.
\eeqa
The regularity condition at the center of the star is
\beq
  {\p \Th_3 \over \p \xi} (\xi=0) =0,
\eeq
and the normalization of $\Th_3$ at the center of the star is
\beq
  \Th_3 (\xi=0) =0,
\eeq
because we take $\Th_0 (\xi=0)=1$.

%%%%%%%%%%%%%%%%%%%%%%
\subsection{4th order}
%%%%%%%%%%%%%%%%%%%%%%
The equation for the velocity potential is
\beq
  \t{\D} \t{\Phi}_4 =-{n \over \Th_0} \Bigl[ (\t{\n} \t{\Phi}_4) \cdot
  (\t{\n} \Th_0) -{1 \over \xi_1} (\t{\bf \O} \times \bxi)_{fig} \cdot
  (\t{\n} \Th_3) \Bigr],
\eeq
and its boundary condition is
\beq
  \Bigl[ (\t{\n} \t{\Phi}_4) \cdot
  (\t{\n} \Th_0) -{1 \over \xi_1} (\t{\bf \O} \times \bxi)_{fig} \cdot
  (\t{\n} \Th_3) \Bigr] \Bigl|_{\xi_1} =0.
\eeq
The equation for determination of the figure is
\beq
  \t{\D} \Th_4 =-n\Th_0^{n-1} \Th_4,
\eeq
and its boundary conditions at the stellar surface are
\beqa
  &&c_4 +{\mu \over 1+p} P_3(\sin \th \cos \vp) =-{\k_0 \over \xi_1^2} S_4
  +\sum_{l,m} {\k_{4:lm} \over \xi_1^{l+1}} Y_l^m, \\
  &&S_4 {d^2 \Th_0 \over d\xi^2}(\xi_1) +{\p \Th_4 \over \p \xi}(\xi_1)
  +{3\mu \over \xi_1(1+p)} P_3(\sin \th \cos \vp) ={2\k_0 \over \xi_1^3}
  S_4 -\sum_{l,m} (l+1) {\k_{4:lm} \over \xi_1^{l+2}} Y_l^m.
\eeqa
The regularity condition at the center of the star is
\beq
  {\p \Th_4 \over \p \xi} (\xi=0) =0,
\eeq
and the normalization of $\Th_4$ at the center of the star is
\beq
  \Th_4 (\xi=0) =0,
\eeq
because we take $\Th_0 (\xi=0)=1$.

%%%%%%%%%%%%%%%%%%%%%%
\subsection{5th order}
%%%%%%%%%%%%%%%%%%%%%%
The equation for the velocity potential is
\beq
  \t{\D} \t{\Phi}_5 =-{n \over \Th_0} \Bigl[ (\t{\n} \t{\Phi}_5) \cdot
  (\t{\n} \Th_0) -{1 \over \xi_1} (\t{\bf \O} \times \bxi)_{fig} \cdot
  (\t{\n} \Th_4) \Bigr],
\eeq
and its boundary condition is
\beq
  \Bigl[ (\t{\n} \t{\Phi}_5) \cdot
  (\t{\n} \Th_0) -{1 \over \xi_1} (\t{\bf \O} \times \bxi)_{fig} \cdot
  (\t{\n} \Th_4) \Bigr] \Bigl|_{\xi_1} =0.
\eeq
The equation for determination of the figure is
\beq
  \t{\D} \Th_5 =-n\Th_0^{n-1} \Th_5,
\eeq
and its boundary conditions at the stellar surface are
\beqa
  &&c_5 -{\mu \over 1+p} P_4(\sin \th \cos \vp) =-{\k_0 \over \xi_1^2} S_5
  +\sum_{l,m} {\k_{5:lm} \over \xi_1^{l+1}} Y_l^m, \\
  &&S_5 {d^2 \Th_0 \over d\xi^2}(\xi_1) +{\p \Th_5 \over \p \xi}(\xi_1)
  -{4\mu \over \xi_1(1+p)} P_4(\sin \th \cos \vp) ={2\k_0 \over \xi_1^3}
  S_5 -\sum_{l,m} (l+1) {\k_{5:lm} \over \xi_1^{l+2}} Y_l^m.
\eeqa
The regularity condition at the center of the star is
\beq
  {\p \Th_5 \over \p \xi} (\xi=0) =0,
\eeq
and the normalization of $\Th_5$ at the center of the star is
\beq
  \Th_5 (\xi=0) =0,
\eeq
because we take $\Th_0 (\xi=0)=1$.

%%%%%%%%%%%%%%%%%%%%%%
\subsection{6th order}
%%%%%%%%%%%%%%%%%%%%%%
The equation for the velocity potential is
\beq
  \t{\D} \t{\Phi}_6 =-{n \over \Th_0} \Bigl[ (\t{\n} \t{\Phi}_6) \cdot
  (\t{\n} \Th_0) -{1 \over \xi_1} (\t{\bf \O} \times \bxi)_{fig} \cdot
  (\t{\n} \Th_5) \Bigr],
\eeq
and its boundary condition is
\beq
  \Bigl[ (\t{\n} \t{\Phi}_6) \cdot
  (\t{\n} \Th_0) -{1 \over \xi_1} (\t{\bf \O} \times \bxi)_{fig} \cdot
  (\t{\n} \Th_5) \Bigr] \Bigl|_{\xi_1} =0.
\eeq
The equation for determination of the figure is
\beq
  \t{\D} \Th_6 =-n\Th_0^{n-1} \Th_6 -{1 \over 2} n(n-1) \Th_0^{n-2} \Th_3^2
  +{\mu \over \xi_1} \t{\D} {\p \t{\Phi}_4 \over \p \vp},
\eeq
and its boundary conditions at the stellar surface are
\beqa
  &&c_6 -{3\mu \b{\bI}_{11}' \over 2M_{tot} a_0^2}
  -{\mu \d \over 2(1+p)^2} -{2\mu \over \xi_1 (1+p)} S_3
  P_2(\sin \th \cos \vp) +{\mu \over 1+p} P_5(\sin \th \cos \vp)
  -{\mu \over \xi_1} {\p \t{\Phi}_4 \over \p \vp}(\xi_1) \nonumber \\
  &&\hspace{30pt}={\k_0 \over \xi_1^2} \Bigl( {S_3^2 \over \xi_1} -S_6 \Bigr)
  -\sum_{l,m} (l+1) {\k_{3:lm} \over \xi_1^{l+2}} S_3 Y_l^m
  +\sum_{l,m} {\k_{6:lm} \over \xi_1^{l+1}} Y_l^m, \\
%%%%%
  &&S_6 {d^2 \Th_0 \over d\xi^2}(\xi_1) +{1 \over 2} S_3^2
  {d^3 \Th_0 \over d\xi^3}(\xi_1) +S_3 {\p^2 \Th_3 \over \p \xi^2} (\xi_1)
  +{\p \Th_6 \over \p \xi}(\xi_1) -{2\mu \over \xi_1^2 (1+p)} S_3
  P_2(\sin \th \cos \vp) +{5\mu \over \xi_1 (1+p)} P_5(\sin \th \cos \vp)
  \nonumber \\
  &&\hspace{30pt}-{\mu \over \xi_1} {\p^2 \t{\Phi}_4 \over \p \xi \p \vp}
  (\xi_1) =-{\k_0 \over \xi_1^3} \Bigl( {3S_3^2 \over \xi_1} -2S_6
  \Bigr) +\sum_{l=0}^{\infty} (l+1)(l+2) {\k_{3:lm} \over \xi_1^{l+3}}
  S_3 Y_l^m -\sum_{l=0}^{\infty} (l+1) {\k_{6:lm} \over \xi_1^{l+2}}
  Y_l^m,
\eeqa
where
\beq
  \d \equiv {9 \over 2a_0^2} \Bigl[ {\b{\bI}_{11} \over M_1}
  +\Bigl( {a_0' \over a_0} \Bigr)^3 {\b{\bI}_{11}' \over M_2} \Bigr].
\eeq
The regularity condition at the center of the star is
\beq
  {\p \Th_6 \over \p \xi} (\xi=0) =0,
\eeq
and the normalization of $\Th_6$ at the center of the star is
\beq
  \Th_6 (\xi=0) =0,
\eeq
because we take $\Th_0 (\xi=0)=1$.

%%%%%%%%%%%%%%%%%%%%%%%%%%%%%%%%%%%
\section{Solutions at zeroth order}
%%%%%%%%%%%%%%%%%%%%%%%%%%%%%%%%%%%

Since the configuration of star 1 is spherical at 0th order,
the equation for determination of the figure becomes
{\it Lane-Emden equation},
\beq
  {1 \over \xi^2} {d \over d\xi} \Bigl( \xi^2 {d\Th_0 \over d\xi}
  \Bigr) =-\Th_0^n. \label{Eq:Lane-Emden}
\eeq
For example, we obtain analytic solutions,
\beqa
  \Th_0 &=&1-{\xi^2 \over 6}~~~~~({\rm for~} n=0), \\
  \Th_0 &=&{\sin \xi \over \xi}~~~~~({\rm for~} n=1),
\eeqa
or numerical solutions by solving the ordinary differential equation
(\ref{Eq:Lane-Emden}).

The velocity field at 0th order is the same as the
orbital motion so that the velocity potential is determined by
\beq
  \n \Phi_0 =\O R (\t{\bf \O} \times \bxi)_{orb}.
\eeq
Then, the solution should be
\beq
  \Phi_0={\O R \over 1+p} r \sin \th \sin \vp.
\eeq
{}From this equation, we find that it is convenient to normalize
$\Phi_0$ as
\beq
  \t{\Phi}_0 ={1 \over 1+p} \xi \sin \th \sin \vp.
\eeq
In the following, we use the normalized velocity potential defined by
\beq
  \t{\Phi} ={\e \Phi \over \O \a a_0}.
\eeq

Here we show the useful relations:
\beqa
  &&(\t{\n} \t{\Phi}_0)^2 ={1 \over (1+p)^2}, \\
  &&(\t{\n} \t{\Phi}_0) \cdot (\t{\bf \O} \times \bxi)_{orb} ={1 \over
  (1+p)^2}, \\
  &&(\t{\n} \t{\Phi}_0) \cdot (\t{\bf \O} \times \bxi)_{fig} ={1 \over 
  1+p} \xi \sin \th \cos \vp.
\eeqa

The orbital angular velocity can be calculated from Eq. (\ref{Eq:firstTV})
as
\beq
  0=\int_{star~1} d^3 x \r \Bigl( -{M_2 \over R^2} +{\O_0^2 R \over 1+p}
  \Bigr).
\eeq
Therefore, we have
\beq
  \O_0^2 ={M_{tot} \over R^3},
\eeq
where $M_{tot} =M_1 +M_2$.

For 0th order, we have the internal and external gravitational potentials
and their derivatives as
\beqa
  \t{U}_{int}(\xi) &=&\Th_0 +c_0, \\
  {\p \t{U}_{int} \over \p \xi}(\xi) &=& {d\Th_0 \over d\xi}, \\
  \t{U}_{ext}(\xi) &=& {\k_0 \over \xi}, \\
  {\p \t{U}_{ext} \over \p \xi}(\xi) &=&-{\k_0 \over \xi^2}.
\eeqa
They become at the stellar surface as
\beqa
  \t{U}_{int} (\xi_1) &=&c_0, \\
  {\p \t{U}_{int} \over \p \xi} (\xi_1) &=&{d\Th_0 \over d\xi}(\xi_1), \\
  \t{U}_{ext} (\xi_1) &=& {\k_0 \over \xi_1}, \\
  {\p \t{U}_{ext} \over \p \xi} (\xi_1) &=& -{\k_0 \over \xi_1^2}.
\eeqa
Therefore, we can decide two constants at 0th order from the boundary
conditions at the stellar surface as
\beqa
  \k_0 &=& -\xi_1^2 {d\Th_0 \over d\xi}(\xi_1), \\
  c_0 &=& -\xi_1 {d\Th_0 \over d\xi}(\xi_1).
\eeqa

%%%%%%%%%%%%%%%%%%%%%%%%%%%%%%%%%%
\section{Solutions at first order}
%%%%%%%%%%%%%%%%%%%%%%%%%%%%%%%%%%

The equation for the velocity potential is
\beqa
  \t{\D} \t{\Phi}_1 &=&-n \Bigl[ (\t{\n} \t{\Phi}_1) -{1 \over \xi_1}
  (\t{\bf \O} \times \bxi)_{fig} \Bigr] \cdot {\t{\n} \Th_0 \over
  \Th_0}, \\
  &=&-n (\t{\n} \t{\Phi}_1) \cdot {\t{\n} \Th_0 \over \Th_0},
\eeqa
where we have used the fact that $\t{\n} \Th_0$ has only $\xi$ component
and $(\t{\bf \O} \times \bxi)_{fig}$ has only $\vp$ component.
The boundary condition is
\beq
  (\t{\n} \t{\Phi}_1) \cdot (\t{\n} \Th_0) \bigl|_{\xi_1} =0.
\eeq
When we expand $\t{\Phi}_1$ as
\beq
  \t{\Phi}_1 =\sum_{l,m} \one \phi_{lm}(\xi) \t{Y}_l^m(\th,\vp),
\eeq
the boundary conditions are that $\one \phi_{lm}$ is regular at $\xi=0$
and 
\beq
  {d \one \phi_{lm} \over d\xi} (\xi_1) =0.
\eeq
We cannot determine the absolute value of $\t{\Phi}_1$ from the
equation and the boundary condition. Therefore, $\t{\Phi}_1$ should be 
zero. Note that $\t{\Phi}_1= {\rm constant}$ is a solution. However,
since the physical value is a velocity field $\t{\n} \t{\Phi}_1$, the
constant solution is meaningless.

Next, we consider the deformation of the figure at $O(\e)$. The
equation is written as
\beq
  \t{\D} \Th_1 =-n \Th_0^{n-1} \Th_1.
\eeq
We expand $\Th_1$ as
\beq
  \Th_1 =\sum_{l,m} \one \psi_{lm}(\xi) Y_l^m(\th,\vp).
\eeq
The surface equation up to $O(\e)$ is written as
\beq
  \Xi =\xi_1 +\e S_1(\th,\vp),
\eeq
where
\beq
  S_1=-{\Th_1 (\xi_1) \over \Th_{0,\xi}(\xi_1)}.
\eeq

The gravitational potentials and their derivatives up to $O(\e)$ are
\beqa
  \t{U}_{int} &=& \Th -\e {\mu (3+2p) \over 2(1+p)^2} +c_0 +\e c_1, \\
  {\p \t{U}_{int} \over \p \xi} &=&{\p \Th \over \p \xi}, \\
  \t{U}_{ext} &=&{\k_0 \over \xi} +\e \sum_{l,m} {\k_{1:lm} \over
  \xi^{l+1}} Y_l^m, \\
  {\p \t{U}_{ext} \over \p \xi} &=& -{\k_0 \over \xi^2} -\e \sum_{l,m}
  (l+1) {\k_{1:lm} \over \xi^{l+2}} Y_l^m.
\eeqa
They become at the stellar surface as
\beqa
  \t{U}_{int} (\Xi) &=&c_0 +\e \Bigl[ c_1 -{\mu (3+2p) \over 2(1+p)^2} 
  \Bigr], \\
%%%%%
  {\p \t{U}_{int} \over \p \xi} (\Xi) &=&{d\Th_0 \over d\xi}(\xi_1)
  +\e \Bigl[ S_1 {d^2 \Th_0 \over d\xi^2}(\xi_1)
  +{\p \Th_1 \over \p \xi}(\xi_1) \Bigr], \\
%%%%%
  \t{U}_{ext} (\Xi) &=& {\k_0 \over \xi_1} \Bigl( 1-\e {S_1 \over \xi_1} 
  \Bigr) +\e \sum_{l,m} {\k_{1:lm} \over \xi_1^{l+1}} Y_l^m, \\
%%%%%
  {\p \t{U}_{ext} \over \p \xi} (\Xi) &=& -{\k_0 \over \xi_1^2} \Bigl(
  1-\e {2S_1 \over \xi_1} \Bigr) -\e \sum_{l,m} (l+1) {\k_{1:lm} \over
  \xi_1^{l+2}} Y_l^m.
\eeqa
Then, the boundary conditions for $O(\e)$ become
\beqa
  &&c_1 -{\mu (3+2p) \over 2(1+p)^2} =-{\k_0 \over \xi_1^2} S_1 +\sum_{l,m}
  {\k_{1:lm} \over \xi_1^{l+1}} Y_l^m, \\
  &&S_1 {d^2 \Th_0 \over d\xi^2}(\xi_1) +{\p \Th_1 \over \p \xi}(\xi_1)
  ={2\k_0 \over \xi_1^3} S_1 -\sum_{l,m} (l+1) {\k_{1:lm}
  \over \xi_1^{l+2}} Y_l^m,
\eeqa
and also $\Th_1$ should be satisfied with the conditions at the center of
the star,
\beqa
  {\p \Th_1 \over \p \xi} (0) &=&0, \\
  \Th_1 (0) &=&0.
\eeqa

%%%%%%%%%%%%%%%%%%%%%%%%%%%%%%
\subsection{The case of $n=0$}
%%%%%%%%%%%%%%%%%%%%%%%%%%%%%%
In this case, since we can express $\one \psi_{lm} =\one A_{lm} \xi^l$
by using a constant $\one A_{lm}$, the boundary conditions can be written
as
\beqa
  &&c_1 -{\mu (3+2p) \over 2(1+p)^2} =\sum_{l,m} \Bigl( -\one A_{lm} \xi_1^l
  +{\k_{1:lm} \over \xi_1^{l+1}} \Bigr) Y_l^m, \\
  &&0=\sum_{l,m} \Bigl[ (3-l) \one A_{lm} \xi_1^{l-1}
  -(l+1) {\k_{1:lm} \over \xi_1^{l+2}} \Bigr] Y_l^m.
\eeqa
Therefore, we obtain
\beqa
  c_1 &=& {\mu (3+2p) \over 2(1+p)^2}, \\
  \k_{1:lm} &=& 0, \\
  \Th_1 &=&0.
\eeqa

%%%%%%%%%%%%%%%%%%%%%%%%%%%%%%%
\subsection{The case of $n >0$}
%%%%%%%%%%%%%%%%%%%%%%%%%%%%%%%
In this case, we have the relation;
\beq
  {d^2 \Th_0 \over d\xi^2}(\xi_1) =-{2 \over \xi_1} {d\Th_0 \over d\xi}
  (\xi_1), \label{Eq:ap_relation}
\eeq
from the Lane-Emden equation at the stellar surface.
Then, the boundary conditions can be written as
\beqa
  &&c_1 -{\mu (3+2p) \over 2(1+p)^2} =\sum_{l,m} \Bigl[ -\one \psi_{lm}
  (\xi_1) +{\k_{1:lm} \over \xi_1^{l+1}} \Bigr] Y_l^m, \\
  &&0=\sum_{l,m} \Bigl[ -{d\one \psi_{lm} \over d\xi}(\xi_1)
  -(l+1) {\k_{1:lm} \over \xi_1^{l+2}} \Bigr] Y_l^m.
\eeqa
Therefore, we obtain
\beqa
  c_1 &=& {\mu (3+2p) \over 2(1+p)^2}, \\
  \k_{1:lm} &=& 0, \\
  \Th_1 &=&0.
\eeqa

%%%%%%%%%%%%%%%%%%%%%%%%%%%%%%%%%%%
\section{Solutions at second order}
%%%%%%%%%%%%%%%%%%%%%%%%%%%%%%%%%%%

The equation for the velocity potential is
\beq
  \t{\D} \t{\Phi}_2 =-n (\t{\n} \t{\Phi}_2) \cdot {\t{\n} \Th_0 \over
  \Th_0}.
\eeq
The boundary condition is
\beq
  (\t{\n} \t{\Phi}_2) \cdot (\t{\n} \Th_0) \bigl|_{\xi_1} =0.
\eeq
Then, $\t{\Phi}_2$ should be zero as in the case of $\t{\Phi}_1$.

Next, we consider the deformation of the figure at $O(\e^2)$. The
equation is written as
\beq
  \t{\D} \Th_2 =-n \Th_0^{n-1} \Th_2.
\eeq
We expand $\Th_2$ as
\beq
  \Th_2 =\sum_{l,m} \two \psi_{lm}(\xi) Y_l^m(\th,\vp).
\eeq
The surface equation up to $O(\e^2)$ is written as
\beq
  \Xi =\xi_1 +\e^2 S_2(\th,\vp),
\eeq
where
\beq
  S_2=-{\Th_2 (\xi_1) \over \Th_{0,\xi}(\xi_1)}.
\eeq

The gravitational potentials and their derivatives up to $O(\e^2)$ are
\beqa
  \t{U}_{int} &=& \Th +c_0 +\e^2 c_2, \\
  {\p \t{U}_{int} \over \p \xi} &=&{\p \Th \over \p \xi}, \\
  \t{U}_{ext} &=&{\k_0 \over \xi} +\e^2 \sum_{l,m} {\k_{2:lm} \over
  \xi^{l+1}} Y_l^m, \\
  {\p \t{U}_{ext} \over \p \xi} &=& -{\k_0 \over \xi^2} -\e^2 \sum_{l,m}
  (l+1) {\k_{2:lm} \over \xi^{l+2}} Y_l^m.
\eeqa
They become at the stellar surface as
\beqa
  \t{U}_{int} (\Xi) &=&c_0 +\e^2 c_2, \\
%%%%%
  {\p \t{U}_{int} \over \p \xi} (\Xi) &=&{d\Th_0 \over d\xi}(\xi_1)
  +\e^2 \Bigl[ S_2 {d^2 \Th_0 \over d\xi^2}(\xi_1)
  +{\p \Th_2 \over \p \xi}(\xi_1) \Bigr], \\
%%%%%
  \t{U}_{ext} (\Xi) &=& {\k_0 \over \xi_1} \Bigl( 1-\e^2 {S_2 \over \xi_1} 
  \Bigr) +\e^2 \sum_{l,m} {\k_{2:lm} \over \xi_1^{l+1}} Y_l^m, \\
%%%%%
  {\p \t{U}_{ext} \over \p \xi} (\Xi) &=& -{\k_0 \over \xi_1^2} \Bigl(
  1-\e^2 {2S_2 \over \xi_1} \Bigr) -\e^2 \sum_{l,m} (l+1) {\k_{2:lm} \over
  \xi_1^{l+2}} Y_l^m.
\eeqa
Then, the boundary conditions for $O(\e^2)$ become
\beqa
  &&c_2 =-{\k_0 \over \xi_1^2} S_2 +\sum_{l,m}
  {\k_{2:lm} \over \xi_1^{l+1}} Y_l^m, \\
  &&S_2 {d^2 \Th_0 \over d\xi^2}(\xi_1) +{\p \Th_2 \over \p \xi}(\xi_1)
  ={2\k_0 \over \xi_1^3} S_2 -\sum_{l,m} (l+1) {\k_{2:lm} \over 
  \xi_1^{l+2}} Y_l^m,
\eeqa
and also $\Th_2$ should be satisfied with the conditions at the center of
the star,
\beqa
  {\p \Th_2 \over \p \xi} (0) &=&0, \\
  \Th_2 (0) &=&0.
\eeqa

%%%%%%%%%%%%%%%%%%%%%%%%%%%%%%
\subsection{The case of $n=0$}
%%%%%%%%%%%%%%%%%%%%%%%%%%%%%%
In this case, since we can express $\two \psi_{lm} =\two A_{lm} \xi^l$
by using a constant $\two A_{lm}$, the boundary conditions can be written
as
\beqa
  &&c_2 =\sum_{l,m} \Bigl( -\two A_{lm} \xi_1^l
  +{\k_{2:lm} \over \xi_1^{l+1}} \Bigr) Y_l^m, \\
  &&0=\sum_{l,m} \Bigl[ (3-l) \two A_{lm} \xi_1^{l-1}
  -(l+1) {\k_{2:lm} \over \xi_1^{l+2}} \Bigr] Y_l^m.
\eeqa
Therefore, we obtain
\beqa
  c_2 &=& 0, \\
  \k_{2:lm} &=&0, \\
  \Th_2 &=&0.
\eeqa

%%%%%%%%%%%%%%%%%%%%%%%%%%%%%%%
\subsection{The case of $n >0$}
%%%%%%%%%%%%%%%%%%%%%%%%%%%%%%%
In this case, the boundary conditions can be written as
\beqa
  &&c_2 =\sum_{l,m} \Bigl[ -\two \psi_{lm} (\xi_1)
  +{\k_{2:lm} \over \xi_1^{l+1}} \Bigr] Y_l^m, \\
  &&0=\sum_{l,m} \Bigl[ -{d\two \psi_{lm} \over d\xi}(\xi_1)
  -(l+1) {\k_{2:lm} \over \xi_1^{l+2}} \Bigr] Y_l^m.
\eeqa
Therefore, we obtain
\beqa
  c_2 &=&0, \\
  \k_{2:lm} &=&0, \\
  \Th_2 &=&0.
\eeqa

%%%%%%%%%%%%%%%%%%%%%%%%%%%%%%%%%%
\section{Solutions at third order}
%%%%%%%%%%%%%%%%%%%%%%%%%%%%%%%%%%

The equation for the velocity potential is
\beq
  \t{\D} \t{\Phi}_3 =-n (\t{\n} \t{\Phi}_3) \cdot {\t{\n} \Th_0 \over
  \Th_0}.
\eeq
The boundary condition is
\beq
  (\t{\n} \t{\Phi}_3) \cdot (\t{\n} \Th_0) \bigl|_{\xi_1} =0.
\eeq
Then, $\t{\Phi}_3$ should be zero as in the cases of $\t{\Phi}_1$ and
$\t{\Phi}_2$.

Next, we consider the deformation of the figure at $O(\e^3)$. The
equation is written as
\beq
  \t{\D} \Th_3 =-n \Th_0^{n-1} \Th_3.
\eeq
We expand $\Th_3$ as
\beq
  \Th_3 =\sum_{l,m} \three \psi_{lm}(\xi) Y_l^m(\th,\vp).
\eeq
The surface equation up to $O(\e^3)$ is written as
\beq
  \Xi =\xi_1 +\e^3 S_3(\th,\vp),
\eeq
where
\beq
  S_3=-{\Th_3 (\xi_1) \over \Th_{0,\xi}(\xi_1)}.
\eeq

The gravitational potentials and their derivatives up to $O(\e^3)$ are
\beqa
  \t{U}_{int} &=& \Th +c_0 +\e^3 \Bigl[ c_3 -{\mu \over \xi_1^2 (1+p)} 
  \xi^2 P_2 (\sin \th \cos \vp) \Bigr], \\
%%%%%
  {\p \t{U}_{int} \over \p \xi} &=&{\p \Th \over \p \xi} -\e^3 {2\mu
  \over \xi_1^2 (1+p)} \xi P_2 (\sin \th \cos \vp), \\
%%%%%
  \t{U}_{ext} &=&{\k_0 \over \xi} +\e^3 \sum_{l,m} {\k_{3:lm} \over
  \xi^{l+1}} Y_l^m, \\
%%%%%
  {\p \t{U}_{ext} \over \p \xi} &=& -{\k_0 \over \xi^2} -\e^3 \sum_{l,m}
  (l+1) {\k_{3:lm} \over \xi^{l+2}} Y_l^m.
\eeqa
They become at the stellar surface as
\beqa
  \t{U}_{int} (\Xi) &=&c_0 +\e^3 \Bigl[ c_3 -{\mu \over 1+p} P_2 (\sin 
  \th \cos \vp) \Bigr], \\
%%%%%
  {\p \t{U}_{int} \over \p \xi} (\Xi) &=&{d\Th_0 \over d\xi}(\xi_1)
  +\e^3 \Bigl[ S_3 {d^2 \Th_0 \over d\xi^2}(\xi_1)
  +{\p \Th_3 \over \p \xi}(\xi_1) -{2\mu \over \xi_1 (1+p)} P_2 (\sin
  \th \cos \vp) \Bigr], \\
%%%%%
  \t{U}_{ext} (\Xi) &=& {\k_0 \over \xi_1} \Bigl( 1-\e^3 {S_3 \over \xi_1} 
  \Bigr) +\e^3 \sum_{l,m} {\k_{3:lm} \over \xi_1^{l+1}} Y_l^m, \\
%%%%%
  {\p \t{U}_{ext} \over \p \xi} (\Xi) &=& -{\k_0 \over \xi_1^2} \Bigl(
  1-\e^3 {2S_3 \over \xi_1} \Bigr) -\e^3 \sum_{l,m} (l+1) {\k_{3:lm} \over
  \xi_1^{l+2}} Y_l^m.
\eeqa
Then, the boundary conditions for $O(\e^3)$ become
\beqa
  &&c_3 -{\mu \over 1+p} P_2 (\sin \th \cos \vp) =-{\k_0 \over
  \xi_1^2} S_3 +\sum_{l,m} {\k_{3:lm} \over \xi_1^{l+1}} Y_l^m, \\
  &&S_3 {d^2 \Th_0 \over d\xi^2}(\xi_1) +{\p \Th_3 \over \p \xi}(\xi_1)
  -{2\mu \over \xi_1 (1+p)} P_2 (\sin \th \cos
  \vp) ={2\k_0 \over \xi_1^3} S_3 -\sum_{l,m} (l+1) {\k_{3:lm}
  \over \xi_1^{l+2}} Y_l^m,
\eeqa
and also $\Th_3$ should be satisfied with the conditions at the center of
the star,
\beqa
  {\p \Th_3 \over \p \xi} (0) &=&0, \\
  \Th_3 (0) &=&0.
\eeqa

%%%%%%%%%%%%%%%%%%%%%%%%%%%%%%
\subsection{The case of $n=0$}
%%%%%%%%%%%%%%%%%%%%%%%%%%%%%%
In this case, since we can express $\three \psi_{lm} =\three A_{lm} \xi^l$
by using a constant $\three A_{lm}$,
the boundary conditions can be written as
\beqa
  &&c_3 -{\mu \over 1+p} P_2 (\sin \th \cos \vp) =\sum_{l,m} \Bigl( -\three
  A_{lm} \xi_1^l +{\k_{3:lm} \over \xi_1^{l+1}} \Bigr) Y_l^m, \\
  &&-{2\mu \over \xi_1 (1+p)} P_2 (\sin \th \cos \vp) =\sum_{l,m} \Bigl[
  (3-l) \three A_{lm} \xi_1^{l-1}
  -(l+1) {\k_{3:lm} \over \xi_1^{l+2}} \Bigr] Y_l^m.
\eeqa
Therefore, we obtain
\beqa
  c_3 &=&0, \\
  \Th_3 &=& \three A_2 \xi^2 P_2 (\sin \th \cos \vp),
\eeqa
where
\beqa
  \three A_2 &=&{5\mu \over 2(1+p) \xi_1^2}, \\
  \k_{3:2} &=&{3\mu \xi_1^3 \over 2(1+p)}.
\eeqa

%%%%%%%%%%%%%%%%%%%%%%%%%%%%%%
\subsection{The case of $n=1$}
%%%%%%%%%%%%%%%%%%%%%%%%%%%%%%
In this case, since we can express $\three \psi_{lm} =\three B_{lm} j_l$
by using a constant $\three B_{lm}$ and a spherical Bessel function $j_l$,
the boundary conditions can be written as
\beqa
  &&c_3 -{\mu \over 1+p} P_2 (\sin \th \cos \vp) =\sum_{l,m} \Bigl[ -\three
  B_{lm} j_l(\xi_1) +{\k_{3:lm} \over \xi_1^{l+1}} \Bigr] Y_l^m, \\
  &&-{2\mu \over \xi_1 (1+p)} P_2 (\sin \th \cos \vp) =\sum_{l,m}
  \Bigl[ -\three
  B_{lm} {dj_l \over d\xi}(\xi_1) -(l+1) {\k_{3:lm} \over \xi_1^{l+2}}
  \Bigr] Y_l^m.
\eeqa
Therefore, we obtain
\beqa
  c_3 &=&0, \\
  \Th_3 &=& \three B_2 j_2 (\xi) P_2 (\sin \th \cos \vp),
\eeqa
where
\beqa
  \three B_2 &=& {5\mu \over 1+p}, \\
  \k_{3:2} &=& {\mu \xi_1 \over 1+p} (15-\xi_1^2), \\
  j_2(\xi) &=& {1 \over \xi^3} \Bigl[ (3 -\xi^2) \sin \xi -3\xi \cos \xi
  \Bigr].
\eeqa

%%%%%%%%%%%%%%%%%%%%%%%%%%%%%%%
\subsection{The case of $n >0$}
%%%%%%%%%%%%%%%%%%%%%%%%%%%%%%%
In this case, the boundary conditions can be written as
\beqa
  &&c_3 -{\mu \over 1+p} P_2 (\sin \th \cos \vp) =\sum_{l,m} \Bigl[ -\three
  \psi_{lm}(\xi_1) +{\k_{3:lm} \over \xi_1^{l+1}} \Bigr] Y_l^m, \\
  &&-{2\mu \over \xi_1 (1+p)} P_2 (\sin \th \cos \vp) =\sum_{l,m} \Bigl[
  -{d \three \psi_{lm} \over d\xi}(\xi_1)
  -(l+1) {\k_{3:lm} \over \xi_1^{l+2}} \Bigr] Y_l^m.
\eeqa
Therefore,
\beqa
  c_3 &=&0, \\
  \Th_3 &=& \three \psi_2 (\xi) P_2 (\sin \th \cos \vp),
\eeqa
and the ordinary differential equation for $\three \psi_2$ is written as
\beq
  \Bigl[ {1 \over \xi^2} {d \over d\xi} \Bigl( \xi^2 {d \over d\xi} \Bigr)
  -{6 \over \xi^2} \Bigr] \three \psi_2 = -n \Th_0^{n-1} \three \psi_2.
\eeq
The coefficient of $\three \psi_2$ and the constant $\k_{3:2}$
are determined from two equations;
\beqa
  &&3 \three \psi_2(\xi_1) +\xi_1 {d\three \psi_2 \over d\xi}(\xi_1)
  ={5\mu \over 1+p}, \\
  &&\k_{3:2}= \xi_1^3 \Bigl[ \three \psi_2(\xi_1) -{\mu \over 1+p} \Bigr]
  ={\xi_1^3 \over 3} \Bigl[ {2\mu \over 1+p} -\xi_1 {d\three \psi_2 \over
  d\xi}(\xi_1) \Bigr].
\eeqa

%%%%%%%%%%%%%%%%%%%%%%%%%%%%%%%%%%%%%%%
\subsection{Reduced quadrupole moments}
%%%%%%%%%%%%%%%%%%%%%%%%%%%%%%%%%%%%%%%
At this order, we obtain the reduced quadrupole moments as
\beqa
  \bI_{11} &=&{2 \over 3} \int_{star~1} d^3 x \r r^2
  P_2 (\sin \th \cos \vp), \\
  &=&\left\{
	\begin{array}{@{\,}ll}
	  \e^3 \displaystyle{8\pi \over 5} \r_c \a^5 \xi_1^5 \three A_2,
	  ~~~~~(n=0) \\
	  \e^3 \displaystyle{8\pi \over 15} \r_c \a^5 \xi_1^3
          \Bigl( {15 \over \xi_1^2} -1 \Bigr) \three B_2,~~~~~(n=1) \\
	  \e^3 \displaystyle{8\pi \over 15} \r_c \a^5 \int_0^{\xi_1} d
  	  \xi n \Th_0^{n-1} (\xi) \three \psi_2 (\xi) \xi^4.
	  ~~~~~(n >0)
	\end{array}
     \right.
\eeqa

%%%%%%%%%%%%%%%%%%%%%%%%%%%%%%%%%%%
\section{Solutions at fourth order} \label{order:4th}
%%%%%%%%%%%%%%%%%%%%%%%%%%%%%%%%%%%

The equation for the velocity potential is
\beqa
  \t{\D} \t{\Phi}_4 &=&-{n \over \Th_0} \Bigl[ (\t{\n} \t{\Phi}_4) \cdot
  (\t{\n} \Th_0) -{1 \over \xi_1} (\t{\bf \O} \times \bxi)_{fig} \cdot 
  (\t{\n} \Th_3) \Bigr], \\
  &=&-{n \over \Th_0} \Bigl[ (\t{\n} \t{\Phi}_4) \cdot (\t{\n} \Th_0)
  +{1 \over 2\xi_1} \three \psi_2 P_2^2 (\cos \th) \sin 2\vp \Bigr].
\eeqa
Therefore, we have the form of $\t{\Phi}_4$ and the equation for it as
\beqa
  &&\t{\Phi}_4 =\four \phi_2 (\xi) P_2^2 (\cos \th) \sin 2\vp, \\
  &&\Bigl[ {1 \over \xi^2} {d \over d\xi} \Bigl( \xi^2 {d \over d\xi}
  \Bigr) -{6 \over \xi^2} +{n \over \Th_0} {d\Th_0 \over d\xi} {d
  \over d\xi} \Bigr] \four \phi_2 =-{n \over 2\xi_1 \Th_0} \three
  \psi_2.
\eeqa
The boundary condition is
\beq
  {d\four \phi_2 \over d\xi}(\xi_1) {d\Th_0 \over d\xi}(\xi_1)
  +{1 \over 2\xi_1} \three \psi_2 (\xi_1) =0.
\eeq

Next, we consider the deformation of the figure at $O(\e^4)$. The
equation is written as
\beq
  \t{\D} \Th_4 =-n \Th_0^{n-1} \Th_4.
\eeq
We expand $\Th_4$ as
\beq
  \Th_4 =\sum_{l,m} \four \psi_{lm}(\xi) Y_l^m(\th,\vp).
\eeq
The surface equation up to $O(\e^4)$ is written as
\beq
  \Xi =\xi_1 +\e^3 S_3(\th,\vp) +\e^4 S_4(\th,\vp),
\eeq
where
\beq
  S_4=-{\Th_4 (\xi_1) \over \Th_{0,\xi}(\xi_1)}.
\eeq

The gravitational potentials and their derivatives up to $O(\e^4)$ are
\beqa
  \t{U}_{int} &=& \Th +c_0 -\e^3 {\mu \over \xi_1^2 (1+p)} 
  \xi^2 P_2 (\sin \th \cos \vp) +\e^4 \Bigl[ c_4 +{\mu \over \xi_1^3
  (1+p)} \xi^3 P_3 (\sin \th \cos \vp) \Bigr], \\
%%%%%
  {\p \t{U}_{int} \over \p \xi} &=&{\p \Th \over \p \xi} -\e^3 {2\mu
  \over \xi_1^2 (1+p)} \xi P_2 (\sin \th \cos \vp) +\e^4 {3\mu \over
  \xi_1^3 (1+p)} \xi^2 P_3 (\sin \th \cos \vp), \\
%%%%%
  \t{U}_{ext} &=&{\k_0 \over \xi} +\e^3 {\k_{3:2} \over \xi^3}
  P_2 (\sin \th \cos \vp)
  +\e^4 \sum_{l,m} {\k_{4:lm} \over \xi^{l+1}} Y_l^m, \\
%%%%%
  {\p \t{U}_{ext} \over \p \xi} &=& -{\k_0 \over \xi^2} -\e^3
  {3\k_{3:2} \over \xi^4} P_2 (\sin \th \cos \vp) -\e^4 \sum_{l,m} (l+1)
  {\k_{4:lm} \over \xi^{l+2}} Y_l^m.
\eeqa
They become at the stellar surface as
\beqa
  \t{U}_{int} (\Xi) &=&c_0 -\e^3 {\mu \over 1+p} P_2 (\sin 
  \th \cos \vp) +\e^4 \Bigl[ c_4 +{\mu \over 1+p} P_3 (\sin \th \cos
  \vp) \Bigr], \\
%%%%%
  {\p \t{U}_{int} \over \p \xi} (\Xi) &=&{d\Th_0 \over d\xi}(\xi_1)
  +\e^3 \Bigl[ S_3 {d^2 \Th_0 \over d\xi^2}(\xi_1)
  +{\p \Th_3 \over \p \xi}(\xi_1) -{2\mu \over \xi_1 (1+p)} P_2 (\sin
  \th \cos \vp) \Bigr] \nonumber \\
  &&+\e^4 \Bigl[ S_4 {d^2 \Th_0 \over d\xi^2}(\xi_1)
  +{\p \Th_4 \over \p \xi}(\xi_1)
  +{3 \mu \over \xi_1(1+p)} P_3 (\sin \th \cos \vp) \Bigr], \\
%%%%%
  \t{U}_{ext} (\Xi) &=& {\k_0 \over \xi_1} \Bigl( 1-\e^3 {S_3 \over \xi_1} 
  -\e^4 {S_4 \over \xi_1} \Bigr) +\e^3 {\k_{3:2} \over
  \xi_1^3} P_2 (\sin \th \cos \vp)
  +\e^4 \sum_{l,m} {\k_{4:lm} \over \xi_1^{l+1}} Y_l^m, \\
%%%%%
  {\p \t{U}_{ext} \over \p \xi} (\Xi) &=& -{\k_0 \over \xi_1^2} \Bigl(
  1-\e^3 {2S_3 \over \xi_1} -\e^4 {2S_4 \over \xi_1} \Bigr) -\e^3
  {3\k_{3:2} \over \xi_1^4} P_2 (\sin \th \cos \vp) -\e^4 \sum_{l,m}
  (l+1) {\k_{4:lm} \over \xi_1^{l+2}} Y_l^m.
\eeqa
Then, the boundary conditions for $O(\e^4)$ become
\beqa
  &&c_4 +{\mu \over 1+p} P_3 (\sin \th \cos \vp) =-{\k_0 \over
  \xi_1^2} S_4 +\sum_{l,m} {\k_{4:lm} \over \xi_1^{l+1}} Y_l^m, \\
  &&S_4 {d^2 \Th_0 \over d\xi^2}(\xi_1) +{\p \Th_4 \over \p \xi}(\xi_1)
  +{3\mu \over \xi_1 (1+p)} P_3 (\sin \th \cos
  \vp) ={2\k_0 \over \xi_1^3} S_4 -\sum_{l,m} (l+1) {\k_{4:lm}
  \over \xi_1^{l+2}} Y_l^m,
\eeqa
and also $\Th_4$ should be satisfied with the conditions at the center of
the star,
\beqa
  {\p \Th_4 \over \p \xi} (0) &=&0, \\
  \Th_4 (0) &=&0.
\eeqa

%%%%%%%%%%%%%%%%%%%%%%%%%%%%%%
\subsection{The case of $n=0$}
%%%%%%%%%%%%%%%%%%%%%%%%%%%%%%
In this case, since we can express $\four \psi_{lm} =\four A_{lm} \xi^l$
by using a constant $\four A_{lm}$,
the boundary conditions can be written as
\beqa
  &&c_4 +{\mu \over 1+p} P_3 (\sin \th \cos \vp) =\sum_{l,m} \Bigl( -\four
  A_{lm} \xi_1^l +{\k_{4:lm} \over \xi_1^{l+1}} \Bigr) Y_l^m, \\
  &&{3\mu \over \xi_1 (1+p)} P_3 (\sin \th \cos \vp) =\sum_{l,m} \Bigl[
  (3-l) \four A_{lm} \xi_1^{l-1}
  -(l+1) {\k_{4:lm} \over \xi_1^{l+2}} \Bigr] Y_l^m.
\eeqa
Therefore, we obtain
\beqa
  c_4 &=&0, \\
  \Th_4 &=& \four A_3 \xi^3 P_3 (\sin \th \cos \vp),
\eeqa
where
\beqa
  \four A_3 &=&-{7\mu \over 4(1+p) \xi_1^3}, \\
  \k_{4:3} &=&-{3\mu \xi_1^4 \over 4(1+p)}.
\eeqa

%%%%%%%%%%%%%%%%%%%%%%%%%%%%%%
\subsection{The case of $n=1$}
%%%%%%%%%%%%%%%%%%%%%%%%%%%%%%
In this case, since we can express $\four \psi_{lm} =\four B_{lm} j_l$
by using a constant $\four B_{lm}$,
the boundary conditions can be written as
\beqa
  &&c_4 +{\mu \over 1+p} P_3 (\sin \th \cos \vp) =\sum_{l,m} \Bigl[ -\four
  B_{lm} j_l(\xi_1) +{\k_{4:lm} \over \xi_1^{l+1}} \Bigr] Y_l^m, \\
  &&{3\mu \over \xi_1 (1+p)} P_3 (\sin \th \cos \vp) =\sum_{l,m} \Bigl[
  -\four B_{lm} {dj_l \over d\xi}(\xi_1) -(l+1) {\k_{4:lm} \over
  \xi_1^{l+2}} \Bigr] Y_l^m.
\eeqa
Therefore, we obtain
\beqa
  c_4 &=&0, \\
  \Th_4 &=& \four B_3 j_3 (\xi) P_3 (\sin \th \cos \vp),
\eeqa
where
\beqa
  \four B_3 &=& -{7\mu \xi_1 \over 3(1+p)}, \\
  \k_{4:3} &=& -{5\mu \xi_1^2 \over 3(1+p)} (21-2\xi_1^2), \\
  j_3(\xi) &=& {1 \over \xi^4} \Bigl[ (15 -6\xi^2) \sin \xi -\xi
  (15 -\xi^2) \cos \xi \Bigr].
\eeqa

%%%%%%%%%%%%%%%%%%%%%%%%%%%%%%%
\subsection{The case of $n >0$}
%%%%%%%%%%%%%%%%%%%%%%%%%%%%%%%
In this case, the boundary conditions can be written as
\beqa
  &&c_4 +{\mu \over 1+p} P_3 (\sin \th \cos \vp) =\sum_{l,m} \Bigl[ -\four
  \psi_{lm}(\xi_1) +{\k_{4:lm} \over \xi_1^{l+1}} \Bigr] Y_l^m, \\
  &&{3\mu \over \xi_1 (1+p)} P_3 (\sin \th \cos \vp) =\sum_{l,m}
  \Bigl[ -{d\four
  \psi_{lm} \over d\xi}(\xi_1) -(l+1) {\k_{4:lm} \over \xi_1^{l+2}}
  \Bigr] Y_l^m.
\eeqa
Therefore, we obtain
\beqa
  c_4 &=&0, \\
  \Th_4 &=& \four \psi_3 (\xi) P_3 (\sin \th \cos \vp),
\eeqa
and the ordinary differential equation for $\four \psi_3$ is written as
\beq
  \Bigl[ {1 \over \xi^2} {d \over d\xi} \Bigl( \xi^2 {d \over d\xi} \Bigr)
  -{12 \over \xi^2} \Bigr] \four \psi_3 =-n \Th_0^{n-1} \four \psi_3.
\eeq
The coefficient of $\four \psi_3$ and the constant $\k_{4:3}$
are determined from two equations;
\beqa
  &&4 \four \psi_3(\xi_1) +\xi_1 {d\four \psi_3 \over d\xi}(\xi_1)
  =-{7\mu \over 1+p}, \\
  &&\k_{4:3}= \xi_1^4 \Bigl[ \four \psi_3(\xi_1) +{\mu \over 1+p} \Bigr]
  =-{\xi_1^4 \over 4} \Bigl[ {3\mu \over 1+p} +\xi_1 {d\four \psi_3 \over
  d\xi}(\xi_1) \Bigr].
\eeqa

%%%%%%%%%%%%%%%%%%%%%%%%%%%%%
\subsection{Octupole moments}
%%%%%%%%%%%%%%%%%%%%%%%%%%%%%
At this order, we can obtain the octupole moments, but we do not
represent them here.

%%%%%%%%%%%%%%%%%%%%%%%%%%%%%%%%%%
\section{Solutions at fifth order}
%%%%%%%%%%%%%%%%%%%%%%%%%%%%%%%%%%

The equation for the velocity potential is
\beqa
  \t{\D} \t{\Phi}_5 &=&-{n \over \Th_0} \Bigl[ (\t{\n} \t{\Phi}_5) \cdot
  (\t{\n} \Th_0) -{1 \over \xi_1} (\t{\bf \O} \times \bxi)_{fig} \cdot 
  (\t{\n} \Th_4) \Bigr], \\
  &=&-{n \over \Th_0} \Bigl[ (\t{\n} \t{\Phi}_5) \cdot (\t{\n} \Th_0)
  -{1 \over 4\xi_1} \four \psi_3 \Bigl\{ P_3^1 (\cos \th) \sin \vp -{1 
  \over 2} P_3^3 (\cos \th) \sin 3\vp \Bigr\} \Bigr].
\eeqa
Therefore, we have the form of $\t{\Phi}_5$ and the equation for it as
\beqa
  &&\t{\Phi}_5 =\five \phi_3 (\xi) \Bigl[ P_3^1 (\cos \th) \sin \vp
  -{1 \over 2} P_3^3 (\cos \th) \sin 3\vp \Bigr], \\
  &&\Bigl[ {1 \over \xi^2} {d \over d\xi} \Bigl( \xi^2 {d \over d\xi}
  \Bigr) -{12 \over \xi^2} +{n \over \Th_0} {d\Th_0 \over d\xi} {d
  \over d\xi} \Bigr] \five \phi_3 ={n \over 4\xi_1 \Th_0} \four
  \psi_3.
\eeqa
The boundary condition is
\beq
  {d\five \phi_3 \over d\xi}(\xi_1) {d\Th_0 \over d\xi}(\xi_1)
  -{1 \over 4\xi_1} \four \psi_3 (\xi_1) =0.
\eeq

Next, we consider the deformation of the figure at $O(\e^5)$. The
equation is written as
\beq
  \t{\D} \Th_5 =-n \Th_0^{n-1} \Th_5.
\eeq
We expand $\Th_5$ as
\beq
  \Th_5 =\sum_{l,m} \five \psi_{lm}(\xi) Y_l^m(\th,\vp).
\eeq
The surface equation up to $O(\e^5)$ is written as
\beq
  \Xi =\xi_1 +\e^3 S_3(\th,\vp) +\e^4 S_4(\th,\vp) +\e^5 S_5(\th,\vp),
\eeq
where
\beq
  S_5=-{\Th_5 (\xi_1) \over \Th_{0,\xi}(\xi_1)}.
\eeq

The gravitational potentials and their derivatives up to $O(\e^5)$ are
\beqa
  \t{U}_{int} &=& \Th +c_0 -\e^3 {\mu \over \xi_1^2 (1+p)} 
  \xi^2 P_2 (\sin \th \cos \vp) +\e^4 {\mu \over \xi_1^3
  (1+p)} \xi^3 P_3 (\sin \th \cos \vp) \nonumber \\
  &&+\e^5 \Bigl[ c_5 -{\mu \over
  \xi_1^4 (1+p)} \xi^4 P_4 (\sin \th \cos \vp) \Bigr], \\
%%%%%
  {\p \t{U}_{int} \over \p \xi} &=&{\p \Th \over \p \xi} -\e^3 {2\mu
  \over \xi_1^2 (1+p)} \xi P_2 (\sin \th \cos \vp) +\e^4 {3\mu \over
  \xi_1^3 (1+p)} \xi^2 P_3 (\sin \th \cos \vp) \nonumber \\
  &&-\e^5 {4\mu \over
  \xi_1^4 (1+p)} \xi^3 P_4 (\sin \th \cos \vp), \\
%%%%%
  \t{U}_{ext} &=&{\k_0 \over \xi} +\e^3 {\k_{3:2} \over
  \xi^3} P_2 (\sin \th \cos \vp)
  +\e^4 {\k_{4:3} \over \xi^4} P_3 (\sin \th \cos \vp)
  +\e^5 \sum_{l,m} {\k_{5:lm} \over \xi^{l+1}} Y_l^m, \\
%%%%%
  {\p \t{U}_{ext} \over \p \xi} &=& -{\k_0 \over \xi^2} -\e^3
  {3\k_{3:2} \over \xi^4} P_2 (\sin \th \cos \vp) -\e^4
  {4\k_{4:3} \over \xi^5} P_3 (\sin \th \cos \vp)
  -\e^5 \sum_{l,m} (l+1) {\k_{5:lm} \over \xi^{l+2}} Y_l^m.
\eeqa
They become at the stellar surface as
\beqa
  \t{U}_{int} (\Xi) &=&c_0 -\e^3 {\mu \over 1+p} P_2 (\sin 
  \th \cos \vp) +\e^4 {\mu \over 1+p} P_3 (\sin \th \cos \vp)
  +\e^5 \Bigl[ c_5 -{\mu \over 1+p} P_4 (\sin \th \cos \vp) \Bigr], \\
%%%%%
  {\p \t{U}_{int} \over \p \xi} (\Xi) &=&{d\Th_0 \over d\xi}(\xi_1)
  +\e^3 \Bigl[ S_3 {d^2 \Th_0 \over d\xi^2}(\xi_1)
  +{\p \Th_3 \over \p \xi}(\xi_1) -{2\mu \over \xi_1 (1+p)} P_2 (\sin
  \th \cos \vp) \Bigr] \nonumber \\
  &&+\e^4 \Bigl[ S_4 {d^2 \Th_0 \over d\xi^2}(\xi_1)
  +{\p \Th_4 \over \p \xi}(\xi_1) +{3 \mu \over \xi_1(1+p)}
  P_3 (\sin \th \cos \vp) \Bigr] \nonumber \\
  &&+\e^5 \Bigl[ S_5 {d^2 \Th_0 \over d\xi^2}(\xi_1)
  +{\p \Th_5 \over \p \xi}(\xi_1) -{4\mu \over \xi_1 (1+p)} P_4 (\sin
  \th \cos \vp) \Bigr], \\
%%%%%
  \t{U}_{ext} (\Xi) &=& {\k_0 \over \xi_1} \Bigl( 1-\e^3 {S_3 \over \xi_1} 
  -\e^4 {S_4 \over \xi_1} -\e^5 {S_5 \over \xi_1} \Bigr) +\e^3
  {\k_{3:2} \over \xi_1^3} P_2 (\sin \th \cos \vp)
  +\e^4 {\k_{4:3} \over \xi_1^4} P_3 (\sin \th \cos \vp)
  +\e^5 \sum_{l,m} {\k_{5:lm} \over \xi_1^{l+1}} Y_l^m, \\
%%%%%
  {\p \t{U}_{ext} \over \p \xi} (\Xi) &=& -{\k_0 \over \xi_1^2} \Bigl(
  1-\e^3 {2S_3 \over \xi_1} -\e^4 {2S_4 \over \xi_1} -\e^5 {2S_5 \over
  \xi_1} \Bigr) -\e^3
  {3\k_{3:2} \over \xi_1^4} P_2 (\sin \th \cos \vp) -\e^4
  {4\k_{4:3} \over \xi_1^5} P_3 (\sin \th \cos \vp) \nonumber \\
  &&-\e^5 \sum_{l,m} (l+1) {\k_{5:lm} \over \xi_1^{l+2}} Y_l^m.
\eeqa
Then, the boundary conditions for $O(\e^5)$ become
\beqa
  &&c_5 -{\mu \over 1+p} P_4 (\sin \th \cos \vp) =-{\k_0 \over
  \xi_1^2} S_5 +\sum_{l,m} {\k_{5:lm} \over \xi_1^{l+1}} Y_l^m, \\
%%%%%
  &&S_5 {d^2 \Th_0 \over d\xi^2}(\xi_1) +{\p \Th_5 \over \p \xi}(\xi_1)
  -{4\mu \over \xi_1 (1+p)} P_4 (\sin \th \cos
  \vp) ={2\k_0 \over \xi_1^3} S_5 -\sum_{l,m} (l+1) {\k_{5:lm}
  \over \xi_1^{l+2}} Y_l^m,
\eeqa
and also $\Th_5$ should be satisfied with the conditions at the center of
the star,
\beqa
  {\p \Th_5 \over \p \xi} (0) &=&0, \\
  \Th_5 (0) &=&0.
\eeqa

%%%%%%%%%%%%%%%%%%%%%%%%%%%%%%
\subsection{The case of $n=0$}
%%%%%%%%%%%%%%%%%%%%%%%%%%%%%%
In this case, since we can express $\five \psi_{lm} =\five A_{lm} \xi^l$
by using a constant $\five A_{lm}$,
the boundary conditions can be written as
\beqa
  &&c_5 -{\mu \over 1+p} P_4 (\sin \th \cos \vp) =\sum_{l,m} \Bigl( -\five
  A_{lm} \xi_1^l +{\k_{5:lm} \over \xi_1^{l+1}} \Bigr) Y_l^m, \\
  &&-{4\mu \over \xi_1 (1+p)} P_4 (\sin \th \cos \vp) =\sum_{l,m} \Bigl[
  (3-l) \five A_{lm} \xi_1^{l-1}
  -(l+1) {\k_{5:lm} \over \xi_1^{l+2}} \Bigr] Y_l^m.
\eeqa
Therefore, we obtain
\beqa
  c_5 &=&0, \\
  \Th_5 &=& \five A_4 \xi^4 P_4 (\sin \th \cos \vp),
\eeqa
where
\beqa
  \five A_4 &=&{3\mu \over 2(1+p) \xi_1^4}, \\
  \k_{5:4} &=&{\mu \xi_1^5 \over 2(1+p)}.
\eeqa

%%%%%%%%%%%%%%%%%%%%%%%%%%%%%%
\subsection{The case of $n=1$}
%%%%%%%%%%%%%%%%%%%%%%%%%%%%%%
In this case, since we can express $\five \psi_{lm} =\five B_{lm} j_l$
by using a constant $\five B_{lm}$,
the boundary conditions can be written as
\beqa
  &&c_5 -{\mu \over 1+p} P_4 (\sin \th \cos \vp) =\sum_{l,m} \Bigl[ -\five
  B_{lm} j_l(\xi_1) +{\k_{5:lm} \over \xi_1^{l+1}} \Bigr] Y_l^m, \\
  &&-{4\mu \over \xi_1 (1+p)} P_4 (\sin \th \cos \vp) =\sum_{l,m} \Bigl[
  -\five B_{lm} {dj_l \over d\xi}(\xi_1) -(l+1) {\k_{5:lm} \over
  \xi_1^{l+2}} \Bigr] Y_l^m.
\eeqa
Therefore, we obtain
\beqa
  c_5 &=&0, \\
  \Th_5 &=& \five B_4 j_4 (\xi) P_4 (\sin \th \cos \vp),
\eeqa
where
\beqa
  \five B_4 &=& {9\mu \xi_1^2 \over (1+p)(15-\xi_1^2)}, \\
  \k_{5:4} &=& {\mu \over 1+p} \Bigl( {\xi_1^4 -105 \xi_1^2 +945 \over 
  \xi_1^2 (15-\xi_1^2)} \Bigr), \\
  j_4(\xi) &=& {1 \over \xi^5} \Bigl[ (105 -45\xi^2 +\xi^4) \sin \xi
  -\xi (105 -10\xi^2) \cos \xi \Bigr].
\eeqa

%%%%%%%%%%%%%%%%%%%%%%%%%%%%%%%
\subsection{The case of $n >0$}
%%%%%%%%%%%%%%%%%%%%%%%%%%%%%%%
In this case,
the boundary conditions can be written as
\beqa
  &&c_5 -{\mu \over 1+p} P_4 (\sin \th \cos \vp) =\sum_{l,m} \Bigl[ -\five
  \psi_{lm}(\xi_1) +{\k_{5:lm} \over \xi_1^{l+1}} \Bigr] Y_l^m, \\
  &&-{4\mu \over \xi_1 (1+p)} P_4 (\sin \th \cos \vp) =\sum_{l,m}
  \Bigl[ -{d\five
  \psi_{lm} \over d\xi}(\xi_1) -(l+1) {\k_{5:lm} \over \xi_1^{l+2}}
  \Bigr] Y_l^m.
\eeqa
Therefore, we obtain
\beqa
  c_5 &=&0, \\
  \Th_5 &=& \five \psi_4 (\xi) P_4 (\sin \th \cos \vp),
\eeqa
and the ordinary differential equation for $\five \psi_4$ is written as
\beq
  \Bigl[ {1 \over \xi^2} {d \over d\xi} \Bigl( \xi^2 {d \over d\xi} \Bigr)
  -{20 \over \xi^2} \Bigr] \five \psi_4 =-n \Th_0^{n-1} \five \psi_4.
\eeq
The coefficient of $\five \psi_4$ and the constant $\k_{5:4}$
are determined from two equations;
\beqa
  &&5 \five \psi_4 (\xi_1) +\xi_1 {d\five \psi_4 \over d\xi}(\xi_1)
  ={9\mu \over 1+p}, \\
  &&\k_{5:4}= \xi_1^5 \Bigl[ \five \psi_4(\xi_1) -{\mu \over 1+p} \Bigr]
  ={\xi_1^5 \over 5} \Bigl[ {4\mu \over 1+p} -\xi_1 {d\five \psi_4 \over
  d\xi}(\xi_1) \Bigr].
\eeqa

%%%%%%%%%%%%%%%%%%%%%%%%%%%%%%%%%%%%%
\subsection{Orbital angular velocity}
%%%%%%%%%%%%%%%%%%%%%%%%%%%%%%%%%%%%%

At this order, the orbital angular velocity has the 5th higher order term
than the monopole term. The total angular velocity is
\beq
  \O^2 ={M_{tot} \over R^3} \Bigl[ 1 +{9 \over 2R^2} \Bigl( \e^3
  {\bar{\bI}_{11} \over M_1} +{\e'}^3 {\bar{\bI}_{11}' \over M_2} \Bigr)
  \Bigr].
\eeq
For the calculational convenience, we rewrite the above equation as
\beq
  \O^2 ={M_{tot} \over R^3} \bigl( 1 +\e^5 \d \bigr),
\eeq
where
\beq
  \d \equiv {9 \over 2a_0^2} \Bigl[ {\b{\bI}_{11} \over M_1}+ \Bigl(
  {a_0' \over a_0} \Bigr)^3 {\b{\bI}_{11}' \over M_2} \Bigr].
\eeq

%%%%%%%%%%%%%%%%%%%%%%%%%%%%%%%%%%
\section{Solutions at sixth order} \label{order:6th}
%%%%%%%%%%%%%%%%%%%%%%%%%%%%%%%%%%

The equation for the velocity potential is
\beqa
  \t{\D} \t{\Phi}_6 &=&-{n \over \Th_0} \Bigl[ (\t{\n} \t{\Phi}_6) \cdot
  (\t{\n} \Th_0) -{1 \over \xi_1} (\t{\bf \O} \times \bxi)_{fig} \cdot 
  (\t{\n} \Th_5) \Bigr], \\
  &=&-{n \over \Th_0} \Bigl[ (\t{\n} \t{\Phi}_6) \cdot (\t{\n} \Th_0)
  -{1 \over 12\xi_1} \five \psi_4 \Bigl\{ P_4^2 (\cos \th) \sin 2\vp -{1 
  \over 4} P_4^4 (\cos \th) \sin 4\vp \Bigr\} \Bigr].
\eeqa
Therefore, we have the form of $\t{\Phi}_6$ and the equation for it as
\beqa
  &&\t{\Phi}_6 =\six \phi_4 (\xi) \Bigl[ P_4^2 (\cos \th) \sin 2\vp
  -{1 \over 4} P_4^4 (\cos \th) \sin 4\vp \Bigr], \\
  &&\Bigl[ {1 \over \xi^2} {d \over d\xi} \Bigl( \xi^2 {d \over d\xi}
  \Bigr) -{20 \over \xi^2} +{n \over \Th_0} {d\Th_0 \over d\xi} {d
  \over d\xi} \Bigr] \six \phi_4 ={n \over 12\xi_1 \Th_0} \five
  \psi_4.
\eeqa
The boundary condition is
\beq
  {d\six \phi_4 \over d\xi}(\xi_1) {d\Th_0 \over d\xi}(\xi_1)
  -{1 \over 12\xi_1} \five \psi_4 (\xi_1) =0.
\eeq

Next, we consider the deformation of the figure at $O(\e^6)$. The
equation is written as
\beq
  \t{\D} \Th_6 =-n \Th_0^{n-1} \Th_6 -{1 \over 2} n(n-1) \Th_0^{n-2}
  \Th_3^2 +{2\mu \over \xi_1} \t{\D} \four \phi_2 (\xi) P_2^2 (\cos
  \th) \cos 2\vp.
\eeq
We expand $\Th_6$ as
\beq
  \Th_6 =\sum_{l,m} \six \psi_{lm} (\xi) Y_l^m(\th,\vp).
\eeq
The surface equation up to $O(\e^6)$ is written as
\beq
  \Xi =\xi_1 +\e^3 S_3(\th,\vp) +\e^4 S_4(\th,\vp) +\e^5 S_5(\th,\vp)
  +\e^6 S_6(\th,\vp),
\eeq
where
\beq
  S_6=-{1 \over \Th_{0,\xi}(\xi_1)} \Bigl[ {S_3^2 \over 2}
  {d^2 \Th_0 \over d\xi^2}(\xi_1) +S_3 {\p \Th_3 \over \p \xi}(\xi_1)
  +\Th_6(\xi_1) \Bigr].
\eeq

The gravitational potentials and their derivatives up to $O(\e^6)$ are
\beqa
  \t{U}_{int} &=& \Th +c_0 -\e^3 {\mu \over \xi_1^2 (1+p)} 
  \xi^2 P_2 (\sin \th \cos \vp) +\e^4 {\mu \over \xi_1^3
  (1+p)} \xi^3 P_3 (\sin \th \cos \vp) \nonumber \\
  &&-\e^5 {\mu \over \xi_1^4 (1+p)} \xi^4 P_4 (\sin \th \cos \vp)
  \nonumber \\
  &&+\e^6 \Bigl[ c_6 -{3\mu \bar{\bI}_{11}' \over 2M_{tot} a_0^2}
  \Bigl( {a_0' \over a_0} \Bigr)^3
  -{\mu \d \over 2(1+p)^2} +{\mu \over \xi_1^5 (1+p)} \xi^5 P_5 (\sin
  \th \cos \vp) -{2\mu \over \xi_1} \four \phi_2 P_2^2 (\cos \th) \cos
  2\vp \Bigr], \\
%%%%%
  {\p \t{U}_{int} \over \p \xi} &=&{\p \Th \over \p \xi} -\e^3 {2\mu
  \over \xi_1^2 (1+p)} \xi P_2 (\sin \th \cos \vp) +\e^4 {3\mu \over
  \xi_1^3 (1+p)} \xi^2 P_3 (\sin \th \cos \vp) -\e^5 {4\mu \over
  \xi_1^4 (1+p)} \xi^3 P_4 (\sin \th \cos \vp) \nonumber \\
  &&+\e^6 \Bigl[ {5\mu \over \xi_1^5 (1+p)} \xi^4 P_5 (\sin \th \cos
  \vp) -{2\mu \over \xi_1} {d\four \phi_2 \over d\xi} P_2^2 (\cos \th) 
  \cos 2\vp \Bigr], \\
%%%%%
  \t{U}_{ext} &=&{\k_0 \over \xi} +\e^3 {\k_{3:2} \over
  \xi^3} P_2 (\sin \th \cos \vp)
  +\e^4 {\k_{4:3} \over \xi^4} P_3 (\sin \th \cos \vp)
  +\e^5 {\k_{5:4} \over \xi^5} P_4 (\sin \th \cos \vp) +\e^6 \sum_{l,m}
  {\k_{6:lm} \over \xi^{l+1}} Y_l^m, \\
%%%%%
  {\p \t{U}_{ext} \over \p \xi} &=& -{\k_0 \over \xi^2} -\e^3
  {3\k_{3:2} \over \xi^4} P_2 (\sin \th \cos \vp) -\e^4
  {4\k_{4:3} \over \xi^5} P_3 (\sin \th \cos \vp) -\e^5
  {5\k_{5:4} \over \xi^6} P_4 (\sin \th \cos \vp) \nonumber \\
  &&-\e^6 \sum_{l,m} (l+1) {\k_{6:lm} \over \xi^{l+2}} Y_l^m.
\eeqa
They become at the stellar surface as
\beqa
  \t{U}_{int} (\Xi) &=&c_0 -\e^3 {\mu \over 1+p} \Bigl( 1
  +{2S_3 \over \xi_1} \e^3 \Bigr) P_2 (\sin \th \cos \vp)
  +\e^4 {\mu \over 1+p} P_3 (\sin \th \cos \vp)
  -\e^5 {\mu \over 1+p} P_4 (\sin \th \cos \vp) \nonumber \\
  &&+\e^6 \Bigl[ c_6 -{3\mu \bar{\bI}_{11}' \over 2M_{tot} a_0^2}
  \Bigl( {a_0' \over a_0} \Bigr)^3
  -{\mu \d \over 2(1+p)^2} +{\mu \over 1+p} P_5 (\sin \th \cos \vp)
  -{2\mu \over \xi_1} \four \phi_2 (\xi_1) P_2^2 (\cos \th) \cos 2\vp
  \Bigr], \\
%%%%%
  {\p \t{U}_{int} \over \p \xi} (\Xi) &=&{d\Th_0 \over d\xi}(\xi_1)
  +\e^3 \Bigl[ S_3 {d^2 \Th_0 \over d\xi^2}(\xi_1) +{\p \Th_3 \over \p
  \xi} (\xi_1) -{2\mu \over \xi_1 (1+p)} \Bigl( 1
  +{S_3 \over \xi_1} \e^3 \Bigr) P_2 (\sin \th \cos \vp) \Bigr]
  \nonumber \\
  &&+\e^4 \Bigl[ S_4 {d^2 \Th_0 \over d\xi^2}(\xi_1)
  +{\p \Th_4 \over \p \xi}(\xi_1) +{3 \mu \over \xi_1(1+p)}
  P_3 (\sin \th \cos \vp) \Bigr] \nonumber \\
  &&+\e^5 \Bigl[ S_5 {d^2 \Th_0 \over d\xi^2}(\xi_1) +{\p \Th_5 \over \p
  \xi} (\xi_1) -{4\mu \over \xi_1 (1+p)} P_4 (\sin
  \th \cos \vp) \Bigr] \nonumber \\
  &&+\e^6 \Bigl[ S_6 {d^2 \Th_0 \over d\xi^2} (\xi_1) +{1 \over 2} S_3^2
  {d^3 \Th_0 \over d\xi^2} (\xi_1) +S_3 {\p^2 \Th_3 \over \p \xi^2} (\xi_1)
  +{\p \Th_6 \over \p \xi} (\xi_1) +{5\mu \over \xi_1 (1+p)} P_5
  (\sin \th \cos \vp) \nonumber \\
  &&\hspace{20pt}-{2\mu \over \xi_1} {d\four \phi_2 \over d\xi}(\xi_1)
  P_2^2 (\cos \th) \cos 2\vp \Bigr], \\
%%%%%
  \t{U}_{ext} (\Xi) &=& {\k_0 \over \xi_1} \Bigl[ 1-\e^3 {S_3 \over \xi_1} 
  -\e^4 {S_4 \over \xi_1} -\e^5 {S_5 \over \xi_1} +\e^6 \Bigl\{ \Bigl(
  {S_3 \over \xi_1} \Bigr)^2 -{S_6 \over \xi_1} \Bigr\} \Bigr]
  +\e^3 \Bigl( 1 -\e^3 {3S_3 \over \xi_1} \Bigr)
  {\k_{3:2} \over \xi_1^3} P_2 (\sin \th \cos \vp) \nonumber \\
  &&+\e^4 {\k_{4:3} \over \xi_1^4} P_3 (\sin \th \cos \vp)
  +\e^5 {\k_{5:4} \over \xi_1^5} P_4 (\sin \th \cos \vp) +\e^6 \sum_{l,m}
  {\k_{6:lm} 
  \over \xi_1^{l+1}} Y_l^m, \\
%%%%%
  {\p \t{U}_{ext} \over \p \xi} (\Xi) &=& -{\k_0 \over \xi_1^2} \Bigl[
  1-\e^3 {2S_3 \over \xi_1} -\e^4 {2S_4 \over \xi_1} -\e^5 {2S_5 \over
  \xi_1} +\e^6 \Bigl\{ 3 \Bigl( {S_3 \over \xi_1} \Bigr)^2 -2 {S_6 \over
  \xi_1} \Bigr\} \Bigr] -\e^3 \Bigl( 1 -\e^3 {4S_3 \over \xi_1} \Bigr)
  {3\k_{3:2} \over \xi_1^4} P_2 (\sin \th \cos \vp)
  \nonumber \\
  &&-\e^4 {4\k_{4:3} \over \xi_1^5} P_3 (\sin \th \cos \vp)
  -\e^5 {5\k_{5:4} \over \xi_1^6} P_4 (\sin \th \cos \vp) -\e^6
  \sum_{l,m} (l+1) {\k_{6:lm} \over \xi_1^{l+2}} Y_l^m.
\eeqa
Then, the boundary conditions for $O(\e^6)$ become
\beqa
  &&\t{c}_6 -{2\mu \over \xi_1 (1+p)} S_3 P_2 (\sin \th \cos \vp) +{\mu
  \over 1+p} P_5 (\sin \th \cos \vp) -{2\mu \over \xi_1} \four \phi_2
  (\xi_1) P_2^2 (\cos \th) \cos 2\vp \nonumber \\
  &&\hspace{30pt}={\k_0 \over
  \xi_1^2} \Bigl( {S_3^2 \over \xi_1} -S_6 \Bigr) -
  {3\k_{3:2} \over \xi_1^4} S_3 P_2 (\sin \th \cos \vp)
  +\sum_{l,m} {\k_{6:lm} \over \xi_1^{l+1}} Y_l^m, \\
%%%%%
  &&S_6 {d^2 \Th_0 \over d\xi^2}(\xi_1) +{1 \over 2} S_3^2
  {d^3 \Th_0 \over d\xi^3}(\xi_1) +S_3 {\p^2 \Th_3 \over \p \xi^2}(\xi_1)
  +{\p \Th_6 \over \p \xi}(\xi_1) -{2\mu \over \xi_1^2 (1+p)} S_3
  P_2 (\sin \th \cos \vp) +{5\mu \over \xi_1
  (1+p)} P_5 (\sin \th \cos \vp) \nonumber \\
  &&-{2\mu \over \xi_1} {d\four \phi_2 \over d\xi}(\xi_1)
  P_2^2 (\cos \th) \cos 2\vp =-{\k_0 \over \xi_1^3} \Bigl(
  {3S_3^2 \over \xi_1} -2S_6 \Bigr) +{12\k_{3:2}
  \over \xi_1^5} S_3 P_2 (\sin \th \cos \vp) -\sum_{l,m} (l+1) {\k_{6:lm}
  \over \xi_1^{l+2}} Y_l^m,
\eeqa
where
\beq
  \t{c}_6 \equiv c_6 -{3\mu \bar{\bI}_{11}' \over 2M_{tot} a_0^2}
  \Bigl( {a_0' \over a_0} \Bigr)^3 -{\mu \d \over 2(1+p)^2}.
\eeq
Also $\Th_6$ should be satisfied with the conditions at the center of
the star,
\beqa
  {\p \Th_6 \over \p \xi} (0) &=&0, \\
  \Th_6 (0) &=&0.
\eeqa

In the calculation of this section, we need the expression of $[P_2(\sin
\th \cos \vp)]^2$. It is written as
\beq
  \bigl[ P_2 (\sin \th \cos \vp) \bigr]^2 ={18 \over 35} P_4 (\sin \th
  \cos \vp) +{2 \over 7} P_2 (\sin \th \cos \vp) +{1 \over 5}.
\eeq

%%%%%%%%%%%%%%%%%%%%%%%%%%%%%%
\subsection{The case of $n=0$}
%%%%%%%%%%%%%%%%%%%%%%%%%%%%%%
In this case, since we can express $\three \psi_2 =\three A_2 \xi^2$
and $\six \psi_{lm} =\six A_{lm} \xi^l$, by using the constants
$\three A_2$ and $\six A_{lm}$, the boundary conditions can be written
as
\beqa
  &&\t{c}_6 +{\mu \over 1+p} P_5 (\sin \th \cos \vp) -{2\mu \over
  \xi_1} \four \phi_2 (\xi_1) P_2^2 (\cos \th) \cos 2\vp \nonumber \\
  &&\hspace{30pt}={\three A_2 \xi_1^2 \over \Th_{0,\xi}(\xi_1)}
  \Bigl[ -{2\mu \over \xi_1 (1+p)} +{1 \over 2} \three A_2 \xi_1
  +{3 \k_{3:2} \over \xi_1^4} \Bigr]
  \Bigl( {18 \over 35} P_4 (\sin \th \cos \vp) +{2 \over 7} P_2 (\sin
  \th \cos \vp) +{1 \over 5} \Bigr) \nonumber \\
  &&\hspace{40pt}+\sum_{l,m} \Bigl( -\six A_{lm} \xi_1^l +{\k_{6:lm} \over
  \xi_1^{l+1}} \Bigr) Y_l^m, \\
%%%%%
  &&{5\mu \over \xi_1 (1+p)} P_5 (\sin \th \cos \vp) -{2\mu \over
  \xi_1} {d\four \phi_2 \over d\xi}(\xi_1) P_2^2 (\cos \th) \cos 2\vp
  \nonumber \\
  &&\hspace{30pt}={\three A_2 \xi_1^2 \over \Th_{0,\xi}(\xi_1)}
  \Bigl[ -{2\mu \over \xi_1^2 (1+p)}
  +{1 \over 2} \three A_2 -{12 \k_{3:2} \over \xi_1^5} \Bigr]
  \Bigl( {18 \over 35} P_4 (\sin 
  \th \cos \vp) +{2 \over 7} P_2 (\sin \th \cos \vp) +{1 \over 5}
  \Bigr) \nonumber \\
  &&\hspace{40pt}+\sum_{l,m} \Bigl[ (3-l) \six A_{lm} \xi_1^{l-1}
  -(l+1) {\k_{6:lm} \over \xi_1^{l+2}} \Bigr] Y_l^m.
\eeqa
Therefore, we obtain
\beqa
  \t{c}_6 &=&{45\mu^2 \over 2(1+p)^2 \xi_1^2}, \\
  \Th_6 &=& \six A_2 \xi^2 P_2 (\sin \th \cos \vp) +\six A_{22} \xi^2
  P_2^2 (\cos \th) \cos 2\vp +\six A_4 \xi^4 P_4 (\sin \th \cos \vp)
  \nonumber \\
  &&+\six A_5 \xi^5 P_5 (\sin \th \cos \vp),
\eeqa
where
\beqa
  \six A_2 &=&{225\mu^2 \over 28(1+p)^2 \xi_1^4}, \\
  \six A_{22} &=&{75\mu^2 \over 8(1+p) \xi_1^4}, \\
  \six A_4 &=&{45\mu^2 \over 14(1+p)^2 \xi_1^6}, \\
  \six A_5 &=& -{11\mu \over 8(1+p) \xi_1^5}.
\eeqa

%%%%%%%%%%%%%%%%%%%%%%%%%%%%%%
\subsection{The case of $n=1$}
%%%%%%%%%%%%%%%%%%%%%%%%%%%%%%

In this case, since we can express $\three \psi_2 =\three B_2 j_2$ and
$\six \psi_{lm} =\six B_{lm}$ except for $\six \psi_{22}$ by using
the constants $\three B_2$ and $\six B_{lm}$,
the boundary conditions can be written as
\beqa
  &&\t{c}_6 +{\mu \over 1+p} P_5 (\sin \th \cos \vp) -{2\mu \over
  \xi_1} \four \phi_2(\xi_1) P_2^2 (\cos \th) \cos 2\vp \nonumber \\
  &&\hspace{30pt}={\three B_2 j_2(\xi_1) \over \Th_{0,\xi}(\xi_1)}
  \Bigl[ -{2\mu \over \xi_1 (1+p)} +\three B_2 {dj_2 \over d\xi}(\xi_1)
  +{3\k_{3:2} \over \xi_1^4} \Bigr]
  \Bigl( {18 \over 35} P_4 (\sin \th \cos \vp) +{2 \over 7} P_2 (\sin
  \th \cos \vp) +{1 \over 5} \Bigr) \nonumber \\
  &&\hspace{40pt}+\sum_{l,m} \Bigl( -\six B_{lm} j_l(\xi_1) +{\k_{6:lm}
  \over \xi_1^{l+1}} \Bigr) Y_l^m, \\
%%%%%
  &&{5\mu \over \xi_1 (1+p)} P_5 (\sin \th \cos \vp) -{2\mu \over
  \xi_1} {d\four \phi_2 \over d\xi}(\xi_1) P_2^2 (\cos \th) \cos 2\vp
  \nonumber \\
  &&\hspace{30pt}={\three B_2 j_2(\xi_1) \over \Th_{0,\xi}(\xi_1)}
  \Bigl[ -{2\mu \over \xi_1^2 (1+p)} +\Bigl( {3 \over \xi_1^2}
  +{\xi_1 \over 2} {d^3 \Th_0 \over d\xi^3}(\xi_1) \Bigr)
  \three B_2 j_2(\xi_1) +\three B_2 {d^2j_2 \over d\xi^2}(\xi_1)
  -{12\k_{3:2} \over \xi_1^5} \Bigr] \nonumber \\
  &&\hspace{60pt} \times \Bigl( {18 \over 35} P_4 (\sin \th \cos \vp) +{2
  \over 7} P_2 (\sin \th \cos \vp) +{1 \over 5} \Bigr) \nonumber \\
  &&\hspace{40pt}+\sum_{l,m} \Bigl[ -\six B_{lm} {dj_l \over d\xi}(\xi_1)
  -(l+1) {\k_{6:lm} \over \xi_1^{l+2}} \Bigr] Y_l^m.
\eeqa
Therefore, we obtain
\beqa
  \t{c}_6 &=&{45\mu^2 \over 2(1+p)^2 \xi_1^2}, \\
  \Th_6 &=& \six B_2 j_2 (\xi) P_2 (\sin \th \cos \vp) +\six \psi_{22} 
  (\xi) P_2^2 (\cos \th) \cos 2\vp +\six B_4 j_4 (\xi) P_4 (\sin \th
  \cos \vp) \nonumber \\
  &&+\six B_5 j_5 (\xi) P_5 (\sin \th \cos \vp),
\eeqa
where
\beqa
  \six B_2 &=& {225\mu^2 \over 7(1+p)^2 \xi_1^2}, \\
  \six B_4 &=& {405\mu^2 \over 7(1+p)^2 (15-\xi_1^2)}, \\
  \six B_5 &=& -{11\mu \xi_1^3 \over (1+p) (105-10\xi_1^2)}, \\
  j_5(\xi) &=& {1 \over \xi^6} \Bigl[ (945 -420\xi^2 +15\xi^4) \sin \xi
  -\xi (945 -105\xi^2 +\xi^4) \cos \xi \Bigr],
\eeqa
and the coefficient of $\six \psi_{22}$ is determined from the equation;
\beq
  3\six \psi_{22}(\xi_1) +\xi_1 {d\six \psi_{22} \over d\xi}(\xi_1)
  ={2\mu \over \xi_1} \Bigl[ 3\four \phi_2(\xi_1) +\xi_1 {d\four \phi_2
  \over d\xi}(\xi_1) \Bigr].
\eeq

%%%%%%%%%%%%%%%%%%%%%%%%%%%%%%%
\subsection{The case of $n >0$}
%%%%%%%%%%%%%%%%%%%%%%%%%%%%%%%
In this case, the boundary conditions can be written as
\beqa
  &&\t{c}_6 +{\mu \over 1+p} P_5 (\sin \th \cos \vp) -{2\mu \over
  \xi_1} \four \phi_2(\xi_1) P_2^2 (\cos \th) \cos 2\vp \nonumber \\
  &&\hspace{30pt}={\three \psi_2(\xi_1) \over \Th_{0,\xi}(\xi_1)}
  \Bigl[ -{2\mu \over \xi_1 (1+p)} +{d\three \psi_2 \over d\xi}(\xi_1)
  +{3\k_{3:2} \over \xi_1^4} \Bigr]
  \Bigl( {18 \over 35} P_4 (\sin \th \cos \vp) +{2 \over 7} P_2 (\sin
  \th \cos \vp) +{1 \over 5} \Bigr) \nonumber \\
  &&\hspace{30pt}+\sum_{l,m} \Bigl[ -\six \psi_{lm}(\xi_1) +{\k_{6:lm}
  \over \xi_1^{l+1}} \Bigr] Y_l^m, \\
%%%%%
  &&{5\mu \over \xi_1 (1+p)} P_5 (\sin \th \cos \vp) -{2\mu \over
  \xi_1} {d\four \phi_2 \over d\xi}(\xi_1) P_2^2 (\cos \th) \cos 2\vp
  \nonumber \\
  &&\hspace{30pt}={\three \psi_2(\xi_1) \over \Th_{0,\xi}(\xi_1)}
  \Bigl[ -{2\mu \over \xi_1^2 (1+p)} +\Bigl( {3 \over \xi_1^2}
  -{1 \over 2\Th_{0,\xi}(\xi_1)} {d^3 \Th_0 \over d\xi^3}(\xi_1)
  \Bigr) \three \psi_2(\xi_1) +{d^2\three \psi_2 \over d\xi^2}(\xi_1)
  -{12 \k_{3:2} \over \xi_1^5} \Bigr] \nonumber \\
  &&\hspace{60pt} \times \Bigl( {18 \over 35} P_4 (\sin \th \cos \vp)
  +{2 \over 7} P_2 (\sin \th \cos \vp) +{1 \over 5} \Bigr) \nonumber \\
  &&\hspace{30pt}+\sum_{l,m} \Bigl[ -{d\six \psi_{lm} \over d\xi}(\xi_1)
  -(l+1) {\k_{6:lm} \over \xi_1^{l+2}} \Bigr] Y_l^m.
\eeqa
Therefore, we obtain
\beqa
  \t{c}_6 &=&-{\xi_1 \over 10\Th_{0,\xi}(\xi_1)} n \Th_0^{n-1}(\xi_1)
  \bigl( \three \psi_2(\xi_1) \bigr)^2 -\six \psi_0(\xi_1)
  -\xi_1 {d\six \psi_0 \over d\xi}(\xi_1), \label{Eq:tilda_c6} \\
%%%%%
  \Th_6 &=& \six \psi_0 (\xi) +\six \psi_2 (\xi) P_2 (\sin \th \cos
  \vp) +\six \psi_{22} (\xi) P_2^2 (\cos \th) \cos 2\vp +\six \psi_4
  (\xi) P_4 (\sin \th \cos \vp) \nonumber \\
  &&+\six \psi_5 (\xi) P_5 (\sin \th \cos \vp),
\eeqa
where the forms of $\six \psi_i$ are determined by the equations
\beqa
  &&{1 \over \xi^2} {d \over d\xi} \Bigl( \xi^2 {d \over d\xi} \Bigr)
  \six \psi_0 =-n\Th_0^{n-1} \six \psi_0 -{1 \over 10} n(n-1) 
  \Th_0^{n-2} (\three \psi_2)^2, \\
%%%%%
  &&\Bigl[ {1 \over \xi^2} {d \over d\xi} \Bigl( \xi^2 {d \over d\xi} \Bigr)
  -{6 \over \xi^2} \Bigr] \six \psi_2 =-n\Th_0^{n-1} \six \psi_2
  -{1 \over 7} n(n-1) \Th_0^{n-2} (\three \psi_2)^2, \\
%%%%%
  &&\Bigl[ {1 \over \xi^2} {d \over d\xi} \Bigl( \xi^2 {d \over d\xi} \Bigr)
  -{6 \over \xi^2} \Bigr] \six \psi_{22} =-n\Th_0^{n-1} \six \psi_{22}
  +{2\mu \over \xi_1} \Bigl[ {1 \over \xi^2} {d \over d\xi}
  \Bigl( \xi^2 {d \over d\xi} \Bigr) -{6 \over \xi^2} \Bigr]
  \four \phi_2, \\
%%%%%
  &&\Bigl[ {1 \over \xi^2} {d \over d\xi} \Bigl( \xi^2 {d \over d\xi} \Bigr)
  -{20 \over \xi^2} \Bigr] \six \psi_4 =-n\Th_0^{n-1} \six \psi_4
  -{9 \over 35} n(n-1) \Th_0^{n-2} (\three \psi_2)^2, \\
%%%%%
  &&\Bigl[ {1 \over \xi^2} {d \over d\xi} \Bigl( \xi^2 {d \over d\xi} \Bigr)
  -{30 \over \xi^2} \Bigr] \six \psi_5 =-n\Th_0^{n-1} \six \psi_5,
\eeqa
and the coefficients of $\six \psi_i$ are determined by the boundary
conditions;
\beqa
  &&3 \six \psi_2(\xi_1) +\xi_1 {d\six \psi_2 \over d\xi}(\xi_1)
  =-{\xi_1 \over 7\Th_{0,\xi}(\xi_1)} n \Th_0^{n-1}(\xi_1)
  \bigl( \three \psi_2 (\xi_1) \bigr)^2, \\
  &&3 \six \psi_{22}(\xi_1) +\xi_1 {d\six \psi_{22} \over d\xi}(\xi_1)
  ={2\mu \over \xi_1} \Bigl[ 3\four \phi_2(\xi_1)
  +\xi_1 {d\four \phi_2 \over d\xi}(\xi_1) \Bigr], \\
  &&5 \six \psi_4(\xi_1) +\xi_1 {d\six \psi_4 \over d\xi}(\xi_1)
  =-{9\xi_1 \over 35\Th_{0,\xi}(\xi_1)} n \Th_0^{n-1}(\xi_1)
  \bigl( \three \psi_2(\xi_1) \bigr)^2, \\
  &&6 \six \psi_5(\xi_1) +\xi_1 {d\six \psi_5 \over d\xi}(\xi_1)
  =-{11\mu \over 1+p}.
\eeqa

%%%%%%%%%%%%%%%%%%%%%%%%%%%%%%%%%%%%%%%%%%%
\section{The change in the central density} \label{sec:change_rho_c}
%%%%%%%%%%%%%%%%%%%%%%%%%%%%%%%%%%%%%%%%%%%

In this section, we present the expression of the change in the central
density at $O(\e^6)$ by using the first tensor virial relation.

First, we divide the first tensor virial relation (\ref{Eq:firstTV}),
\beq
  {3 \over n} \Pi_1 +{3 \over n'} \Pi_2 +(W_{self})_{tot} +(W_{int})_{tot}
  +2T_{tot} =0,
\eeq
into three parts of $O(\e^0)$, $O(\e)$ and $O(\e^6)$.
After that we demand that these three parts should be zero independently.

\vspace{0.2cm}
\noindent
(a) The 0th order of $\e$:
\beq
  {M_1^2 \over 2a_0} \Bigl[ {5-n \over 1+n} {1 \over \xi_1^3
  (\Th_{0,\xi} (\xi_1))^2} \int_0^{\xi_1} d\xi \xi^2 \Th_0^{1+n} -1 \Bigr]
  +{M_2^2 \over 2a_0'} \Bigl[ {5-n' \over 1+n'} {1 \over {\xi_1'}^3
  (\b{\Th}_{0,\xi'} (\xi_1'))^2} \int_0^{\xi_1'} d\xi' {\xi'}^2
  \b{\Th}_0^{1+n'} -1 \Bigr] =0.
\eeq
If we assume that both of the terms concerned with star 1 and star 2
become zero independently, we have two relations;
\beqa
  {5-n \over 1+n} {1 \over \xi_1^3 (\Th_{0,\xi} (\xi_1))^2}
  \int_0^{\xi_1} d\xi \xi^2 \Th_0^{1+n} =1,
  \label{Eq:TVR1} \\
  {5-n' \over 1+n'} {1 \over {\xi_1'}^3 (\b{\Th}_{0,\xi'} (\xi_1'))^2}
  \int_0^{\xi_1'} d\xi' {\xi'}^2 \b{\Th}_0^{1+n'} =1.
  \label{Eq:TVR2}
\eeqa
The above relations are proved using the Lane-Emden equations.

\vspace{0.2cm}
\noindent
(b) The 1st order of $\e$:

It is clear that this term becomes zero.

\vspace{0.2cm}
\noindent
(c) The 6th order of $\e$:

By using the relations;
\beqa
  \t{c}_6 &=&- \Bigl( {3-n \over 2n} \Bigr) \xi_1 \Th_{0,\xi} (\xi_1)
  {\d \r_c \over \r_c} -\six \psi_0 (\xi_1), \\
  \t{c}_6' &=&- \Bigl( {3-n' \over 2n'} \Bigr) \xi_1' \b{\Th}_{0,\xi'}
  (\xi_1') {\d \r_c' \over \r_c'} -\six \b{\psi}_0 (\xi_1'), \\
  \mu &=&-\Bigl( {1+p \over p} \Bigr) \xi_1 \Th_{0,\xi} (\xi_1), \\
  \mu' &=&-(1+p) \xi_1' \b{\Th}_{0,\xi'} (\xi_1'),
\eeqa
we can express the equation at order $\e^6$ as
\beqa
  &&{M_1^2 \over a_0} \biggl[ -{5-n \over 2(1+n)} {1 \over \xi_1^3
  (\Th_{0,\xi} (\xi_1))^2} \Bigl[ {5-n \over 2n} {\d \r_c \over \r_c}
  \int_0^{\xi_1} d\xi
  \xi^2 \Th_0^{1+n} -(1+n) \int_0^{\xi_1} d\xi \xi^2 \Bigl\{ \Th_0^n
  \six \psi_0 +{1 \over 10} n \Th_0^{n-1} (\three \psi_2)^2 \Bigr\}
  \Bigr] \nonumber \\
  &&\hspace{30pt}+{15 \over 4p} {\b{\bI}_{11} \over M_1 a_0^2}
  -{\six \psi_0 (\xi_1) \over 2\xi_1 \Th_{0,\xi} (\xi_1)}
  +{n-1 \over 4n} {\d \r_c \over \r_c} \biggr] \nonumber \\
  &&+{M_2^2 \over a_0'} \Bigl( {a_0' \over a_0} \Bigr)^6 \biggl[
  -{5-n' \over 2(1+n')} {1 \over {\xi_1'}^3 (\b{\Th}_{0,\xi'} (\xi_1'))^2}
  \nonumber \\
  &&\hspace{70pt}\times
  \Bigl[ {5-n' \over 2n'} {\d \r_c' \over \r_c'} \int_0^{\xi_1'} d\xi'
  {\xi'}^2 \b{\Th}_0^{1+n'} -(1+n') \int_0^{\xi_1'} d\xi' {\xi'}^2 \Bigl\{
  \b{\Th}_0^{n'} \six \b{\psi}_0
  +{1 \over 10} n' \b{\Th}_0^{n'-1} (\three \b{\psi}_2)^2 \Bigr\} \Bigr]
  \nonumber \\
  &&\hspace{30pt}+{15 p \over 4} {\b{\bI}_{11}' \over M_2 {a_0'}^2}
  -{\six \b{\psi}_0 (\xi_1') \over 2\xi_1' \b{\Th}_{0,\xi'} (\xi_1')}
  +{n'-1 \over 4n'} {\d \r_c' \over \r_c'} \biggr] =0.
\eeqa
If we assume that both of the terms concerned with star 1 and star 2
become zero independently, we obtain two relations;
\beqa
  {\d \r_c \over \r_c} &=&{2n \over 3-n} \biggl[
  {15 \over 4p} {\b{\bI}_{11} \over M_1 a_0^2}
  -{\six \psi_0 (\xi_1) \over 2\xi_1 \Th_{0,\xi} (\xi_1)}
  +{5-n \over 2\xi_1^3 (\Th_{0,\xi} (\xi_1))^2}
  \int_0^{\xi_1} d\xi \xi^2 \Bigl\{ \Th_0^n \six \psi_0
  +{1 \over 10} n \Th_0^{n-1} (\three \psi_2)^2 \Bigr\} \biggl], \\
%%%%%
  {\d \r_c' \over \r_c'} &=&{2n' \over 3-n'} \biggl[
  {15p \over 4} {\b{\bI}_{11}' \over M_2 {a_0'}^2}
  -{\six \b{\psi}_0 (\xi_1') \over 2\xi_1' \b{\Th}_{0,\xi'} (\xi_1')}
  +{5-n' \over 2{\xi_1'}^3 (\b{\Th}_{0,\xi'} (\xi_1'))^2} \int_0^{\xi_1'}
  d\xi' {\xi'}^2 \Bigl\{ \b{\Th}_0^{n'} \six \b{\psi}_0 +{1 \over 10} n'
  \b{\Th}_0^{n'-1} (\three \b{\psi}_2)^2 \Bigr\} \biggr],
\eeqa 
where we have used Eqs. (\ref{Eq:TVR1}) and (\ref{Eq:TVR2}).

%%%%%%%%%%%%%%%%%%%%%%%%%%%%%%%%%%%%%%%%%%%%%%%%%%%%%%%%%%%%%%%%

%%%%%%%%%%%%%%%%%%%%%%%%%%%%%%%%%%%%%%%%%%%%
\newpage
\begin{center}
  {\Large Tables}
\end{center}

\begin{table}
\caption{The functions and their derivatives to represent the equilibrium
figure for $n=0.5$. (I)
}
 \begin{center}
  \begin{tabular}{l|cccccccc}
  \multicolumn{9}{c}{$n=0.5$} \\ \hline
  $\xi$&$\Th_0$&$d \Th_0/d \xi$&$\three \psi_2$&
  $d \three \psi_2/d \xi$&$\four \psi_3$&$d \four \psi_3/d \xi$&
  $\five \psi_4$&$d \five \psi_4/d \xi$ \\ \hline
  0.00&1.000&0.000&0.000&0.000&0.000&0.000&
	0.000&0.000 \\
  0.10&0.9983&-3.332(-2)&5.371(-3)&0.1074&-1.316(-4)&-3.946(-3)&
	4.003(-6)&1.601(-4) \\
  0.20&0.9933&-6.653(-2)&2.146(-2)&0.2143&-1.052(-3)&-1.576(-2)&
	6.400(-5)&1.279(-3) \\
  0.30&0.9850&-9.955(-2)&4.820(-2)&0.3203&-3.544(-3)&-3.539(-2)&
	3.236(-4)&4.311(-3) \\
  0.40&0.9734&-0.1323&8.548(-2)&0.4249&-8.385(-3)&-6.270(-2)&
	1.021(-3)&1.019(-2) \\
  0.50&0.9586&-0.1646&0.1331&0.5277&-1.634(-2)&-9.755(-2)&
	2.488(-3)&1.985(-2) \\
  0.60&0.9405&-0.1964&0.1909&0.6280&-2.814(-2)&-0.1397&
	5.146(-3)&3.416(-2) \\
  0.70&0.9193&-0.2276&0.2586&0.7255&-4.452(-2)&-0.1890&
	9.505(-3)&5.400(-2) \\
  0.80&0.8950&-0.2581&0.3359&0.8197&-6.617(-2)&-0.2450&
	1.616(-2)&8.017(-2) \\
  0.90&0.8678&-0.2877&0.4224&0.9099&-9.374(-2)&-0.3075&
	2.577(-2)&0.1134 \\
  1.00&0.8375&-0.3165&0.5178&0.9956&-0.1279&-0.3759&
	3.910(-2)&0.1545 \\
  1.10&0.8045&-0.3441&0.6214&1.076&-0.1691&-0.4498&
	5.695(-2)&0.2040 \\
  1.20&0.7687&-0.3707&0.7328&1.151&-0.2180&-0.5287&
	8.020(-2)&0.2624 \\
  1.30&0.7304&-0.3959&0.8514&1.219&-0.2750&-0.6118&
	0.1097&0.3302 \\
  1.40&0.6896&-0.4197&0.9764&1.280&-0.3405&-0.6985&
	0.1466&0.4078 \\
  1.50&0.6465&-0.4420&1.107&1.333&-0.4148&-0.7879&
	0.1916&0.4952 \\
  1.60&0.6013&-0.4626&1.243&1.377&-0.4981&-0.8789&
	0.2459&0.5926 \\
  1.70&0.5540&-0.4814&1.382&1.411&-0.5906&-0.9705&
	0.3105&0.6996 \\
  1.80&0.5051&-0.4982&1.525&1.433&-0.6922&-1.061&
	0.3862&0.8158 \\
  1.90&0.4545&-0.5128&1.668&1.442&-0.8027&-1.149&
	0.4739&0.9404 \\
  2.00&0.4026&-0.5250&1.813&1.436&-0.9219&-1.233&
	0.5745&1.072 \\
  2.10&0.3496&-0.5346&1.955&1.412&-1.049&-1.310&
	0.6885&1.209 \\
  2.20&0.2958&-0.5413&2.094&1.367&-1.183&-1.376&
	0.8164&1.349 \\
  2.30&0.2414&-0.5448&2.228&1.296&-1.324&-1.427&
	0.9582&1.487 \\
  2.40&0.1869&-0.5447&2.352&1.190&-1.468&-1.456&
	1.114&1.617 \\
  2.50&0.1326&-0.5403&2.464&1.037&-1.614&-1.451&
	1.281&1.728 \\
  2.60&7.904(-2)&-0.5306&2.557&0.8061&-1.756&-1.389&
	1.458&1.798 \\
  2.70&2.674(-2)&-0.5139&2.620&0.3959&-1.887&-1.189&
	1.637&1.753 \\
  2.75270&0.000&-0.5000&2.628&-0.3645&-1.942&-0.6776&
	1.724&1.368 \\
  \end{tabular}
 \end{center}
 \label{table1}
\end{table}%

\begin{table}
\caption{The functions and their derivatives to represent the equilibrium
figure for $n=1$. (I)
}
 \begin{center}
  \begin{tabular}{l|cccccccc}
  \multicolumn{9}{c}{$n=1.0$} \\ \hline
  $\xi$&$\Th_0$&$d \Th_0/d \xi$&$\three \psi_2$&
  $d \three \psi_2/d \xi$&$\four \psi_3$&$d \four \psi_3/d \xi$&
  $\five \psi_4$&$d \five \psi_4/d \xi$ \\ \hline
  0.00&1.000&0.000&0.000&0.000&0.000&0.000&
	0.000&0.000 \\
  0.10&0.9983&-3.330(-2)&3.331(-3)&6.657(-2)&-6.977(-5)&-2.092(-3)&
	1.831(-6)&7.324(-5) \\
  0.20&0.9933&-6.640(-2)&1.330(-2)&0.1326&-5.573(-4)&-8.347(-3)&
	2.926(-5)&5.847(-4) \\
  0.30&0.9851&-9.910(-2)&2.981(-2)&0.1974&-1.876(-3)&-1.869(-2)&
	1.478(-4)&1.967(-3) \\
  0.40&0.9735&-0.1312&5.273(-2)&0.2606&-4.428(-3)&-3.302(-2)&
	4.656(-4)&4.639(-3) \\
  0.50&0.9589&-0.1625&8.186(-2)&0.3216&-8.606(-3)&-5.116(-2)&
	1.132(-3)&9.006(-3) \\
  0.60&0.9411&-0.1929&0.1169&0.3797&-1.478(-2)&-7.291(-2)&
	2.336(-3)&1.544(-2) \\
  0.70&0.9203&-0.2221&0.1577&0.4347&-2.330(-2)&-9.804(-2)&
	4.302(-3)&2.431(-2) \\
  0.80&0.8967&-0.2500&0.2038&0.4859&-3.449(-2)&-0.1263&
	7.289(-3)&3.591(-2) \\
  0.90&0.8704&-0.2764&0.2547&0.5329&-4.865(-2)&-0.1572&
	1.158(-2)&5.054(-2) \\
  1.00&0.8415&-0.3012&0.3102&0.5753&-6.602(-2)&-0.1907&
	1.750(-2)&6.842(-2) \\
  1.10&0.8102&-0.3242&0.3696&0.6128&-8.684(-2)&-0.2261&
	2.538(-2)&8.974(-2) \\
  1.20&0.7767&-0.3453&0.4326&0.6450&-0.1113&-0.2632&
	3.557(-2)&0.1147 \\
  1.30&0.7412&-0.3644&0.4984&0.6717&-0.1395&-0.3015&
	4.844(-2)&0.1432 \\
  1.40&0.7039&-0.3814&0.5667&0.6925&-0.1716&-0.3405&
	6.434(-2)&0.1755 \\
  1.50&0.6650&-0.3962&0.6367&0.7074&-0.2076&-0.3798&
	8.367(-2)&0.2115 \\
  1.60&0.6247&-0.4087&0.7080&0.7161&-0.2476&-0.4190&
	0.1068&0.2511 \\
  1.70&0.5833&-0.4189&0.7798&0.7186&-0.2914&-0.4575&
	0.1340&0.2942 \\
  1.80&0.5410&-0.4268&0.8515&0.7148&-0.3390&-0.4949&
	0.1657&0.3405 \\
  1.90&0.4981&-0.4323&0.9225&0.7048&-0.3903&-0.5307&
	0.2022&0.3898 \\
  2.00&0.4546&-0.4354&0.9922&0.6886&-0.4451&-0.5645&
	0.2438&0.4419 \\
  2.10&0.4111&-0.4361&1.060&0.6664&-0.5031&-0.5957&
	0.2907&0.4963 \\
  2.20&0.3675&-0.4345&1.125&0.6382&-0.5642&-0.6240&
	0.3431&0.5527 \\
  2.30&0.3242&-0.4307&1.187&0.6044&-0.6278&-0.6490&
	0.4013&0.6106 \\
  2.40&0.2814&-0.4245&1.246&0.5651&-0.6938&-0.6703&
	0.4653&0.6695 \\
  2.50&0.2394&-0.4162&1.300&0.5207&-0.7618&-0.6875&
	0.5352&0.7289 \\
  2.60&0.1983&-0.4058&1.350&0.4715&-0.8312&-0.7004&
	0.6110&0.7882 \\
  2.70&0.1583&-0.3935&1.394&0.4179&-0.9017&-0.7086&
	0.6928&0.8468 \\
  2.80&0.1196&-0.3792&1.433&0.3604&-0.9728&-0.7119&
	0.7804&0.9041 \\
  2.90&8.250(-2)&-0.3633&1.466&0.2993&-1.044&-0.7100&
	0.8735&0.9595 \\
  3.00&4.704(-2)&-0.3457&1.493&0.2352&-1.115&-0.7030&
	0.9722&1.012 \\
  3.10&1.341(-2)&-0.3266&1.513&0.1686&-1.184&-0.6906&
	1.076&1.062 \\
  3.14159&0.000&-0.3183&1.520&0.1402&-1.213&-0.6838&
	1.120&1.081 \\
  \end{tabular}
 \end{center}
 \label{table2}
\end{table}%

\begin{table}
\caption{The functions and their derivatives to represent the equilibrium
figure for $n=1.5$. (I)
}
 \begin{center}
  \begin{tabular}{l|cccccccc}
  \multicolumn{9}{c}{$n=1.5$} \\ \hline
  $\xi$&$\Th_0$&$d \Th_0/d \xi$&$\three \psi_2$&
  $d \three \psi_2/d \xi$&$\four \psi_3$&$d \four \psi_3/d \xi$&
  $\five \psi_4$&$d \five \psi_4/d \xi$ \\ \hline
  0.00&1.000&0.000&0.000&0.000&0.000&0.000&
	0.000&0.000 \\
  0.10&0.9983&-3.328(-2)&1.994(-3)&3.984(-2)&-3.512(-5)&-1.053(-3)&
	7.821(-7)&3.127(-5) \\
  0.20&0.9934&-6.627(-2)&7.951(-3)&7.917(-2)&-2.802(-4)&-4.194(-3)&
	1.249(-5)&2.494(-4) \\
  0.30&0.9851&-9.866(-2)&1.779(-2)&0.1175&-9.419(-4)&-9.372(-3)&
	6.301(-5)&8.375(-4) \\
  0.40&0.9737&-0.1302&3.140(-2)&0.1543&-2.220(-3)&-1.650(-2)&
	1.982(-4)&1.971(-3) \\
  0.50&0.9591&-0.1605&4.859(-2)&0.1892&-4.303(-3)&-2.547(-2)&
	4.810(-4)&3.815(-3) \\
  0.60&0.9416&-0.1895&6.916(-2)&0.2218&-7.370(-3)&-3.613(-2)&
	9.900(-4)&6.521(-3) \\
  0.70&0.9213&-0.2169&9.286(-2)&0.2517&-1.158(-2)&-4.831(-2)&
	1.818(-3)&1.022(-2) \\
  0.80&0.8983&-0.2424&0.1194&0.2786&-1.708(-2)&-6.184(-2)&
	3.071(-3)&1.503(-2) \\
  0.90&0.8728&-0.2659&0.1485&0.3022&-2.398(-2)&-7.650(-2)&
	4.866(-3)&2.106(-2) \\
  1.00&0.8452&-0.2873&0.1797&0.3225&-3.241(-2)&-9.211(-2)&
	7.326(-3)&2.836(-2) \\
  1.10&0.8155&-0.3064&0.2129&0.3393&-4.243(-2)&-0.1084&
	1.058(-2)&3.701(-2) \\
  1.20&0.7840&-0.3231&0.2475&0.3526&-5.411(-2)&-0.1253&
	1.477(-2)&4.703(-2) \\
  1.30&0.7509&-0.3375&0.2833&0.3623&-6.750(-2)&-0.1424&
	2.004(-2)&5.845(-2) \\
  1.40&0.7166&-0.3494&0.3198&0.3686&-8.260(-2)&-0.1597&
	2.651(-2)&7.126(-2) \\
  1.50&0.6811&-0.3590&0.3569&0.3716&-9.943(-2)&-0.1769&
	3.433(-2)&8.545(-2) \\
  1.60&0.6448&-0.3661&0.3940&0.3716&-0.1180&-0.1939&
	4.364(-2)&0.1010 \\
  1.70&0.6080&-0.3710&0.4311&0.3686&-0.1382&-0.2106&
	5.458(-2)&0.1179 \\
  1.80&0.5707&-0.3737&0.4677&0.3631&-0.1601&-0.2268&
	6.726(-2)&0.1360 \\
  1.90&0.5333&-0.3742&0.5036&0.3553&-0.1835&-0.2424&
	8.182(-2)&0.1554 \\
  2.00&0.4959&-0.3728&0.5387&0.3455&-0.2085&-0.2575&
	9.838(-2)&0.1760 \\
  2.10&0.4588&-0.3696&0.5727&0.3341&-0.2350&-0.2720&
	0.1171&0.1978 \\
  2.20&0.4221&-0.3646&0.6054&0.3214&-0.2629&-0.2859&
	0.1380&0.2207 \\
  2.30&0.3859&-0.3582&0.6369&0.3079&-0.2922&-0.2992&
	0.1613&0.2448 \\
  2.40&0.3505&-0.3504&0.6670&0.2938&-0.3227&-0.3120&
	0.1870&0.2700 \\
  2.50&0.3159&-0.3413&0.6957&0.2795&-0.3546&-0.3244&
	0.2153&0.2965 \\
  2.60&0.2823&-0.3313&0.7229&0.2654&-0.3876&-0.3365&
	0.2463&0.3241 \\
  2.70&0.2497&-0.3204&0.7488&0.2519&-0.4219&-0.3485&
	0.2802&0.3531 \\
  2.80&0.2182&-0.3088&0.7733&0.2393&-0.4573&-0.3606&
	0.3170&0.3836 \\
  2.90&0.1879&-0.2967&0.7967&0.2278&-0.4940&-0.3730&
	0.3569&0.4157 \\
  3.00&0.1589&-0.2843&0.8189&0.2180&-0.5319&-0.3860&
	0.4002&0.4497 \\
  3.10&0.1311&-0.2715&0.8403&0.2102&-0.5712&-0.3999&
	0.4470&0.4859 \\
  3.20&0.1046&-0.2588&0.8611&0.2047&-0.6119&-0.4151&
	0.4975&0.5246 \\
  3.30&7.931(-2)&-0.2460&0.8814&0.2021&-0.6543&-0.4323&
	0.5520&0.5666 \\
  3.40&5.534(-2)&-0.2334&0.9016&0.2031&-0.6985&-0.4519&
	0.6109&0.6124 \\
  3.50&3.262(-2)&-0.2212&0.9221&0.2086&-0.7448&-0.4752&
	0.6746&0.6633 \\
  3.60&1.109(-2)&-0.2094&0.9435&0.2206&-0.7937&-0.5041&
	0.7438&0.7213 \\
  3.65375&0.000&-0.2033&0.9557&0.2318&-0.8213&-0.5240&
	0.7835&0.7575 \\
  \end{tabular}
 \end{center}
 \label{table3}
\end{table}%

\begin{table}
\caption{The functions and their derivatives to represent the equilibrium
figure for $n=2$. (I)
}
 \begin{center}
  \begin{tabular}{l|cccccccc}
  \multicolumn{9}{c}{$n=2.0$} \\ \hline
  $\xi$&$\Th_0$&$d \Th_0/d \xi$&$\three \psi_2$&
  $d \three \psi_2/d \xi$&$\four \psi_3$&$d \four \psi_3/d \xi$&
  $\five \psi_4$&$d \five \psi_4/d \xi$ \\ \hline
  0.00&1.000&0.000&0.000&0.000&0.000&0.000&
	0.000&0.000 \\
  0.10&0.9983&-3.327(-2)&1.125(-3)&2.247(-2)&-1.631(-5)&-4.890(-4)&
	3.016(-7)&1.206(-5) \\
  0.20&0.9934&-6.614(-2)&4.481(-3)&4.456(-2)&-1.301(-4)&-1.945(-3)&
	4.812(-6)&9.607(-5) \\
  0.30&0.9851&-9.822(-2)&1.001(-2)&6.589(-2)&-4.366(-4)&-4.337(-3)&
	2.425(-5)&3.221(-4) \\
  0.40&0.9738&-0.1292&1.762(-2)&8.613(-2)&-1.027(-3)&-7.612(-3)&
	7.617(-5)&7.563(-4) \\
  0.50&0.9594&-0.1586&2.719(-2)&0.1050&-1.986(-3)&-1.170(-2)&
	1.845(-4)&1.460(-3) \\
  0.60&0.9421&-0.1863&3.856(-2)&0.1222&-3.392(-3)&-1.653(-2)&
	3.789(-4)&2.486(-3) \\
  0.70&0.9222&-0.2119&5.156(-2)&0.1375&-5.313(-3)&-2.199(-2)&
	6.941(-4)&3.883(-3) \\
  0.80&0.8998&-0.2352&6.599(-2)&0.1508&-7.807(-3)&-2.799(-2)&
	1.169(-3)&5.689(-3) \\
  0.90&0.8752&-0.2561&8.165(-2)&0.1620&-1.093(-2)&-3.443(-2)&
	1.847(-3)&7.934(-3) \\
  1.00&0.8487&-0.2745&9.833(-2)&0.1712&-1.470(-2)&-4.120(-2)&
	2.771(-3)&1.064(-2) \\
  1.10&0.8204&-0.2904&0.1158&0.1782&-1.917(-2)&-4.821(-2)&
	3.991(-3)&1.383(-2) \\
  1.20&0.7907&-0.3036&0.1339&0.1833&-2.435(-2)&-5.538(-2)&
	5.553(-3)&1.750(-2) \\
  1.30&0.7598&-0.3142&0.1524&0.1864&-3.025(-2)&-6.262(-2)&
	7.507(-3)&2.166(-2) \\
  1.40&0.7279&-0.3224&0.1711&0.1879&-3.688(-2)&-6.986(-2)&
	9.902(-3)&2.632(-2) \\
  1.50&0.6954&-0.3281&0.1899&0.1878&-4.422(-2)&-7.706(-2)&
	1.279(-2)&3.147(-2) \\
  1.60&0.6624&-0.3316&0.2086&0.1863&-5.228(-2)&-8.417(-2)&
	1.621(-2)&3.710(-2) \\
  1.70&0.6291&-0.3330&0.2272&0.1838&-6.105(-2)&-9.117(-2)&
	2.022(-2)&4.323(-2) \\
  1.80&0.5958&-0.3325&0.2454&0.1803&-7.051(-2)&-9.805(-2)&
	2.487(-2)&4.984(-2) \\
  1.90&0.5627&-0.3302&0.2632&0.1762&-8.066(-2)&-0.1048&
	3.021(-2)&5.693(-2) \\
  2.00&0.5298&-0.3263&0.2806&0.1715&-9.147(-2)&-0.1115&
	3.628(-2)&6.453(-2) \\
  2.10&0.4975&-0.3211&0.2975&0.1666&-0.1029&-0.1180&
	4.313(-2)&7.262(-2) \\
  2.20&0.4656&-0.3147&0.3139&0.1616&-0.1151&-0.1245&
	5.082(-2)&8.124(-2) \\
  2.30&0.4345&-0.3073&0.3298&0.1567&-0.1279&-0.1310&
	5.940(-2)&9.039(-2) \\
  2.40&0.4042&-0.2991&0.3452&0.1519&-0.1413&-0.1375&
	6.892(-2)&0.1001 \\
  2.50&0.3747&-0.2902&0.3602&0.1475&-0.1554&-0.1441&
	7.944(-2)&0.1104 \\
  2.60&0.3462&-0.2808&0.3748&0.1435&-0.1701&-0.1509&
	9.102(-2)&0.1214 \\
  2.70&0.3186&-0.2710&0.3889&0.1401&-0.1855&-0.1578&
	0.1037&0.1330 \\
  2.80&0.2920&-0.2610&0.4028&0.1372&-0.2017&-0.1650&
	0.1177&0.1454 \\
  2.90&0.2664&-0.2508&0.4164&0.1351&-0.2185&-0.1724&
	0.1328&0.1586 \\
  3.00&0.2418&-0.2406&0.4298&0.1337&-0.2362&-0.1803&
	0.1494&0.1726 \\
  3.10&0.2183&-0.2305&0.4432&0.1330&-0.2546&-0.1886&
	0.1674&0.1875 \\
  3.20&0.1957&-0.2204&0.4565&0.1332&-0.2739&-0.1973&
	0.1869&0.2035 \\
  3.30&0.1742&-0.2106&0.4698&0.1341&-0.2941&-0.2067&
	0.2081&0.2205 \\
  3.40&0.1536&-0.2010&0.4833&0.1359&-0.3153&-0.2166&
	0.2311&0.2387 \\
  3.50&0.1340&-0.1917&0.4970&0.1386&-0.3374&-0.2273&
	0.2559&0.2582 \\
  3.60&0.1153&-0.1827&0.5111&0.1422&-0.3607&-0.2387&
	0.2828&0.2791 \\
  3.70&9.742(-2)&-0.1740&0.5255&0.1466&-0.3852&-0.2510&
	0.3118&0.3015 \\
  3.80&8.043(-2)&-0.1658&0.5404&0.1519&-0.4110&-0.2641&
	0.3431&0.3255 \\
  3.90&6.425(-2)&-0.1579&0.5559&0.1581&-0.4381&-0.2783&
	0.3769&0.3513 \\
  4.00&4.884(-2)&-0.1504&0.5721&0.1653&-0.4666&-0.2935&
	0.4134&0.3790 \\
  4.10&3.416(-2)&-0.1433&0.5890&0.1733&-0.4968&-0.3098&
	0.4528&0.4087 \\
  4.20&2.016(-2)&-0.1367&0.6068&0.1824&-0.5287&-0.3274&
	0.4952&0.4407 \\
  4.30&6.811(-3)&-0.1304&0.6255&0.1924&-0.5623&-0.3463&
	0.5410&0.4751 \\
  4.35287&0.000&-0.1272&0.6358&0.1980&-0.5809&-0.3569&
	0.5666&0.4944 \\
  \end{tabular}
 \end{center}
 \label{table4}
\end{table}%

\begin{table}
\caption{The functions and their derivatives to represent the equilibrium
figure for $n=0.5$. (II)
}
 \begin{center}
  \begin{tabular}{l|cccccccccc}
  \multicolumn{11}{c}{$n=0.5$} \\ \hline
  $\xi$&$\six \psi_0$&$d \six \psi_0/d \xi$&$\six \psi_2$&
  $d \six \psi_2/d \xi$&$\six \psi_{22}$&$d \six \psi_{22}/d \xi$&
  $\six \psi_4$&$d \six \psi_4/d \xi$&$\six \psi_5$&
  $d \six \psi_5/d \xi$ \\ \hline
  0.00&0.000&0.000&0.000&0.000&0.000&0.000&
	0.000&0.000&0.000&0.000 \\
  0.10&1.720(-10)&1.032(-8)&5.400(-3)&0.1080&9.847(-3)&0.1969&
	2.061(-6)&8.245(-5)&-1.312(-7)&-6.558(-6) \\
  0.20&1.104(-8)&3.317(-7)&2.158(-2)&0.2155&3.935(-2)&0.3929&
	3.299(-5)&6.601(-4)&-4.195(-6)&-1.049(-4) \\
  0.30&1.264(-7)&2.535(-6)&4.846(-2)&0.3220&8.838(-2)&0.5873&
	1.672(-4)&2.231(-3)&-3.183(-5)&-5.301(-4) \\
  0.40&7.150(-7)&1.078(-5)&8.594(-2)&0.4272&0.1567&0.7792&
	5.291(-4)&5.299(-3)&-1.339(-4)&-1.672(-3) \\
  0.50&2.752(-6)&3.330(-5)&0.1338&0.5305&0.2441&0.9678&
	1.294(-3)&1.038(-2)&-4.080(-4)&-4.072(-3) \\
  0.60&8.309(-6)&8.410(-5)&0.1920&0.6315&0.3502&1.152&
	2.689(-3)&1.799(-2)&-1.013(-3)&-8.419(-3) \\
  0.70&2.123(-5)&1.850(-4)&0.2601&0.7297&0.4744&1.331&
	4.994(-3)&2.868(-2)&-2.184(-3)&-1.554(-2) \\
  0.80&4.805(-5)&3.684(-4)&0.3378&0.8247&0.6162&1.505&
	8.547(-3)&4.302(-2)&-4.245(-3)&-2.640(-2) \\
  0.90&9.917(-5)&6.801(-4)&0.4249&0.9159&0.7750&1.671&
	1.374(-2)&6.160(-2)&-7.624(-3)&-4.207(-2) \\
  1.00&1.904(-4)&1.184(-3)&0.5208&1.003&0.9501&1.829&
	2.103(-2)&8.508(-2)&-1.286(-2)&-6.377(-2) \\
  1.10&3.453(-4)&1.968(-3)&0.6253&1.085&1.140&1.978&
	3.094(-2)&0.1141&-2.062(-2)&-9.277(-2) \\
  1.20&5.973(-4)&3.151(-3)&0.7377&1.162&1.345&2.116&
	4.407(-2)&0.1496&-3.170(-2)&-0.1305 \\
  1.30&9.942(-4)&4.895(-3)&0.8575&1.234&1.563&2.243&
	6.109(-2)&0.1922&-4.705(-2)&-0.1783 \\
  1.40&1.602(-3)&7.419(-3)&0.9843&1.299&1.793&2.357&
	8.279(-2)&0.2432&-6.774(-2)&-0.2376 \\
  1.50&2.514(-3)&1.102(-2)&1.117&1.358&2.034&2.456&
	0.1100&0.3036&-9.501(-2)&-0.3101 \\
  1.60&3.856(-3)&1.612(-2)&1.256&1.411&2.284&2.539&
	0.1439&0.3750&-0.1302&-0.3970 \\
  1.70&5.806(-3)&2.329(-2)&1.399&1.456&2.542&2.604&
	0.1855&0.4594&-0.1749&-0.4997 \\
  1.80&8.610(-3)&3.335(-2)&1.547&1.494&2.804&2.648&
	0.2363&0.5593&-0.2307&-0.6194 \\
  1.90&1.261(-2)&4.753(-2)&1.698&1.526&3.070&2.668&
	0.2980&0.6782&-0.2994&-0.7571 \\
  2.00&1.831(-2)&6.766(-2)&1.852&1.551&3.337&2.661&
	0.3727&0.8214&-0.3828&-0.9134 \\
  2.10&2.644(-2)&9.663(-2)&2.008&1.572&3.601&2.621&
	0.4633&0.9966&-0.4827&-1.088 \\
  2.20&3.809(-2)&0.1393&2.166&1.592&3.860&2.544&
	0.5735&1.217&-0.6010&-1.281 \\
  2.30&5.505(-2)&0.2045&2.326&1.620&4.109&2.418&
	0.7089&1.505&-0.7394&-1.489 \\
  2.40&8.032(-2)&0.3100&2.491&1.672&4.342&2.231&
	0.8782&1.906&-0.8992&-1.708 \\
  2.50&0.1197&0.4979&2.663&1.797&4.552&1.956&
	1.097&2.528&-1.081&-1.929 \\
  2.60&0.1867&0.9026&2.857&2.158&4.728&1.538&
	1.402&3.713&-1.284&-2.130 \\
  2.70&0.3305&2.398&3.133&3.903&4.849&0.7890&
	1.917&7.685&-1.504&-2.243 \\
  2.75270&0.6770&$\infty$&3.641&$\infty$&4.868&-0.6130&
	2.888&$\infty$&-1.620&-1.969 \\
  \end{tabular}
 \end{center}
 \label{table5}
\end{table}%

\begin{table}
\caption{The functions and their derivatives to represent the equilibrium
figure for $n=1$. (II)
}
 \begin{center}
  \begin{tabular}{l|cccccccccc}
  \multicolumn{11}{c}{$n=1.0$} \\ \hline
  $\xi$&$\six \psi_0$&$d \six \psi_0/d \xi$&$\six \psi_2$&
  $d \six \psi_2/d \xi$&$\six \psi_{22}$&$d \six \psi_{22}/d \xi$&
  $\six \psi_4$&$d \six \psi_4/d \xi$&$\six \psi_5$&
  $d \six \psi_5/d \xi$ \\ \hline
  0.00&0.000&0.000&0.000&0.000&0.000&0.000&
	0.000&0.000&0.000&0.000 \\
  0.10&0.000&0.000&2.170(-3)&4.336(-2)&4.281(-3)&8.556(-2)&
	1.193(-6)&4.770(-5)&-5.203(-8)&-2.601(-6) \\
  0.20&0.000&0.000&8.660(-3)&8.635(-2)&1.709(-2)&0.1704&
	1.906(-5)&3.808(-4)&-1.663(-6)&-4.155(-5) \\
  0.30&0.000&0.000&1.942(-2)&0.1286&3.831(-2)&0.2538&
	9.627(-5)&1.281(-3)&-1.260(-5)&-2.098(-4) \\
  0.40&0.000&0.000&3.434(-2)&0.1697&6.779(-2)&0.3352&
	3.033(-4)&3.022(-3)&-5.297(-5)&-6.605(-4) \\
  0.50&0.000&0.000&5.332(-2)&0.2094&0.1053&0.4138&
	7.374(-4)&5.866(-3)&-1.611(-4)&-1.605(-3) \\
  0.60&0.000&0.000&7.617(-2)&0.2473&0.1504&0.4890&
	1.521(-3)&1.006(-2)&-3.992(-4)&-3.308(-3) \\
  0.70&0.000&0.000&0.1027&0.2831&0.2029&0.5601&
	2.802(-3)&1.583(-2)&-8.584(-4)&-6.085(-3) \\
  0.80&0.000&0.000&0.1327&0.3165&0.2623&0.6267&
	4.748(-3)&2.339(-2)&-1.664(-3)&-1.030(-2) \\
  0.90&0.000&0.000&0.1659&0.3471&0.3281&0.6880&
	7.546(-3)&3.292(-2)&-2.979(-3)&-1.634(-2) \\
  1.00&0.000&0.000&0.2020&0.3747&0.3997&0.7438&
	1.140(-2)&4.456(-2)&-5.008(-3)&-2.465(-2) \\
  1.10&0.000&0.000&0.2408&0.3992&0.4766&0.7934&
	1.653(-2)&5.845(-2)&-8.000(-3)&-3.568(-2) \\
  1.20&0.000&0.000&0.2817&0.4201&0.5582&0.8365&
	2.317(-2)&7.468(-2)&-1.225(-2)&-4.991(-2) \\
  1.30&0.000&0.000&0.3247&0.4375&0.6437&0.8728&
	3.155(-2)&9.330(-2)&-1.810(-2)&-6.780(-2) \\
  1.40&0.000&0.000&0.3691&0.4511&0.7325&0.9019&
	4.191(-2)&0.1143&-2.595(-2)&-8.986(-2) \\
  1.50&0.000&0.000&0.4147&0.4607&0.8238&0.9237&
	5.450(-2)&0.1378&-3.623(-2)&-0.1165 \\
  1.60&0.000&0.000&0.4611&0.4664&0.9170&0.9380&
	6.954(-2)&0.1636&-4.943(-2)&-0.1483 \\
  1.70&0.000&0.000&0.5079&0.4681&1.011&0.9446&
	8.728(-2)&0.1916&-6.607(-2)&-0.1856 \\
  1.80&0.000&0.000&0.5546&0.4656&1.106&0.9436&
	0.1079&0.2218&-8.673(-2)&-0.2287 \\
  1.90&0.000&0.000&0.6009&0.4591&1.200&0.9350&
	0.1317&0.2539&-0.1120&-0.2781 \\
  2.00&0.000&0.000&0.6463&0.4485&1.292&0.9189&
	0.1588&0.2878&-0.1426&-0.3340 \\
  2.10&0.000&0.000&0.6905&0.4340&1.383&0.8955&
	0.1893&0.3233&-0.1791&-0.3967 \\
  2.20&0.000&0.000&0.7330&0.4157&1.471&0.8649&
	0.2235&0.3600&-0.2221&-0.4663 \\
  2.30&0.000&0.000&0.7735&0.3936&1.556&0.8276&
	0.2614&0.3977&-0.2725&-0.5429 \\
  2.40&0.000&0.000&0.8116&0.3681&1.637&0.7838&
	0.3030&0.4361&-0.3310&-0.6265 \\
  2.50&0.000&0.000&0.8470&0.3391&1.713&0.7341&
	0.3486&0.4748&-0.3981&-0.7170 \\
  2.60&0.000&0.000&0.8793&0.3071&1.783&0.6787&
	0.3980&0.5134&-0.4746&-0.8142 \\
  2.70&0.000&0.000&0.9083&0.2722&1.848&0.6184&
	0.4513&0.5516&-0.5611&-0.9179 \\
  2.80&0.000&0.000&0.9337&0.2347&1.907&0.5537&
	0.5083&0.5889&-0.6584&-1.028 \\
  2.90&0.000&0.000&0.9552&0.1949&1.959&0.4851&
	0.5690&0.6250&-0.7669&-1.143 \\
  3.00&0.000&0.000&0.9726&0.1532&2.004&0.4135&
	0.6332&0.6594&-0.8872&-1.264 \\
  3.10&0.000&0.000&0.9858&0.1098&2.041&0.3394&
	0.7008&0.6917&-1.020&-1.388 \\
  3.14159&0.000&0.000&0.9899&9.134(-2)&2.055&0.3081&
	0.7298&0.7044&-1.079&-1.441 \\
  \end{tabular}
 \end{center}
 \label{table6}
\end{table}%

\begin{table}
\caption{The functions and their derivatives to represent the equilibrium
figure for $n=1.5$. (II)
}
 \begin{center}
  \begin{tabular}{l|cccccccccc}
  \multicolumn{11}{c}{$n=1.5$} \\ \hline
  $\xi$&$\six \psi_0$&$d \six \psi_0/d \xi$&$\six \psi_2$&
  $d \six \psi_2/d \xi$&$\six \psi_{22}$&$d \six \psi_{22}/d \xi$&
  $\six \psi_4$&$d \six \psi_4/d \xi$&$\six \psi_5$&
  $d \six \psi_5/d \xi$ \\ \hline
  0.00&0.000&0.000&0.000&0.000&0.000&0.000&
	0.000&0.000&0.000&0.000 \\
  0.10&-7.107(-11)&-4.264(-9)&9.164(-4)&1.831(-2)&1.737(-3)&3.470(-2)&
	6.423(-7)&2.568(-5)&-1.894(-8)&-9.468(-7) \\
  0.20&-4.536(-9)&-1.359(-7)&3.654(-3)&3.638(-2)&6.926(-3)&6.898(-2)&
	1.024(-5)&2.043(-4)&-6.050(-7)&-1.511(-5) \\
  0.30&-5.142(-8)&-1.025(-6)&8.178(-3)&5.399(-2)&1.550(-2)&0.1024&
	5.152(-5)&6.832(-4)&-4.581(-6)&-7.620(-5) \\
  0.40&-2.870(-7)&-4.283(-6)&1.443(-2)&7.091(-2)&2.737(-2)&0.1346&
	1.614(-4)&1.599(-3)&-1.923(-5)&-2.395(-4) \\
  0.50&-1.085(-6)&-1.292(-5)&2.233(-2)&8.694(-2)&4.237(-2)&0.1652&
	3.898(-4)&3.071(-3)&-5.838(-5)&-5.805(-4) \\
  0.60&-3.207(-6)&-3.170(-5)&3.178(-2)&0.1019&6.034(-2)&0.1939&
	7.976(-4)&5.201(-3)&-1.444(-4)&-1.193(-3) \\
  0.70&-7.988(-6)&-6.741(-5)&4.266(-2)&0.1155&8.107(-2)&0.2203&
	1.455(-3)&8.065(-3)&-3.098(-4)&-2.188(-3) \\
  0.80&-1.755(-5)&-1.290(-4)&5.484(-2)&0.1278&0.1043&0.2442&
	2.437(-3)&1.171(-2)&-5.989(-4)&-3.690(-3) \\
  0.90&-3.502(-5)&-2.276(-4)&6.817(-2)&0.1385&0.1298&0.2655&
	3.824(-3)&1.617(-2)&-1.069(-3)&-5.835(-3) \\
  1.00&-6.476(-5)&-3.767(-4)&8.249(-2)&0.1476&0.1573&0.2840&
	5.697(-3)&2.142(-2)&-1.792(-3)&-8.767(-3) \\
  1.10&-1.126(-4)&-5.916(-4)&9.763(-2)&0.1550&0.1865&0.2996&
	8.135(-3)&2.745(-2)&-2.854(-3)&-1.264(-2) \\
  1.20&-1.859(-4)&-8.897(-4)&0.1134&0.1605&0.2172&0.3123&
	1.121(-2)&3.417(-2)&-4.356(-3)&-1.760(-2) \\
  1.30&-2.939(-4)&-1.290(-3)&0.1297&0.1643&0.2489&0.3222&
	1.499(-2)&4.152(-2)&-6.416(-3)&-2.382(-2) \\
  1.40&-4.479(-4)&-1.812(-3)&0.1462&0.1664&0.2815&0.3293&
	1.953(-2)&4.937(-2)&-9.166(-3)&-3.144(-2) \\
  1.50&-6.611(-4)&-2.478(-3)&0.1629&0.1666&0.3147&0.3337&
	2.488(-2)&5.760(-2)&-1.276(-2)&-4.062(-2) \\
  1.60&-9.490(-4)&-3.310(-3)&0.1795&0.1652&0.3482&0.3357&
	3.106(-2)&6.607(-2)&-1.735(-2)&-5.151(-2) \\
  1.70&-1.329(-3)&-4.330(-3)&0.1959&0.1621&0.3817&0.3355&
	3.809(-2)&7.462(-2)&-2.312(-2)&-6.427(-2) \\
  1.80&-1.822(-3)&-5.563(-3)&0.2119&0.1575&0.4152&0.3332&
	4.598(-2)&8.308(-2)&-3.027(-2)&-7.903(-2) \\
  1.90&-2.450(-3)&-7.032(-3)&0.2273&0.1514&0.4483&0.3293&
	5.470(-2)&9.127(-2)&-3.900(-2)&-9.594(-2) \\
  2.00&-3.237(-3)&-8.764(-3)&0.2421&0.1440&0.4810&0.3239&
	6.422(-2)&9.903(-2)&-4.953(-2)&-0.1151 \\
  2.10&-4.212(-3)&-1.079(-2)&0.2561&0.1353&0.5131&0.3175&
	7.448(-2)&0.1062&-6.211(-2)&-0.1368 \\
  2.20&-5.405(-3)&-1.313(-2)&0.2691&0.1255&0.5445&0.3103&
	8.542(-2)&0.1125&-7.698(-2)&-0.1610 \\
  2.30&-6.849(-3)&-1.582(-2)&0.2811&0.1147&0.5751&0.3027&
	9.695(-2)&0.1178&-9.440(-2)&-0.1880 \\
  2.40&-8.582(-3)&-1.891(-2)&0.2920&0.1030&0.6050&0.2951&
	0.1090&0.1220&-0.1147&-0.2179 \\
  2.50&-1.065(-2)&-2.243(-2)&0.3017&9.043(-2)&0.6342&0.2877&
	0.1213&0.1248&-0.1381&-0.2508 \\
  2.60&-1.309(-2)&-2.646(-2)&0.3101&7.713(-2)&0.6626&0.2810&
	0.1338&0.1259&-0.1649&-0.2871 \\
  2.70&-1.596(-2)&-3.106(-2)&0.3171&6.312(-2)&0.6904&0.2753&
	0.1464&0.1252&-0.1956&-0.3269 \\
  2.80&-1.932(-2)&-3.634(-2)&0.3227&4.839(-2)&0.7177&0.2710&
	0.1588&0.1224&-0.2304&-0.3704 \\
  2.90&-2.325(-2)&-4.242(-2)&0.3268&3.291(-2)&0.7447&0.2683&
	0.1708&0.1170&-0.2698&-0.4181 \\
  3.00&-2.784(-2)&-4.950(-2)&0.3293&1.656(-2)&0.7714&0.2678&
	0.1821&0.1086&-0.3142&-0.4703 \\
  3.10&-3.319(-2)&-5.784(-2)&0.3300&-9.219(-4)&0.7983&0.2698&
	0.1924&9.651(-2)&-0.3641&-0.5275 \\
  3.20&-3.946(-2)&-6.785(-2)&0.3290&-1.994(-2)&0.8255&0.2747&
	0.2013&7.969(-2)&-0.4199&-0.5902 \\
  3.30&-4.684(-2)&-8.021(-2)&0.3260&-4.126(-2)&0.8533&0.2832&
	0.2081&5.650(-2)&-0.4823&-0.6593 \\
  3.40&-5.562(-2)&-9.609(-2)&0.3206&-6.634(-2)&0.8823&0.2959&
	0.2123&2.398(-2)&-0.5520&-0.7357 \\
  3.50&-6.626(-2)&-0.1181&0.3125&-9.850(-2)&0.9127&0.3140&
	0.2124&-2.432(-2)&-0.6298&-0.8209 \\
  3.60&-7.969(-2)&-0.1548&0.3004&-0.1493&0.9453&0.3395&
	0.2062&-0.1100&-0.7166&-0.9174 \\
  3.65375&-8.910(-2)&-0.2311&0.2908&-0.2555&0.9640&0.3585&
	0.1977&-0.3006&-0.7674&-0.9761 \\
  \end{tabular}
 \end{center}
 \label{table7}
\end{table}%

\begin{table}
\caption{The functions and their derivatives to represent the equilibrium
figure for $n=2$. (II)
}
 \begin{center}
  \begin{tabular}{l|cccccccccc}
  \multicolumn{11}{c}{$n=2.0$} \\ \hline
  $\xi$&$\six \psi_0$&$d \six \psi_0/d \xi$&$\six \psi_2$&
  $d \six \psi_2/d \xi$&$\six \psi_{22}$&$d \six \psi_{22}/d \xi$&
  $\six \psi_4$&$d \six \psi_4/d \xi$&$\six \psi_5$&
  $d \six \psi_5/d \xi$ \\ \hline
  0.00&0.000&0.000&0.000&0.000&0.000&0.000&
	0.000&0.000&0.000&0.000 \\
  0.10&-6.032(-11)&-3.618(-9)&3.396(-4)&6.782(-3)&6.308(-4)&1.260(-2)&
	2.728(-7)&1.090(-5)&-6.086(-9)&-3.042(-7) \\
  0.20&-3.839(-9)&-1.149(-7)&1.353(-3)&1.345(-2)&2.513(-3)&2.499(-2)&
	4.339(-6)&8.643(-5)&-1.943(-7)&-4.852(-6) \\
  0.30&-4.331(-8)&-8.611(-7)&3.022(-3)&1.989(-2)&5.616(-3)&3.698(-2)&
	2.175(-5)&2.874(-4)&-1.470(-6)&-2.443(-5) \\
  0.40&-2.401(-7)&-3.564(-6)&5.319(-3)&2.599(-2)&9.890(-3)&4.839(-2)&
	6.779(-5)&6.672(-4)&-6.161(-6)&-7.665(-5) \\
  0.50&-9.002(-7)&-1.063(-5)&8.206(-3)&3.167(-2)&1.527(-2)&5.905(-2)&
	1.626(-4)&1.269(-3)&-1.868(-5)&-1.854(-4) \\
  0.60&-2.633(-6)&-2.573(-5)&1.164(-2)&3.683(-2)&2.167(-2)&6.883(-2)&
	3.300(-4)&2.123(-3)&-4.610(-5)&-3.801(-4) \\
  0.70&-6.478(-6)&-5.384(-5)&1.555(-2)&4.141(-2)&2.900(-2)&7.763(-2)&
	5.962(-4)&3.246(-3)&-9.870(-5)&-6.950(-4) \\
  0.80&-1.404(-5)&-1.012(-4)&1.990(-2)&4.535(-2)&3.716(-2)&8.536(-2)&
	9.883(-4)&4.640(-3)&-1.904(-4)&-1.168(-3) \\
  0.90&-2.759(-5)&-1.750(-4)&2.460(-2)&4.862(-2)&4.604(-2)&9.200(-2)&
	1.533(-3)&6.293(-3)&-3.390(-4)&-1.841(-3) \\
  1.00&-5.018(-5)&-2.832(-4)&2.960(-2)&5.119(-2)&5.553(-2)&9.753(-2)&
	2.255(-3)&8.181(-3)&-5.668(-4)&-2.757(-3) \\
  1.10&-8.566(-5)&-4.342(-4)&3.482(-2)&5.306(-2)&6.551(-2)&0.1020&
	3.176(-3)&1.027(-2)&-9.001(-4)&-3.962(-3) \\
  1.20&-1.387(-4)&-6.365(-4)&4.019(-2)&5.424(-2)&7.589(-2)&0.1054&
	4.313(-3)&1.251(-2)&-1.370(-3)&-5.500(-3) \\
  1.30&-2.149(-4)&-8.980(-4)&4.564(-2)&5.474(-2)&8.656(-2)&0.1079&
	5.681(-3)&1.485(-2)&-2.013(-3)&-7.420(-3) \\
  1.40&-3.206(-4)&-1.226(-3)&5.111(-2)&5.462(-2)&9.743(-2)&0.1095&
	7.286(-3)&1.725(-2)&-2.868(-3)&-9.768(-3) \\
  1.50&-4.626(-4)&-1.628(-3)&5.654(-2)&5.389(-2)&0.1084&0.1103&
	9.130(-3)&1.964(-2)&-3.982(-3)&-1.259(-2) \\
  1.60&-6.487(-4)&-2.108(-3)&6.187(-2)&5.263(-2)&0.1195&0.1105&
	1.121(-2)&2.198(-2)&-5.404(-3)&-1.594(-2) \\
  1.70&-8.870(-4)&-2.671(-3)&6.705(-2)&5.088(-2)&0.1305&0.1103&
	1.352(-2)&2.420(-2)&-7.189(-3)&-1.986(-2) \\
  1.80&-1.186(-3)&-3.321(-3)&7.204(-2)&4.870(-2)&0.1415&0.1096&
	1.605(-2)&2.627(-2)&-9.396(-3)&-2.440(-2) \\
  1.90&-1.554(-3)&-4.059(-3)&7.678(-2)&4.615(-2)&0.1524&0.1087&
	1.877(-2)&2.815(-2)&-1.209(-2)&-2.961(-2) \\
  2.00&-2.001(-3)&-4.888(-3)&8.125(-2)&4.329(-2)&0.1633&0.1077&
	2.167(-2)&2.978(-2)&-1.534(-2)&-3.556(-2) \\
  2.10&-2.535(-3)&-5.807(-3)&8.543(-2)&4.017(-2)&0.1740&0.1067&
	2.472(-2)&3.116(-2)&-1.923(-2)&-4.229(-2) \\
  2.20&-3.165(-3)&-6.817(-3)&8.928(-2)&3.684(-2)&0.1846&0.1059&
	2.789(-2)&3.224(-2)&-2.383(-2)&-4.986(-2) \\
  2.30&-3.901(-3)&-7.919(-3)&9.279(-2)&3.336(-2)&0.1952&0.1053&
	3.116(-2)&3.301(-2)&-2.923(-2)&-5.834(-2) \\
  2.40&-4.752(-3)&-9.113(-3)&9.595(-2)&2.976(-2)&0.2057&0.1050&
	3.448(-2)&3.345(-2)&-3.553(-2)&-6.781(-2) \\
  2.50&-5.727(-3)&-1.040(-2)&9.874(-2)&2.607(-2)&0.2162&0.1051&
	3.783(-2)&3.354(-2)&-4.283(-2)&-7.833(-2) \\
  2.60&-6.835(-3)&-1.178(-2)&0.1012&2.234(-2)&0.2267&0.1057&
	4.118(-2)&3.327(-2)&-5.124(-2)&-9.000(-2) \\
  2.70&-8.087(-3)&-1.326(-2)&0.1032&1.857(-2)&0.2373&0.1069&
	4.448(-2)&3.264(-2)&-6.087(-2)&-0.1029 \\
  2.80&-9.491(-3)&-1.484(-2)&0.1049&1.478(-2)&0.2481&0.1087&
	4.769(-2)&3.161(-2)&-7.186(-2)&-0.1171 \\
  2.90&-1.106(-2)&-1.652(-2)&0.1062&1.099(-2)&0.2591&0.1112&
	5.078(-2)&3.019(-2)&-8.434(-2)&-0.1328 \\
  3.00&-1.280(-2)&-1.832(-2)&0.1071&7.195(-3)&0.2704&0.1145&
	5.372(-2)&2.835(-2)&-9.847(-2)&-0.1500 \\
  3.10&-1.473(-2)&-2.024(-2)&0.1076&3.392(-3)&0.2820&0.1185&
	5.644(-2)&2.608(-2)&-0.1144&-0.1690 \\
  3.20&-1.685(-2)&-2.228(-2)&0.1078&-4.267(-4)&0.2941&0.1234&
	5.892(-2)&2.334(-2)&-0.1323&-0.1897 \\
  3.30&-1.919(-2)&-2.447(-2)&0.1075&-4.275(-3)&0.3067&0.1292&
	6.109(-2)&2.012(-2)&-0.1524&-0.2125 \\
  3.40&-2.175(-2)&-2.681(-2)&0.1069&-8.172(-3)&0.3200&0.1360&
	6.292(-2)&1.637(-2)&-0.1749&-0.2374 \\
  3.50&-2.455(-2)&-2.933(-2)&0.1059&-1.214(-2)&0.3339&0.1437&
	6.435(-2)&1.206(-2)&-0.2000&-0.2647 \\
  3.60&-2.762(-2)&-3.203(-2)&0.1045&-1.621(-2)&0.3487&0.1524&
	6.531(-2)&7.129(-3)&-0.2279&-0.2945 \\
  3.70&-3.097(-2)&-3.495(-2)&0.1026&-2.042(-2)&0.3645&0.1621&
	6.575(-2)&1.528(-3)&-0.2590&-0.3270 \\
  3.80&-3.462(-2)&-3.810(-2)&0.1004&-2.480(-2)&0.3812&0.1730&
	6.559(-2)&-4.813(-3)&-0.2934&-0.3626 \\
  3.90&-3.860(-2)&-4.152(-2)&9.766(-2)&-2.940(-2)&0.3991&0.1850&
	6.476(-2)&-1.197(-2)&-0.3316&-0.4014 \\
  4.00&-4.293(-2)&-4.522(-2)&9.448(-2)&-3.427(-2)&0.4182&0.1982&
	6.317(-2)&-2.004(-2)&-0.3738&-0.4438 \\
  4.10&-4.765(-2)&-4.926(-2)&9.080(-2)&-3.945(-2)&0.4388&0.2127&
	6.072(-2)&-2.911(-2)&-0.4205&-0.4900 \\
  4.20&-5.280(-2)&-5.366(-2)&8.658(-2)&-4.502(-2)&0.4608&0.2284&
	5.731(-2)&-3.930(-2)&-0.4720&-0.5404 \\
  4.30&-5.840(-2)&-5.846(-2)&8.178(-2)&-5.103(-2)&0.4845&0.2456&
	5.282(-2)&-5.074(-2)&-0.5287&-0.5954 \\
  4.35287&-6.156(-2)&-6.119(-2)&7.899(-2)&-5.442(-2)&0.4977&0.2552&
	4.996(-2)&-5.735(-2)&-0.5610&-0.6264 \\
  \end{tabular}
 \end{center}
 \label{table8}
\end{table}%

\begin{table}
\caption{The velocity potentials and their derivatives for $n=0.5$.
}
 \begin{center}
  \begin{tabular}{l|cccccc}
  \multicolumn{7}{c}{$n=0.5$} \\ \hline
  $\xi$&$\four \phi_2$&$d \four \phi_2/d \xi$&$\five \phi_3$&
  $d \five \phi_3/d \xi$&$\six \phi_4$&$d \six \phi_4/d \xi$ \\ \hline
  0.00&0.000&0.000&0.000&0.000&0.000&0.000 \\
  0.10&1.646(-3)&3.291(-2)&1.439(-5)&4.324(-4)&-1.148(-7)&-4.623(-6) \\
  0.20&6.583(-3)&6.584(-2)&1.149(-4)&1.725(-3)&-1.828(-6)&-3.662(-5) \\
  0.30&1.481(-2)&9.878(-2)&3.879(-4)&3.881(-3)&-9.248(-6)&-1.234(-4) \\
  0.40&2.634(-2)&0.1318&9.196(-4)&6.901(-3)&-2.923(-5)&-2.926(-4) \\
  0.50&4.117(-2)&0.1648&1.797(-3)&1.079(-2)&-7.139(-5)&-5.717(-4) \\
  0.60&5.930(-2)&0.1978&3.106(-3)&1.555(-2)&-1.481(-4)&-9.888(-4) \\
  0.70&8.074(-2)&0.2310&4.936(-3)&2.119(-2)&-2.746(-4)&-1.572(-3) \\
  0.80&0.1055&0.2642&7.373(-3)&2.771(-2)&-4.689(-4)&-2.350(-3) \\
  0.90&0.1336&0.2975&1.051(-2)&3.511(-2)&-7.519(-4)&-3.351(-3) \\
  1.00&0.1650&0.3309&1.442(-2)&4.341(-2)&-1.147(-3)&-4.605(-3) \\
  1.10&0.1998&0.3644&1.922(-2)&5.262(-2)&-1.682(-3)&-6.141(-3) \\
  1.20&0.2379&0.3981&2.498(-2)&6.275(-2)&-2.386(-3)&-7.991(-3) \\
  1.30&0.2794&0.4318&3.180(-2)&7.379(-2)&-3.292(-3)&-1.019(-2) \\
  1.40&0.3243&0.4658&3.977(-2)&8.578(-2)&-4.435(-3)&-1.276(-2) \\
  1.50&0.3725&0.4999&4.899(-2)&9.872(-2)&-5.856(-3)&-1.574(-2) \\
  1.60&0.4242&0.5343&5.955(-2)&0.1126&-7.597(-3)&-1.916(-2) \\
  1.70&0.4794&0.5688&7.155(-2)&0.1275&-9.705(-3)&-2.307(-2) \\
  1.80&0.5380&0.6036&8.509(-2)&0.1435&-1.223(-2)&-2.749(-2) \\
  1.90&0.6001&0.6386&0.1003&0.1604&-1.522(-2)&-3.246(-2) \\
  2.00&0.6658&0.6740&0.1172&0.1784&-1.874(-2)&-3.804(-2) \\
  2.10&0.7349&0.7097&0.1360&0.1975&-2.285(-2)&-4.425(-2) \\
  2.20&0.8077&0.7458&0.1567&0.2178&-2.761(-2)&-5.115(-2) \\
  2.30&0.8841&0.7823&0.1796&0.2392&-3.310(-2)&-5.880(-2) \\
  2.40&0.9642&0.8192&0.2046&0.2618&-3.940(-2)&-6.723(-2) \\
  2.50&1.048&0.8568&0.2320&0.2858&-4.658(-2)&-7.653(-2) \\
  2.60&1.136&0.8949&0.2618&0.3111&-5.473(-2)&-8.676(-2) \\
  2.70&1.227&0.9339&0.2943&0.3380&-6.396(-2)&-9.802(-2) \\
  2.75270&1.277&0.9548&0.3125&0.3528&-6.930(-2)&-0.1044 \\
  \end{tabular}
 \end{center}
 \label{table9}
\end{table}%

\begin{table}
\caption{The velocity potentials and their derivatives for $n=1$.
}
 \begin{center}
  \begin{tabular}{l|cccccc}
  \multicolumn{7}{c}{$n=1.0$} \\ \hline
  $\xi$&$\four \phi_2$&$d \four \phi_2/d \xi$&$\five \phi_3$&
  $d \five \phi_3/d \xi$&$\six \phi_4$&$d \six \phi_4/d \xi$ \\ \hline
  0.00&0.000&0.000&0.000&0.000&0.000&0.000 \\
  0.10&9.819(-4)&1.964(-2)&7.695(-6)&2.313(-4)&-5.453(-8)&-2.195(-6) \\
  0.20&3.929(-3)&3.930(-2)&6.150(-5)&9.233(-4)&-8.684(-7)&-1.740(-5) \\
  0.30&8.844(-3)&5.901(-2)&2.077(-4)&2.079(-3)&-4.396(-6)&-5.871(-5) \\
  0.40&1.573(-2)&7.878(-2)&4.927(-4)&3.702(-3)&-1.391(-5)&-1.394(-4) \\
  0.50&2.460(-2)&9.863(-2)&9.636(-4)&5.797(-3)&-3.401(-5)&-2.728(-4) \\
  0.60&3.546(-2)&0.1186&1.668(-3)&8.371(-3)&-7.068(-5)&-4.729(-4) \\
  0.70&4.832(-2)&0.1387&2.654(-3)&1.143(-2)&-1.313(-4)&-7.539(-4) \\
  0.80&6.320(-2)&0.1589&3.971(-3)&1.499(-2)&-2.246(-4)&-1.130(-3) \\
  0.90&8.012(-2)&0.1794&5.668(-3)&1.905(-2)&-3.610(-4)&-1.618(-3) \\
  1.00&9.908(-2)&0.2000&7.799(-3)&2.364(-2)&-5.523(-4)&-2.232(-3) \\
  1.10&0.1201&0.2209&1.041(-2)&2.876(-2)&-8.121(-4)&-2.989(-3) \\
  1.20&0.1433&0.2420&1.357(-2)&3.444(-2)&-1.155(-3)&-3.908(-3) \\
  1.30&0.1685&0.2634&1.732(-2)&4.069(-2)&-1.600(-3)&-5.007(-3) \\
  1.40&0.1960&0.2851&2.173(-2)&4.753(-2)&-2.164(-3)&-6.306(-3) \\
  1.50&0.2256&0.3072&2.685(-2)&5.499(-2)&-2.868(-3)&-7.827(-3) \\
  1.60&0.2574&0.3296&3.275(-2)&6.310(-2)&-3.737(-3)&-9.591(-3) \\
  1.70&0.2915&0.3525&3.949(-2)&7.188(-2)&-4.795(-3)&-1.162(-2) \\
  1.80&0.3279&0.3758&4.714(-2)&8.136(-2)&-6.072(-3)&-1.395(-2) \\
  1.90&0.3667&0.3996&5.579(-2)&9.159(-2)&-7.597(-3)&-1.661(-2) \\
  2.00&0.4079&0.4240&6.549(-2)&0.1026&-9.405(-3)&-1.961(-2) \\
  2.10&0.4515&0.4490&7.633(-2)&0.1145&-1.153(-2)&-2.301(-2) \\
  2.20&0.4977&0.4745&8.841(-2)&0.1272&-1.402(-2)&-2.684(-2) \\
  2.30&0.5464&0.5008&0.1018&0.1408&-1.692(-2)&-3.113(-2) \\
  2.40&0.5979&0.5279&0.1166&0.1555&-2.026(-2)&-3.593(-2) \\
  2.50&0.6520&0.5558&0.1329&0.1712&-2.412(-2)&-4.130(-2) \\
  2.60&0.7090&0.5845&0.1509&0.1880&-2.854(-2)&-4.728(-2) \\
  2.70&0.7690&0.6143&0.1706&0.2061&-3.360(-2)&-5.394(-2) \\
  2.80&0.8319&0.6451&0.1922&0.2255&-3.936(-2)&-6.135(-2) \\
  2.90&0.8980&0.6771&0.2157&0.2463&-4.590(-2)&-6.957(-2) \\
  3.00&0.9674&0.7104&0.2415&0.2687&-5.330(-2)&-7.871(-2) \\
  3.10&1.040&0.7450&0.2695&0.2927&-6.167(-2)&-8.884(-2) \\
  3.14159&1.071&0.7599&0.2819&0.3032&-6.546(-2)&-9.337(-2) \\
  \end{tabular}
 \end{center}
 \label{table10}
\end{table}%

\begin{table}
\caption{The velocity potentials and their derivatives for $n=1.5$.
}
 \begin{center}
  \begin{tabular}{l|cccccc}
  \multicolumn{7}{c}{$n=1.5$} \\ \hline
  $\xi$&$\four \phi_2$&$d \four \phi_2/d \xi$&$\five \phi_3$&
  $d \five \phi_3/d \xi$&$\six \phi_4$&$d \six \phi_4/d \xi$ \\ \hline
  0.00&0.000&0.000&0.000&0.000&0.000&0.000 \\
  0.10&5.484(-4)&1.097(-2)&3.754(-6)&1.129(-4)&-2.308(-8)&-9.293(-7) \\
  0.20&2.195(-3)&2.196(-2)&3.001(-5)&4.507(-4)&-3.677(-7)&-7.372(-6) \\
  0.30&4.943(-3)&3.301(-2)&1.014(-4)&1.016(-3)&-1.863(-6)&-2.490(-5) \\
  0.40&8.798(-3)&4.412(-2)&2.409(-4)&1.812(-3)&-5.902(-6)&-5.923(-5) \\
  0.50&1.377(-2)&5.533(-2)&4.717(-4)&2.844(-3)&-1.446(-5)&-1.162(-4) \\
  0.60&1.987(-2)&6.666(-2)&8.177(-4)&4.118(-3)&-3.010(-5)&-2.021(-4) \\
  0.70&2.711(-2)&7.813(-2)&1.303(-3)&5.641(-3)&-5.603(-5)&-3.232(-4) \\
  0.80&3.550(-2)&8.978(-2)&1.954(-3)&7.422(-3)&-9.613(-5)&-4.865(-4) \\
  0.90&4.507(-2)&0.1016&2.797(-3)&9.473(-3)&-1.550(-4)&-6.993(-4) \\
  1.00&5.583(-2)&0.1137&3.858(-3)&1.181(-2)&-2.379(-4)&-9.695(-4) \\
  1.10&6.782(-2)&0.1260&5.168(-3)&1.443(-2)&-3.510(-4)&-1.306(-3) \\
  1.20&8.105(-2)&0.1386&6.755(-3)&1.737(-2)&-5.015(-4)&-1.717(-3) \\
  1.30&9.555(-2)&0.1515&8.653(-3)&2.064(-2)&-6.972(-4)&-2.213(-3) \\
  1.40&0.1114&0.1648&1.089(-2)&2.425(-2)&-9.473(-4)&-2.806(-3) \\
  1.50&0.1285&0.1784&1.352(-2)&2.824(-2)&-1.262(-3)&-3.506(-3) \\
  1.60&0.1471&0.1925&1.655(-2)&3.262(-2)&-1.653(-3)&-4.328(-3) \\
  1.70&0.1670&0.2070&2.005(-2)&3.741(-2)&-2.132(-3)&-5.285(-3) \\
  1.80&0.1885&0.2219&2.405(-2)&4.266(-2)&-2.715(-3)&-6.394(-3) \\
  1.90&0.2114&0.2374&2.860(-2)&4.838(-2)&-3.416(-3)&-7.671(-3) \\
  2.00&0.2360&0.2535&3.375(-2)&5.462(-2)&-4.255(-3)&-9.136(-3) \\
  2.10&0.2621&0.2701&3.954(-2)&6.140(-2)&-5.250(-3)&-1.081(-2) \\
  2.20&0.2900&0.2874&4.605(-2)&6.878(-2)&-6.425(-3)&-1.271(-2) \\
  2.30&0.3196&0.3054&5.332(-2)&7.679(-2)&-7.802(-3)&-1.487(-2) \\
  2.40&0.3511&0.3241&6.143(-2)&8.547(-2)&-9.409(-3)&-1.732(-2) \\
  2.50&0.3845&0.3436&7.044(-2)&9.488(-2)&-1.128(-2)&-2.008(-2) \\
  2.60&0.4199&0.3639&8.043(-2)&0.1051&-1.344(-2)&-2.318(-2) \\
  2.70&0.4573&0.3851&9.148(-2)&0.1161&-1.592(-2)&-2.666(-2) \\
  2.80&0.4969&0.4073&0.1037&0.1280&-1.878(-2)&-3.057(-2) \\
  2.90&0.5388&0.4305&0.1171&0.1409&-2.205(-2)&-3.493(-2) \\
  3.00&0.5830&0.4547&0.1319&0.1548&-2.579(-2)&-3.981(-2) \\
  3.10&0.6298&0.4801&0.1481&0.1698&-3.003(-2)&-4.523(-2) \\
  3.20&0.6791&0.5066&0.1659&0.1860&-3.485(-2)&-5.127(-2) \\
  3.30&0.7311&0.5343&0.1854&0.2035&-4.031(-2)&-5.797(-2) \\
  3.40&0.7860&0.5634&0.2066&0.2223&-4.647(-2)&-6.539(-2) \\
  3.50&0.8438&0.5938&0.2299&0.2424&-5.341(-2)&-7.360(-2) \\
  3.60&0.9048&0.6256&0.2552&0.2641&-6.122(-2)&-8.266(-2) \\
  3.65375&0.9389&0.6433&0.2697&0.2764&-6.580(-2)&-8.790(-2) \\
  \end{tabular}
 \end{center}
 \label{table11}
\end{table}%

\begin{table}
\caption{The velocity potentials and their derivatives for $n=2$.
}
 \begin{center}
  \begin{tabular}{l|cccccc}
  \multicolumn{7}{c}{$n=2.0$} \\ \hline
  $\xi$&$\four \phi_2$&$d \four \phi_2/d \xi$&$\five \phi_3$&
  $d \five \phi_3/d \xi$&$\six \phi_4$&$d \six \phi_4/d \xi$ \\ \hline
  0.00&0.000&0.000&0.000&0.000&0.000&0.000 \\
  0.10&2.801(-4)&5.604(-3)&1.632(-6)&4.909(-5)&-8.494(-9)&-3.420(-7) \\
  0.20&1.121(-3)&1.123(-2)&1.306(-5)&1.962(-4)&-1.354(-7)&-2.716(-6) \\
  0.30&2.527(-3)&1.689(-2)&4.416(-5)&4.430(-4)&-6.869(-7)&-9.190(-6) \\
  0.40&4.502(-3)&2.261(-2)&1.050(-4)&7.917(-4)&-2.179(-6)&-2.190(-5) \\
  0.50&7.052(-3)&2.842(-2)&2.060(-4)&1.246(-3)&-5.347(-6)&-4.310(-5) \\
  0.60&1.019(-2)&3.432(-2)&3.577(-4)&1.809(-3)&-1.116(-5)&-7.519(-5) \\
  0.70&1.392(-2)&4.034(-2)&5.715(-4)&2.487(-3)&-2.083(-5)&-1.207(-4) \\
  0.80&1.826(-2)&4.650(-2)&8.591(-4)&3.286(-3)&-3.584(-5)&-1.825(-4) \\
  0.90&2.323(-2)&5.283(-2)&1.233(-3)&4.212(-3)&-5.797(-5)&-2.637(-4) \\
  1.00&2.883(-2)&5.933(-2)&1.706(-3)&5.276(-3)&-8.933(-5)&-3.675(-4) \\
  1.10&3.510(-2)&6.605(-2)&2.293(-3)&6.485(-3)&-1.324(-4)&-4.978(-4) \\
  1.20&4.205(-2)&7.300(-2)&3.008(-3)&7.850(-3)&-1.899(-4)&-6.587(-4) \\
  1.30&4.971(-2)&8.020(-2)&3.868(-3)&9.384(-3)&-2.653(-4)&-8.548(-4) \\
  1.40&5.810(-2)&8.767(-2)&4.891(-3)&1.110(-2)&-3.622(-4)&-1.091(-3) \\
  1.50&6.725(-2)&9.545(-2)&6.095(-3)&1.301(-2)&-4.850(-4)&-1.373(-3) \\
  1.60&7.720(-2)&0.1036&7.500(-3)&1.513(-2)&-6.386(-4)&-1.708(-3) \\
  1.70&8.798(-2)&0.1120&9.128(-3)&1.748(-2)&-8.285(-4)&-2.101(-3) \\
  1.80&9.962(-2)&0.1209&1.100(-2)&2.008(-2)&-1.061(-3)&-2.561(-3) \\
  1.90&0.1122&0.1301&1.315(-2)&2.294(-2)&-1.343(-3)&-3.097(-3) \\
  2.00&0.1257&0.1398&1.560(-2)&2.609(-2)&-1.683(-3)&-3.717(-3) \\
  2.10&0.1401&0.1500&1.838(-2)&2.955(-2)&-2.090(-3)&-4.432(-3) \\
  2.20&0.1557&0.1607&2.152(-2)&3.335(-2)&-2.573(-3)&-5.252(-3) \\
  2.30&0.1723&0.1718&2.506(-2)&3.751(-2)&-3.144(-3)&-6.192(-3) \\
  2.40&0.1901&0.1836&2.904(-2)&4.206(-2)&-3.816(-3)&-7.263(-3) \\
  2.50&0.2090&0.1960&3.349(-2)&4.703(-2)&-4.602(-3)&-8.480(-3) \\
  2.60&0.2293&0.2089&3.846(-2)&5.245(-2)&-5.517(-3)&-9.860(-3) \\
  2.70&0.2508&0.2226&4.399(-2)&5.835(-2)&-6.579(-3)&-1.142(-2) \\
  2.80&0.2738&0.2369&5.014(-2)&6.477(-2)&-7.808(-3)&-1.318(-2) \\
  2.90&0.2983&0.2520&5.697(-2)&7.175(-2)&-9.222(-3)&-1.515(-2) \\
  3.00&0.3242&0.2678&6.451(-2)&7.932(-2)&-1.085(-2)&-1.737(-2) \\
  3.10&0.3518&0.2844&7.285(-2)&8.753(-2)&-1.270(-2)&-1.985(-2) \\
  3.20&0.3811&0.3019&8.204(-2)&9.641(-2)&-1.483(-2)&-2.261(-2) \\
  3.30&0.4122&0.3202&9.216(-2)&0.1060&-1.724(-2)&-2.570(-2) \\
  3.40&0.4452&0.3395&0.1033&0.1164&-1.998(-2)&-2.912(-2) \\
  3.50&0.4802&0.3596&0.1155&0.1275&-2.307(-2)&-3.291(-2) \\
  3.60&0.5172&0.3808&0.1288&0.1395&-2.657(-2)&-3.711(-2) \\
  3.70&0.5564&0.4029&0.1434&0.1524&-3.051(-2)&-4.174(-2) \\
  3.80&0.5978&0.4261&0.1593&0.1663&-3.494(-2)&-4.685(-2) \\
  3.90&0.6416&0.4503&0.1767&0.1812&-3.990(-2)&-5.246(-2) \\
  4.00&0.6879&0.4756&0.1956&0.1971&-4.544(-2)&-5.861(-2) \\
  4.10&0.7368&0.5020&0.2161&0.2140&-5.164(-2)&-6.535(-2) \\
  4.20&0.7883&0.5296&0.2384&0.2322&-5.853(-2)&-7.270(-2) \\
  4.30&0.8427&0.5583&0.2626&0.2515&-6.620(-2)&-8.073(-2) \\
  4.35287&0.8726&0.5739&0.2762&0.2622&-7.059(-2)&-8.525(-2) \\
  \end{tabular}
 \end{center}
 \label{table12}
\end{table}%

\begin{table}
\caption{The normalized radii for spherical stars, the orbital separations
at the contact point, the energies, the reduced quadrupole moments, and
the changes in the central density for $n=0.5$, $1$, $1.5$ and $2$.
}
 \begin{center}
  \begin{tabular}{c|cc|ccc|cc}
  $n$&$\xi_1$&$R_c/a_0$&$\t{E}_{self}$&$\t{E}_{point}$&$\t{E}_{quad}$&
  $\b{\bI}_{11}/(Ma_0^2)$&$\d \r_c/\r_c$ \\ \hline
  0.5&2.75270&2.496&-1.111&-0.5000&1.962&0.6691&2.069 \\
  1.0&3.14159&2.426&-1.000&-0.5000&1.040&0.3465&2.280 \\
  1.5&3.65375&2.384&-0.8571&-0.5000&0.5749&0.1915&2.158 \\
  2.0&4.35287&2.358&-0.6667&-0.5000&0.2958&9.861(-2)&1.924
  \\
  \end{tabular}
 \end{center}
 \label{table13}
\end{table}%

\begin{table}
\caption{The individual energies for $n=0.5$, $1$, $1.5$ and $2$.
}
 \begin{center}
  \begin{tabular}{c|cccccccc}
  $n$&$\t{\Pi}_{self}$&$\t{\Pi}_{quad}$&$(\t{W}_{self})_{self}$&
  $(\t{W}_{self})_{quad}$&$(\t{W}_{int})_{point}$&$(\t{W}_{int})_{quad}$&
  $\t{T}_{point}$&$\t{T}_{quad}$ \\ \hline
  0.5&0.2222&-0.9946&-1.333&1.953&-1.000&-2.007&0.5000&3.011 \\
  1.0&0.5000&-1.300&-1.500&1.819&-1.000&-1.040&0.5000&1.559 \\
  1.5&0.8571&-1.437&-1.714&1.724&-1.000&-0.5746&0.5000&0.8620 \\
  2.0&1.333&-1.479&-2.000&1.627&-1.000&-0.2958&0.5000&0.4437 
 \\
  \end{tabular}
 \end{center}
 \label{table14}
\end{table}%

\begin{table}
\caption{The total energy, the total angular momentum and the orbital
angular velocity along the equilibrium sequences of the binary system.
}
 \begin{center}
  \begin{tabular}{l|lll}
  $R/a_0$&$E/(M^2/a_0)$&$J/(M^3a_0)^{1/2}$&$\O/(\pi \r_0)^{1/2}$ \\ \hline
  \multicolumn{4}{c}{$n=0.5$} \\ \hline
    7.0&-1.183&1.871&8.819(-2) \\
    6.5&-1.188&1.803&9.857(-2) \\
    6.0&-1.194&1.733&0.1112 \\ 
    5.5&-1.202&1.659&0.1267 \\
    5.0&-1.211&1.583&0.1462 \\
    4.5&-1.222&1.502&0.1713 \\
    4.0&-1.236&1.418&0.2047 \\
    3.5&-1.253&1.330&0.2508 \\
    3.0&-1.275&1.240&0.3181 \\
    2.5&-1.303&1.152&0.4257 \\
    2.496&-1.303&1.151&0.4269 \\ \hline
  \multicolumn{4}{c}{$n=1$} \\ \hline
    7.0&-1.071&1.871&8.818(-2) \\
    6.5&-1.077&1.803&9.855(-2) \\
    6.0&-1.083&1.732&0.1111 \\ 
    5.5&-1.091&1.659&0.1266 \\
    5.0&-1.100&1.582&0.1461 \\
    4.5&-1.111&1.501&0.1712 \\
    4.0&-1.125&1.416&0.2044 \\
    3.5&-1.142&1.327&0.2501 \\
    3.0&-1.165&1.233&0.3163 \\
    2.5&-1.196&1.136&0.4197 \\
    2.426&-1.201&1.122&0.4400 \\ \hline
  \multicolumn{4}{c}{$n=1.5$} \\ \hline
    7.0&-0.9286&1.871&8.818(-2) \\
    6.5&-0.9341&1.803&9.855(-2) \\
    6.0&-0.9405&1.732&0.1111 \\
    5.5&-0.9480&1.659&0.1266 \\
    5.0&-0.9571&1.582&0.1461 \\
    4.5&-0.9682&1.501&0.1711 \\
    4.0&-0.9820&1.415&0.2043 \\
    3.5&-0.9997&1.325&0.2498 \\
    3.0&-1.023&1.229&0.3154 \\
    2.5&-1.055&1.128&0.4167 \\
    2.384&-1.064&1.104&0.4487 \\ \hline
  \multicolumn{4}{c}{$n=2$} \\ \hline
    7.0&-0.7381&1.871&8.818(-2) \\
    6.5&-0.7436&1.803&9.854(-2) \\
    6.0&-0.7500&1.732&0.1111 \\
    5.5&-0.7576&1.658&0.1266 \\
    5.0&-0.7666&1.581&0.1461 \\
    4.5&-0.7777&1.500&0.1711 \\
    4.0&-0.7916&1.415&0.2042 \\
    3.5&-0.8094&1.324&0.2496 \\
    3.0&-0.8329&1.227&0.3148 \\
    2.5&-0.8654&1.123&0.4150 \\
    2.358&-0.8770&1.092&0.4539 \\
  \end{tabular}
 \end{center}
 \label{table15}
\end{table}%

%%%%%%%%%%%%%%%%%%%%%%%%%%

\newpage
\begin{center}
  {\Large Figures}
\end{center}

%%%%% Fig.1 %%%%%

%%%%% Fig.2 %%%%%

%%%%% Fig.3 %%%%%

%%%%% Fig.4 %%%%%

%%%%% Fig.5 %%%%%

\end{document}